\documentclass[aps,prd,a4paper,twocolumn]{revtex4}

\usepackage{graphicx}
\usepackage{bm}
\usepackage{amsfonts}
\usepackage{amsmath}

\usepackage[russian,ngerman,english]{babel}

\newcommand{\ve}[1]{\mbox{\boldmath$#1$}}

\let\oldbibitem\bibitem
\renewcommand\bibitem[2][]{\oldbibitem{#2}}

\begin{document}

\title{Post-linear metric of a compact source of matter}   

\author{Sven Zschocke}

\affiliation{Institute of Planetary Geodesy - Lohrmann Observatory,
Dresden Technical University, Helmholtzstrasse 10, D-01069 Dresden, Germany}

\begin{abstract}
The Multipolar Post-Minkowskian (MPM) formalism represents an approach for determining the metric density in the exterior 
of a compact source of matter. In the MPM formalism the metric density is given in harmonic coordinates and in terms 
of symmetric tracefree (STF) multipoles. In this investigation, the post-linear metric density of this formalism is used in 
order to determine the post-linear metric tensor in the exterior of a compact source of matter. The metric tensor is given 
in harmonic coordinates and in terms of STF multipoles. The post-linear metric coefficients are associated with an integration  
procedure. The integration of these post-linear metric coefficients is performed explicitly for the case of a stationary source, 
where the first multipoles (monopole and quadrupole) of the source are taken into account. These studies are a requirement for  
further investigations in the theory of light propagation aiming at highly precise astrometric measurements in the solar system,  
where the post-linear coefficients of the metric tensor of solar system bodies become relevant.  
\end{abstract}


\pacs{04.20.−q, 04.25.Nx, 95.10.Jk, 95.30.Sf}

\maketitle


\section{Introduction}\label{Section0}

The field equations of gravity \cite{Einstein1,Einstein2} represent a set of ten coupled non-linear partial differential equations for the ten  
components of the metric tensor $g_{\alpha \beta}$ which governs the geometry of space-time.  
Despite of their complicated mathematical structure, exact solutions of the 
field equations have been obtained for gravitational systems which have a symmetry,  
for instance \cite{Book_Exact_Solutions}: the Schwarzschild solution for a spherically symmetric body \cite{Schwarzschild}, the Kerr solution  
for a spherically symmetric body in uniform rotational motion \cite{Kerr}, the Weyl-Levi-Civita-Erez-Rosen solution for an axially symmetric body
\cite{Weyl1,Weyl2,Levi_Civita,Erez_Rosen,Young_Coulter}, the Reissner-Nordstr\"om solution for an electrically charged spherically symmetric body
\cite{Reissner,Nordstrom}, and the Kerr-Newman solution for an electrically charged spherically symmetric body in uniform rotational motion \cite{Newman}.  
However, for a body of arbitrary shape and inner structure and which can also be in arbitrary rotational motions and oscillations, the field equations 
of gravity can only be solved within some approximation scheme.  

For an asymptotically flat space-time it is convenient to decompose the metric tensor into the flat Minkowski 
metric $\eta_{\alpha \beta}$ and a metric perturbation $h_{\alpha \beta}$,  
\begin{eqnarray}
g_{\alpha \beta}\left(t,\ve{x}\right) = \eta_{\alpha \beta} + h_{\alpha \beta}\left(t,\ve{x}\right).  
\label{expansion_metric_1a}
\end{eqnarray}

\noindent 
The post-Minkowskian scheme is certainly one of the most important approximations in the theory of gravity, which states that for  
weak gravitational fields, $\left|h_{\alpha\beta}\right| \ll 1$, the metric perturbation can be series expanded in powers of the gravitational constant,   
\begin{eqnarray}
h_{\alpha \beta}\left(t,\ve{x}\right) = G^1 h_{\alpha \beta}^{\left({\rm 1PM}\right)}\!\left(t,\ve{x}\right)
+ G^2 h_{\alpha \beta}^{\left({\rm 2PM}\right)}\!\left(t,\ve{x}\right) + {\cal O}\left(G^3\right),  
\nonumber\\ 
\label{expansion_metric_1}
\end{eqnarray}

\noindent
where $h_{\alpha\beta}^{\left({\rm 1PM}\right)}$ is the linear term and $h_{\alpha\beta}^{\left({\rm 2PM}\right)}$ is the post-linear term  
of the metric tensor, and $G^2 |h_{\alpha\beta}^{\left({\rm 2PM}\right)}| \ll G^1 |h_{\alpha\beta}^{\left({\rm 1PM}\right)}| \ll 1$.  

Our motivation to consider the post-linear term in the metric perturbation (\ref{expansion_metric_1}) is triggered by the rapid progress in astrometric science,
which has recently made the impressive advancement from the milli-arcsecond level \cite{Hipparcos,Hipparcos1,Hipparcos2} to the
micro-arcsecond level \cite{GAIA,GAIA1,GAIA2,GAIA4,GAIA_DR2_1} in angular measurements of celestial objects like stars and quasars.  
A prerequisite of astrometric measurements is the precise modeling of the trajectory of a light signal which propagates from the celestial object
through the curved space-time of the solar system towards the observer. And because the trajectory of a light signal depends on the geometry of space-time,  
it becomes obvious why the metric perturbation (\ref{expansion_metric_1}) is of fundamental importance in the theory of light propagation and astrometry.  
Already at the micro-arcsecond level in positional measurements of celestial objects, the linear term of the metric perturbation (\ref{expansion_metric_1})
is not sufficient for modeling the positional observations performed within the solar system  
\cite{Klioner_2003,KlionerKopeikin1992,Article_Zschocke1,AshbyBertotti2010,Teyssandier,Minazzoli2,Deng_2015,Deng_Xie,HBL2014b,Xie_Huang,Zschocke3,Zschocke4,Zschocke5}.  
Meanwhile, there
are several mission proposals aiming at the sub-micro-arcsecond and even the nano-arcsecond scale of accuracy \cite{Gaia_NIR,Theia,NEAT1,NEAT2,NEAT3}.
That is why post-linear effects of the metric perturbation (\ref{expansion_metric_1}) are coming more and more into focus of astronomers and in the theory of
light propagation
\cite{Xu_Wu,Xu_Gong_Wu_Soffel_Klioner,MC2009,2PN_Light_PropagationA,KS,Conference_Cambridge,Talk_Klioner}.

The Multipolar Post-Minkowskian formalism is based on the Landau-Lifschitz formulation of Einstein's theory.  
In this approach, instead of determining directly the metric tensor,  
one operates with the gothic metric density, ${\overline g}^{\alpha\beta} = \sqrt{-g}\,g^{\alpha\beta}$ where $g$ is the determinant of the metric tensor.  
Like in case of the metric tensor, for an asymptotically flat space-time it is appropriate to decompose the gothic metric density into the  
flat Minkowskian metric $\eta^{\alpha\beta}$ and a gothic metric perturbation $\overline{h}^{\alpha \beta}\left(t,\ve{x}\right)$, 
\begin{eqnarray}
\overline{g}^{\alpha\beta}\left(t,\ve{x}\right) = \eta^{\alpha\beta} - \overline{h}^{\alpha \beta}\left(t,\ve{x}\right). 
\label{expansion_metric_2a}
\end{eqnarray}

\noindent
For weak gravitational fields, $|\overline{h}^{\alpha \beta}| \ll 1$, the corresponding post-Minkowskian series expansion  
of the perturbation of the gothic metric density reads as follows,  
\begin{eqnarray}
\overline{h}^{\alpha \beta}\left(t,\ve{x}\right) = G^1 \overline{h}_{\left({\rm 1PM}\right)}^{\alpha\beta}\!\left(t,\ve{x}\right) 
+ G^2 \overline{h}_{\left({\rm 2PM}\right)}^{\alpha\beta}\!\left(t,\ve{x}\right) + {\cal O}\left(G^3\right)\!,  
\nonumber\\ 
\label{expansion_metric_2}
\end{eqnarray}

\noindent
where $\overline{h}_{\left({\rm 1PM}\right)}^{\alpha \beta}$ is the linear term and $\overline{h}_{\left({\rm 2PM}\right)}^{\alpha \beta}$ is 
the post-linear term of the gothic metric, and  
$G^2 |\overline{h}_{\left({\rm 2PM}\right)}^{\alpha \beta}| \ll G^1 |\overline{h}_{\left({\rm 1PM}\right)}^{\alpha \beta}| \ll 1$. 
The knowledge of the contravariant components of the gothic metric perturbation (\ref{expansion_metric_2}) allows to determine  
the covariant components of the metric perturbation (\ref{expansion_metric_1}); relations between the gothic metric and the metric tensor are 
given in Appendix \ref{Appendix3}.  

The Multipolar Post-Minkowskian (MPM) formalism has been developed within a series of 
articles \cite{Blanchet_Damour1,Blanchet_Damour2,Blanchet_Damour3,Blanchet_Damour4,2PN_Metric1,Multipole_Damour_2} and provides a robust framework 
in order to determine the gothic metric perturbation (\ref{expansion_metric_2}) of compact sources of matter. 
In the MPM formalism, the gothic metric density is expressed in terms of 
so-called symmetric and trace-free (STF) multipoles, allowing for arbitrary shape, inner structure, oscillations and rotational motions of the source.    
The MPM approach was mainly intended for theoretical understanding of the generation of gravitational waves  
by some isolated source of matter, like inspiralling binary stars which consist of  
compact objects like black holes or neutron stars. The compact source of matter can of course also be interpreted as some massive solar system body, 
being of arbitrary shape and inner structure, and which can be in arbitrarily rotational motions and oscillations.
 
Within the MPM approach the linear term and the post-linear term of the gothic metric perturbation (\ref{expansion_metric_2}) have been
determined long time ago for the case of a compact source of matter.  
Accordingly, the aim of this investigation is to give the linear and the post-linear term of the metric perturbation (\ref{expansion_metric_1}) for 
a compact source of matter.  

The determination of post-linear metric coefficients involves quite ambitious computations and the results of the MPM approach become rather cumbersome  
already for the very first few multipoles beyond the simple monopole term \cite{2PN_Metric2}.  
However, for many applications, for instance in the theory of light propagation, it is sufficient to consider the stationary case, where 
the gravitational fields generated by the body become time-independent, hence the post-Minkowskian expansion (\ref{expansion_metric_1}) simplifies as follows,  
\begin{eqnarray}
h_{\alpha\beta}\left(\ve{x}\right) = G^1 h_{\alpha \beta}^{\left({\rm 1PM}\right)}\!\left(\ve{x}\right)
+ G^2 h_{\alpha\beta}^{\left({\rm 2PM}\right)}\!\left(\ve{x}\right) + {\cal O}\left(G^3\right). 
\label{expansion_metric_3}
\end{eqnarray}

\noindent
In the stationary case the computations of the MPM formalism are considerably simpler than in the case of time-dependent gravitational fields.  
In the theory of light propagation in the solar system, the impact of post-linear terms of the metric tensor on the light propagation is 
only known for the monopole term, but not for higher multipoles.  
It is, therefore, a further aim of this investigation to determine, in a transparent manner, the post-linear 
metric including the quadrupole structure of a compact source, which can be considered as some massive solar system body.  
 
The manuscript is organized as follows: In Section \ref{Section1} the exact field equations of gravity in harmonic gauge are given.  
The residual harmonic gauge freedom is considered in Section \ref{Section_Gauge}.  
The post-Minkowskian expansion and some fundamental results of the MPM formalism which are relevant for our considerations 
are summarized in Section \ref{Section2} and Section \ref{Section3}.  
The gothic metric density in the linear and post-linear approximation for time-dependent sources is given in Section \ref{Section4}.
The metric tensor in the linear and post-linear approximation for time-dependent sources is given in Section \ref{Section5}. 
Finally, in Section \ref{Section6} the metric tensor in the linear and post-linear approximation is given  
explicitly for the case of a source with time-independent monopole and spin and quadrupole structure. A summary can be found in Section \ref{Section8}.  
The notations as well as details of the calculations are relegated to several Appendices.

\section{The exact field equations of gravity}\label{Section1}  

The field equations of gravity \cite{Einstein1,Einstein2} relate the metric tensor $g_{\alpha\beta}$ of the physical space-time ${\cal M}$ to the  
stress-energy tensor of matter $T_{\alpha\beta}$, which can be written in the following form (\S 17.1 in \cite{MTW}),  
\begin{eqnarray}
R_{\alpha\beta} - \frac{1}{2}\,g_{\alpha\beta}\,R = \frac{8\,\pi\,G}{c^4}\,T_{\alpha\beta}\,, 
\label{exact_field_equations}
\end{eqnarray}
 
\noindent
where $R_{\alpha \beta} = \Gamma^{\rho}_{\alpha\beta,\rho} - \Gamma^{\rho}_{\alpha\rho,\beta} 
+ \Gamma^{\rho}_{\sigma\rho}\,\Gamma^{\sigma}_{\alpha\beta}
- \Gamma^{\rho}_{\sigma\beta}\,\Gamma^{\sigma}_{\alpha\rho}\,$ is the Ricci tensor (cf. Eq.~(8.47) in \cite{MTW}),  
\begin{eqnarray}
\Gamma^{\alpha}_{\mu\nu} = \frac{1}{2}\,g^{\alpha\beta} \left(g_{\beta\mu\,,\,\nu} + g_{\beta\nu\,,\,\mu} - g_{\mu\nu\,,\,\beta}\right),  
\label{Christoffel_Symbols}
\end{eqnarray}

\noindent 
are the Christoffel symbols, and $R = R^{\mu}_{\mu}$ is the Ricci scalar. 

The field equations (\ref{exact_field_equations}) represent a set of ten coup\-led non-linear partial differential equations for the ten components of the 
metric tensor. Because of the contracted Bianchi identities (cf. Eq.~(13.52) in \cite{MTW}) there are only six field equations which are independent  
of each other \cite{Footnote1}.  
These six field equations determine the ten components of the metric tensor up to a coordinate transformation 
which involves four arbitrary functions $x^{\prime\,\mu} = x^{\prime\,\mu}\left(x^{\nu}\right)$. This freedom in choosing the coordinate system  
is called general covariance of the field equations of gravity. 

For practical calculations in celestial mechanics and in the theory of light propagation it is  
very convenient to chose concrete reference systems instead of keeping the covariance of the field equations. 
A powerful tool is to use harmonic coordinates $x^{\mu} = \left(c t,\ve{x}\right)$,             
which are introduced by the harmonic gauge condition \cite{MTW,Thorne,Fock,Kopeikin_Efroimsky_Kaplan,Carroll,Poisson_Lecture_Notes,Will_Wiseman}
(cf. Eq.~(67.02) in \cite{Fock}, Eq.~(5.2a) in \cite{Thorne})
\begin{eqnarray}
\left(\sqrt{- g}\,g^{\alpha \beta}\right)_{,\,\beta} = 0\,, 
\label{harmonic_gauge_condition_1}
\end{eqnarray}

\noindent
where 
\begin{eqnarray}
{\overline g}^{\alpha\beta} = \sqrt{-g}\,g^{\alpha\beta} \,,
\label{gothic_metric}
\end{eqnarray}

\noindent 
is the gothic metric density \cite{MTW,Thorne,Fock,Kopeikin_Efroimsky_Kaplan,Carroll,Poisson_Lecture_Notes,Will_Wiseman}, with  
$g$ being the determinant of the covariant components of the metric tensor.  
It is very useful to operate with the gothic metric density ${\overline g}^{\alpha\beta}$ rather than the metric tensor $g_{\alpha\beta}$, 
because the field equations in harmonic coordinates become considerably simpler in terms of the gothic metric density.  

It should not be surprising that (\ref{harmonic_gauge_condition_1}) is not a general-covariant relation,  
because this condition just selects a specific type of reference system, namely the (class of) harmonic reference systems.  
Although the harmonic coordinate condition (8) is not general-covariant, it is Lorentz-covariant  
in the slightly generalized meaning of linear orthogonal transformations in curvilinear harmonic coordinates \cite{Fock}. 
The choice of harmonic reference systems is in line with the philosophy of general relativity that one may adopt concrete reference systems, 
while observables (coordinate-independent scalars)
are determined as the final step in the calculations.
The harmonic gauge condition (\ref{harmonic_gauge_condition_1}) is called de Donder gauge in honor of its inventor for the exact field equations \cite{Donder}.
The harmonic reference system for the exact field equations has also been introduced independently by Lanczos \cite{Lanczos}, while
the harmonic gauge condition to first order (linearized gravity) was originally introduced by Einstein \cite{Einstein3,Einstein4}
(cf. Eq.~(4) in \cite{Einstein3}, Eq.~(5) in \cite{Einstein4}).

An alternative form for the definition of harmonic coordinates via the gauge condition (\ref{harmonic_gauge_condition_1}) is given by  
the condition (cf. Eq.~(93.03) in \cite{Fock}, Eq.~(3.270) in \cite{Kopeikin_Efroimsky_Kaplan})  
\begin{eqnarray}
\square_g x^{\mu} = 0 \,,  
\label{alternative_gauge_condition}
\end{eqnarray}

\noindent 
where
\begin{eqnarray}
\square_g &=& \frac{1}{\sqrt{-g}}\,\partial_{\alpha} \left(\sqrt{-g}\,g^{\alpha\beta}\right)\partial_{\beta} 
\label{Alembert}
\\
&=& g^{\alpha\beta}\,\partial_{\alpha}\,\partial_{\beta}  
\label{Alembert_in_harmonic_coordinates}
\end{eqnarray}

\noindent
is the covariant d'Alembert operator, in (\ref{Alembert}) given in arbitrary curvilinear four-coordinates,  
while in (\ref{Alembert_in_harmonic_coordinates}) given in terms of harmonic curvilinear four-coordinates. 
It is crucial to realize that the four functions $x^{\mu}$ in (\ref{alternative_gauge_condition}) are just functions, not components of a vector.  
A function which obeys the homogeneous d'Alembert equation, $\square_g f = 0$, is called harmonic function.
That evident similarity is the reason of why coordinates $x^{\mu}$ are called harmonic coordinates. The harmonic four-coordinates 
$\left(ct,\ve{x}\right)$ provide the closest approximation to rectilinear four-coordinates that one can have in curved space-time and that is why 
they are often called Cartesian-like coordinates. 

Besides of the harmonic gauge (\ref{harmonic_gauge_condition_1}) also the decomposition (\ref{expansion_metric_2a}) is used,
which implies that the gothic metric perturbation $\overline{h}^{\alpha\beta}\left(t,\ve{x}\right)$ propagates as dynamical field on the  
flat background space-time ${\cal M}_0$. Then, the exact field equations of gravity (\ref{exact_field_equations}) read  
\cite{MTW,Thorne,Kopeikin_Efroimsky_Kaplan,Poisson_Lecture_Notes,Will_Wiseman}  
(cf. Eq.~(5.2b) in \cite{Thorne}, Eq.~(1.6.1) in \cite{Poisson_Lecture_Notes}, Eqs.~(2.4) - (2.6) in \cite{Will_Wiseman})  
\begin{eqnarray}
\square\,\overline{h}^{\alpha\beta}\left(x\right) = - \frac{16\,\pi\,G}{c^4}\,\left(\tau^{\alpha \beta}\left(x\right) + t^{\alpha \beta}\left(x\right)\right),
\label{Field_Equations_10}
\end{eqnarray}

\noindent
where $x = \left(ct,\ve{x}\right)$ are curvilinear harmonic coordinates on the flat background space-time  
and $\square = \eta^{\mu\nu}\,\partial_{\mu} \partial_{\nu}$ is the flat d'Alembert operator given in terms of these curvilinear  
harmonic coordinates \cite{Footnote2}. The field equations (\ref{Field_Equations_10}) are called Landau-Lifschitz formulation of Einstein's theory of gravity 
in harmonic coordinates.  
The exact field equations of gravity (\ref{exact_field_equations}) are general-covariant, while the exact field equations in harmonic coordinate 
systems (\ref{Field_Equations_10}) are only Lorentz-covariant.  
The terms on the r.h.s. in (\ref{Field_Equations_10}) are given by
\begin{eqnarray}
&& \tau^{\alpha \beta}\left(x\right) = \left(- g\left(x\right)\right)\,T^{\alpha \beta}\left(x\right)\,,
\label{metric_35}
\\
\nonumber\\
&& t^{\alpha \beta}\left(x\right) = \left(- g\left(x\right)\right)\,t_{\rm LL}^{\alpha \beta}\left(x\right) 
\nonumber\\ 
&& \hspace{0.35cm} + \frac{c^4}{16\,\pi\,G}
\left(\overline{h}^{\alpha\mu}_{\;\;\;,\;\nu}\left(x\right)\;\overline{h}^{\beta \nu}_{\;\;\;,\;\mu}\left(x\right) 
- \overline{h}^{\alpha\beta}_{\;\;\;,\;\mu\nu}\left(x\right)\;\overline{h}^{\mu \nu}\left(x\right)\right),
\nonumber\\ 
\label{metric_40}
\end{eqnarray}

\noindent
where $T^{\alpha \beta}$ is the stress-energy tensor of matter, while $t^{\alpha \beta}$ is the stress-energy pseudotensor of the gravitational field,  
and $t_{\rm LL}^{\alpha \beta}$ is the Landau-Lifschitz pseudotensor  
of gravitational field, in explicit form given by Eq.~(20.22) in \cite{MTW} and by Eq.~(101.7) in \cite{Landau_Lifschitz}. 

It has already been emphasized that the usage of the harmonic gauge condition, either in the form (\ref{harmonic_gauge_condition_1}) 
or in the form (\ref{alternative_gauge_condition}), implies the loss of the general covariance. That is why the expressions  
$\tau^{\alpha \beta}$ and $t^{\alpha \beta}$ are not general-covariant tensors, but they are Lorentz-covariant tensors.  
The vanishing of the covariant derivative of stress-energy tensor  
of matter, $T^{\alpha \beta}_{\;\;\;\;\;;\,\beta} = 0$, implies \cite{Thorne,Kopeikin_Efroimsky_Kaplan,MTW,Will_Wiseman,Poisson_Lecture_Notes,Landau_Lifschitz} 
(cf. Eq.~(5.4) in \cite{Thorne}, Eq.~(2.8) in \cite{Will_Wiseman})  
\begin{eqnarray}
&& \hspace{-0.75cm} \left(\tau^{\alpha\beta} + t^{\alpha\beta}\right)_{\,,\,\beta} = 0 \; \Longrightarrow 
\; \left[\left(-g\right) \left(T^{\alpha\beta} + t_{\rm LL}^{\alpha\beta}\right)\right]_{\,,\,\beta} = 0  
\label{Momentum_Conservation}
\end{eqnarray}

\noindent
which represents a local conservation law and admits the formulation of a global conservation law for the four-momentum of the entire gravitational system; 
cf. Eqs.~(20.23a) - (20.23c) in \cite{MTW} or Eqs.~(1.1.7) and (1.2.1) in \cite{Poisson_Lecture_Notes}. 
 
The gravitational system is assumed to be spatially compact, meaning that there exists a three-dimensional sphere of finite radius $R$ which completely  
contains the source of matter, so that the stress-energy tensor of matter $T^{\alpha \beta}\left(t,\ve{x}\right) = 0$ when $\left|\ve{x}\right| > R$.  
Furthermore, the gravitational system is assumed to be isolated, that means flatness of the metric at spatial  
infinity and the constraint of no-incoming gravitational radiation are imposed  
\cite{Fock,Kopeikin_Efroimsky_Kaplan,KlionerKopeikin1992,IAU_Resolution1,Radiation_Condition,Zschocke_Multipole_Expansion},
\begin{eqnarray}
&& \hspace{-0.5cm} \lim_{r \rightarrow \infty \atop t  + \frac{r}{c} = {\rm const}}\,\overline{h}^{\mu \nu}\left(t,\ve{x}\right) = 0\,,
\label{Asymptotic_1}
\\
\nonumber\\
&& \hspace{-0.5cm} \lim_{r \rightarrow \infty \atop t + \frac{r}{c} = {\rm const}}
\left(\frac{\partial}{\partial r} r\,\overline{h}^{\mu \nu}\left(t,\ve{x}\right)
+ \frac{\partial}{\partial ct} \,r\,\overline{h}^{\mu \nu}\left(t,\ve{x}\right)\right) = 0\,,
\label{Asymptotic_2}
\end{eqnarray}

\noindent
where $r = \left|\ve{x}\right|$. These conditions are called Fock-Sommerfeld boundary conditions.  
The formal solution of the exact field equations (\ref{Field_Equations_10}) for an isolated system  
is given by \cite{MTW,Kopeikin_Efroimsky_Kaplan,Poisson_Lecture_Notes,Will_Wiseman} (e.g. Eq.~(36.38) in \cite{MTW}),
\begin{eqnarray}
\overline{h}^{\alpha \beta}\left(t,\ve{x}\right) = - \frac{16\,\pi\,G}{c^4}\,  
\left(\square_{\rm R}^{-1} \left(\tau^{\alpha\beta} + t^{\alpha\beta}\right)\right)\left(t,\ve{x}\right),  
\label{metric_45}
\end{eqnarray}

\noindent
where the inverse d'Alembert operator 
reads \cite{Blanchet_Damour1,Blanchet_Damour2,Blanchet_Damour3,2PN_Metric1,2PN_Metric2,DSX1,DSX2,Book_Gravitational_Waves}  
\begin{eqnarray}
\left(\square^{-1}_{\rm R} f \right)\left(t,\ve{x}\right) =  - \frac{1}{4\,\pi} \int d^3 x^{\prime}\; 
\frac{1}{\left|\ve{x}-\ve{x}^{\prime}\right|}\,f\left(u\,,\,\ve{x}^{\prime}\right).   
\label{Inverse_d_Alembert_1}
\end{eqnarray}

\noindent 
The time of retardation between the source point $\ve{x}^{\prime}$, for instance located inside the source of matter, 
and the field point $\ve{x}$, for instance located outside of matter, is
\begin{eqnarray}
u = t - \frac{\left|\ve{x} - \ve{x}^{\prime}\right|}{c}\,,  
\label{Retarded_Time_2}
\end{eqnarray}

\noindent
where the natural constant $c$ is the speed of gravitational action which equals the speed of light in vacuum \cite{MTW,Fock}. 

The spatial integral in (\ref{metric_45}) runs over the entire three-dimensional space, that means it gets support inside and outside of the matter source,  
because the integrand depends on the metric perturbation which extends to the entire three-dimensional spatial space.  
It should be emphasized that (\ref{Field_Equations_10}) are the exact field equations of gravity and (\ref{metric_45}) 
represents an exact solution of the field equations, because the only requirements to get these equations have been the  
harmonic gauge and the Fock-Sommerfeld boundary conditions. However, the exact solution (\ref{metric_45}) is an implicit  
integro-differential equation, because the metric perturbation appears on both sides of Eq.~(\ref{metric_45}).

\section{The residual gauge freedom}\label{Section_Gauge}  

\begin{figure}[!ht]
\begin{center}
\includegraphics[scale=0.12]{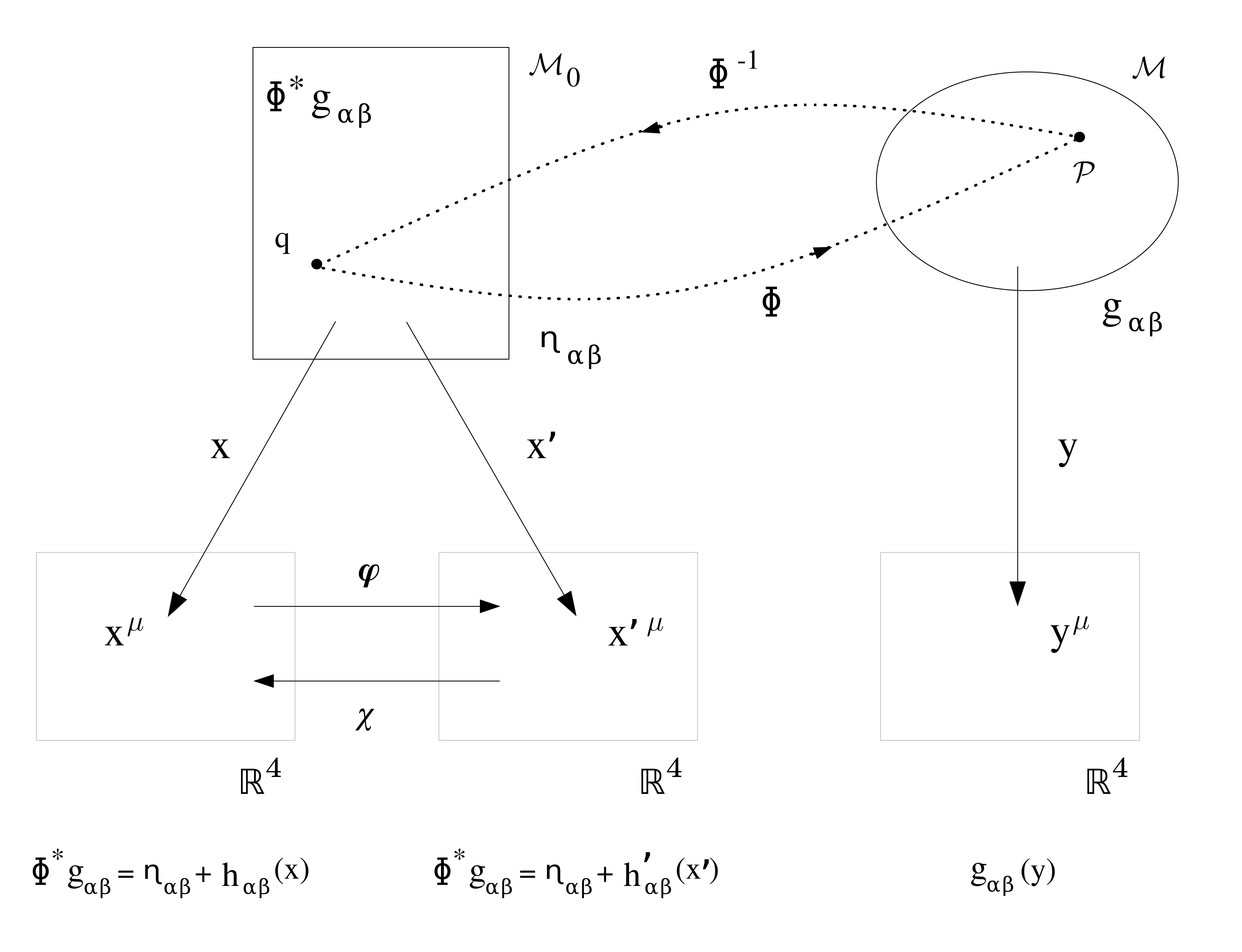}
\end{center}
	\caption{A geometrical representation of the Lorentz-covariant gauge transformation (\ref{harmonic_gauge_condition_2})  
        and its inverse (\ref{harmonic_gauge_condition_inverse}). The physical manifold ${\cal M}$ is covered by coordinates $\{y\}$ and is endowed
        with the metric tensor $g_{\alpha\beta}\left(y\right)$ which is a solution of the exact field equations (\ref{exact_field_equations}).
        The flat background mani\-fold ${\cal M}_0$ is covered by Minkowskian coordinates $\{x\}$ and is endowed
        with the metric tensor $\eta_{\alpha\beta}$.
        The diffeomorphism $\Phi: {\cal M}_0 \rightarrow {\cal M}$ maps the flat background manifold to the physical manifold, e.g. a point
        $q \in {\cal M}_0$ to a point ${\cal P} \in {\cal M}$. Its inverse diffeomorphism $\Phi^{-1}: {\cal M} \rightarrow {\cal M}_0$ maps the physical manifold
        to the flat background manifold, e.g. a point ${\cal P} \in {\cal M}$ to a point $q \in {\cal M}_0$.
        The metric tensor $g_{\alpha\beta}\left(y\right)$ of the physical manifold ${\cal M}$ is pulled back on the flat background manifold ${\cal M}_0$
	(active coordinate transformation as given by Eq.~(A.9) in \cite{Carroll}). The pulled back metric is denoted by $\Phi^{\ast}\,g_{\alpha\beta}$
        and is defined by $g_{\alpha\beta}\left(x\right) = \eta_{\alpha\beta} + h_{\alpha\beta}\left(x\right)$.
        The pulled back metric $g_{\alpha\beta}\left(x\right)$ on ${\cal M}_0$ is physically equivalent to the metric $g_{\alpha\beta}\left(y\right)$
        on ${\cal M}$, that means: if the metric $g_{\alpha\beta}\left(y\right)$ is a solution of the exact field equations (\ref{exact_field_equations}) 
	on the physical manifold
	${\cal M}$, then $h_{\alpha\beta} = \Phi^{\ast}\,g_{\alpha\beta} - \eta_{\alpha\beta}$ (for relations between metric and metric density 
	see Appendix \ref{Appendix3}) will be a solution of the  
        exact field equations (\ref{Field_Equations_10}) in the flat background manifold ${\cal M}_0$ (cf. text below Eq.~(7.51) in \cite{Hawking_Ellis}).  
        The background manifold can also be co\-vered by another harmonic coordinate system $\{x^{\prime} \}$, which is related to the
        Minkowskian coordinate system $\{x\}$ by (\ref{harmonic_gauge_condition_2}) with its inverse (\ref{harmonic_gauge_condition_inverse}).
        The pulled back metric in coor\-dinate system $\{x^{\prime} \}$ is defined by
        $g^{\prime}_{\alpha\beta}\left(x^{\prime}\right)=\eta_{\alpha\beta}+h^{\prime}_{\alpha\beta}\left(x^{\prime}\right)$.
        The relation between these pulled back metric tensors, $g_{\alpha\beta}\left(x\right)$ and $g^{\prime}_{\alpha\beta}\left(x^{\prime}\right)$, 
	in ${\cal M}_0$ is given by (29) where a series expansion of the argument of
	$g^{\prime}_{\alpha\beta}\left(x^{\prime}\right)$ around $x$ has been performed.}  
\label{Diagram1}
\end{figure}

In order to solve the field equations of gravity (\ref{exact_field_equations}) so-called harmonic coordinates have been imposed 
by (\ref{alternative_gauge_condition}) which have simplified the field equations in the form given by (\ref{Field_Equations_10}). 
This coordinate condition (\ref{alternative_gauge_condition}) does not uniquely determine the coordinate system but selects  
a class of infinitely many harmonic reference systems, and permits  
a coordinate transformation from the old harmonic system $\{x^{\alpha}\}$ to a new harmonic system $\{x^{\prime\,\alpha}\}$ 
(cf. Box 18.2 in \cite{MTW} or Eq.~(11.5) in \cite{Thorne} or Eq.~(3.521) in \cite{Kopeikin_Efroimsky_Kaplan}) \cite{Footnote3},  
\begin{eqnarray}
x^{\prime\,\alpha} = x^{\alpha} + \varphi^{\alpha}\left(x\right),  
\label{harmonic_gauge_condition_2}
\end{eqnarray}

\noindent
where $\varphi^{\alpha}\left(x\right)$ is a vector field; see Figure~\ref{Diagram1}. 

The four-coordinates in both systems refer to one and the same point ${\cal P}$ of the 
physical manifold ${\cal M}$, that means $x^{\prime} = x^{\prime}\left({\cal P}\right)$ and $x = x \left({\cal P}\right)$  
denote the four-coordinates in both systems but of one  
and the same point ${\cal P}$ of the physical mani\-fold, which is arbitrary: $\forall\;{\cal P} \in {\cal M}$.  
It is implicitly assumed that the coordinate transformation (\ref{harmonic_gauge_condition_2}) is infinitesimal in the sense that the derivatives
of the functions $\varphi^{\alpha}$ with respect to space and time are of the same order as the metric perturbation,
$\varphi^{\alpha}_{\;,\,\mu} = {\cal O}\left(h^{\alpha}_{\mu}\right)$ hence
$|\varphi^{\alpha}_{\;,\,\mu}| \ll 1$.
 
For later purposes we note the Jacobian matrix of the coordinate transformation (\ref{harmonic_gauge_condition_2}),  
\begin{eqnarray}
A^{\alpha}_{\mu}\left(x\right) = \left(\frac{\partial x^{\prime\,\alpha}}{\partial x^{\mu}}\right)
= \delta^{\alpha}_{\mu} + \varphi^{\alpha}_{\;\;,\,\mu}\left(x\right).  
\label{Jacobi_Matrix}
\end{eqnarray}
 
\noindent 
One may conclude from (\ref{alternative_gauge_condition}) that the coordinate transformation (\ref{harmonic_gauge_condition_2}) preserves 
the harmonic coordinate condition (\ref{alternative_gauge_condition}) if the functions  
$\varphi^{\alpha}$ obey the homogeneous Laplace-Beltrami equation in the old coordinate system $\{x^{\alpha}\}$ 
(cf. Eq.~(3.522) in \cite{Kopeikin_Efroimsky_Kaplan}),  
\begin{eqnarray}
g^{\mu\nu}\left(x\right)\;\varphi^{\alpha}_{\;\;,\,\mu \nu} \left(x\right) = 0\,,   
\label{Laplace_Beltrami_Equation}  
\end{eqnarray}

\noindent 
where $g^{\mu\nu}\left(x\right)$ is the old metric tensor in the old coordinate system. 
The exact field equations of gravity in harmonic gauge (\ref{Field_Equations_10}) are invariant 
under a gauge transformation (\ref{harmonic_gauge_condition_2}) if the functions obey the homogeneous Laplace-Beltrami equation (\ref{Laplace_Beltrami_Equation}).  
The functions $\varphi^{\alpha}$ in (\ref{harmonic_gauge_condition_2}) are nothing more than a change of coordinates and, therefore, they contain no  
physical information about the gravitational system. They are called gauge vector and the 
coordinate transformation (\ref{harmonic_gauge_condition_2}) with (\ref{Laplace_Beltrami_Equation}) is called residual gauge transformation.  
These gauge functions $\varphi^{\alpha}$ are obtained by solving the differential equation (\ref{Laplace_Beltrami_Equation}). 

The coordinate transformation (\ref{harmonic_gauge_condition_2}) is a passively constructed diffeomorphism, that means there is 
a differentiable inverse transformation from a new harmonic system $\{x^{\prime\,\alpha}\}$ to the old harmonic system $\{x^{\alpha}\}$,  
\begin{eqnarray}
x^{\alpha} = x^{\prime\,\alpha} + \chi^{\alpha}\left(x^{\prime}\right), 
\label{harmonic_gauge_condition_inverse}
\end{eqnarray}

\noindent
where $\chi^{\alpha}\left(x^{\prime}\right)$ is a vector field; see Figure~\ref{Diagram1}.  
The four-coordinates in both systems refer to one and the same point ${\cal P}$ of the  
physical manifold ${\cal M}$, that means $x = x\left({\cal P}\right)$ and $x^{\prime} = x^{\prime}\left({\cal P}\right)$ 
denote the four-coordinates in both systems but of one 
and the same point ${\cal P}$ of the physical mani\-fold, which is arbitrary: $\forall\;{\cal P} \in {\cal M}$.

The Jacobian matrix of the inverse coordinate transformation (\ref{harmonic_gauge_condition_inverse}) is given by 
\begin{eqnarray}
B^{\alpha}_{\mu}\left(x^{\prime}\right) = \left(\frac{\partial x^{\alpha}}{\partial x^{\prime\,\mu}}\right)
= \delta^{\alpha}_{\mu} + \chi^{\alpha}_{\;\;,\,\mu}\left(x^{\prime}\right).
\label{Jacobi_Matrix_inverse}
\end{eqnarray}

\noindent
The gauge functions $\chi^{\alpha}$ obey the homogeneous Laplace-Beltrami equation in the new harmonic coordinate system $\{x^{\prime\,\alpha}\}$
\begin{eqnarray}
g^{\prime\,\mu\nu}\left(x^{\prime}\right)\;\chi^{\alpha}_{\;\;,\,\mu \nu} \left(x^{\prime}\right) = 0\,, 
\label{Laplace_Beltrami_Equation_inverse}
\end{eqnarray}
 
\noindent
where $g^{\prime\,\mu\nu}\left(x^{\prime}\right)$ is the new metric tensor in the new harmonic coordinate system.  
The inverse coordinate transformation (\ref{harmonic_gauge_condition_inverse}) is frequently used in the literature. Here it is emphasized  
that the gauge functions $\varphi^{\alpha}\left(x\right)$ in the old harmonic system $\{x^{\alpha}\}$ have to be distinguished from the 
gauge functions $\chi^{\alpha}\left(x^{\prime}\right)$ in the new harmonic system $\{x^{\prime\,\alpha}\}$. However, the gauge-independent terms of the 
metric tensor remain unaffected by a coordinate transformation, that means one is free in choosing either (\ref{harmonic_gauge_condition_2}) 
or (\ref{harmonic_gauge_condition_inverse}), albeit one has to state clearly which of them is used. Here, throughout  
this investigation, the coordinate transformation (\ref{harmonic_gauge_condition_2}) is used and the 
inverse coordinate transformation (\ref{harmonic_gauge_condition_inverse}) will not be applied.  

Let us now consider how the metric tensor and the gothic metric density transform under an infinitesimal gauge transformation (\ref{harmonic_gauge_condition_2}). 

\subsection{The residual gauge transformation of the metric tensor}\label{GaugeA}  

The covariant components of the metric tensor transform as follows \cite{MTW,Thorne,Fock,Kopeikin_Efroimsky_Kaplan,Poisson_Lecture_Notes} 
(e.g. Eq.~(11.10) in \cite{Thorne})  
\begin{eqnarray}
g_{\alpha\beta}\left(x\right)
= \frac{\partial x^{\prime\,\mu}}{\partial x^{\alpha}}\, \frac{\partial x^{\prime\,\nu}}{\partial x^{\beta}}\,
g^{\,\prime}_{\mu\nu}\left(x^{\prime}\right). 
\label{transformation_metric_tensor_A1}
\end{eqnarray}

\noindent
The arguments on the l.h.s. and r.h.s. in Eq.~(\ref{transformation_metric_tensor_A1}) refer to one and the same point ${\cal P}$ of the  
physical manifold ${\cal M}$, that means $x^{\prime} = x^{\prime}\left({\cal P}\right)$ and $x = x \left({\cal P}\right)$ denote the four-coordinates  
in both systems but of one and the same point ${\cal P}$ of the physical manifold, which is arbitrary: $\forall\;{\cal P}\in {\cal M}$.  
By inserting (\ref{harmonic_gauge_condition_2}) into (\ref{transformation_metric_tensor_A1}) and performing a series expansion 
(recall that the residual gauge transformation is infinitesimal) of the metric tensor on the r.h.s.  
around the old coordinates $\{x\}$ of the same point ${\cal P}$ of the physical manifold, one obtains (cf. Eqs.~(11.11a) - (11.11c) in \cite{Thorne})  
\begin{eqnarray}
&& g_{\alpha\beta} = g^{\prime}_{\alpha\beta} + \varphi^{\mu}_{\;\; ,\,\alpha}\;g^{\prime}_{\mu\beta}
+ \varphi^{\nu}_{\;\;,\,\beta}\;g^{\prime}_{\nu\alpha}
+ \varphi^{\mu}_{\;\;,\,\alpha}\,\varphi^{\nu}_{\;\;,\,\beta}\;g^{\prime}_{\mu\nu}
\nonumber\\
&& + \left(\delta^{\mu}_{\alpha} + \varphi^{\mu}_{\;,\,\alpha}\right)\left(\delta^{\nu}_{\beta} + \varphi^{\nu}_{\;,\,\beta}\right)
\sum\limits_{n=1}^{\infty} \frac{1}{n!}\;g^{\prime}_{\mu\nu\,,\,\mu_1 \dots \mu_n} \;\varphi^{\mu_1} \dots \varphi^{\mu_n}\,,
\nonumber\\ 
\label{transformation_metric_tensor_A2}
\end{eqnarray}

\noindent
where all expressions are functions of one and the same argument $x = \left(ct,\ve{x}\right)$. 
It should be noticed that this relation is not general-covariant but Lorentz-covariant, in line with the fact that the general-covariance of the 
field equations (\ref{exact_field_equations}) is lost when they are expressed in harmonic reference systems: the exact field equations (\ref{Field_Equations_10})  
are only Lorentz-covariant. For some reflections about the general-covariant gauge transformation of the metric tensor see Section \ref{GaugeC}.  

The harmonic coordinates $x^{\prime\,\alpha}$ on the l.h.s. in (\ref{harmonic_gauge_condition_2}) are curvilinear harmonic coordinates
in the flat background space-time, while the harmonic coordinates $x^{\alpha}$ on the r.h.s. in (\ref{harmonic_gauge_condition_2}) are
chosen as Minkowskian coordinates in the flat background space-time,
hence the partial derivatives in (\ref{transformation_metric_tensor_A2}) are just flat-space partial derivatives of Minkowskian coordinates.
The partial derivatives in (\ref{transformation_metric_tensor_A2}) would have to be replaced by flat-space covariant derivatives
if one would use curvilinear coordinates in the flat background space-time; cf. text above Eq.~(1.1) in \cite{Thorne},     
text above Eq.~(1.13a) in \cite{Thorne} as well as text below Eqs.~(11.11a) - (11.11c) in \cite{Thorne} and see also
Box 18.2 D in \cite{MTW}. Further mathematical insights can be found in Section 7.1 in \cite{Carroll}.  

The gauge dependent degrees of freedom in (\ref{transformation_metric_tensor_A2}), i.e. all those terms which depend on the gauge functions, are redundant in  
the sense that they have no impact on physical observables. That means, the two different metric tensors $g^{\prime}_{\alpha\beta}\left(x\right)$ and 
$g_{\alpha\beta}\left(x\right)$ in (\ref{transformation_metric_tensor_A2}) describe one and the same gravitational system. Accordingly, the residual gauge freedom  
(\ref{harmonic_gauge_condition_2}) permits to identify and to isolate non-physical degrees of freedom hidden in the old metric tensor 
$g_{\alpha\beta}\left(x\right)$ which allows to arrive at considerably simpler form for the new metric tensor $g^{\prime}_{\alpha\beta}\left(x\right)$.

\subsection{The residual gauge transformation of the metric density}\label{GaugeB} 

The gothic metric (\ref{gothic_metric}) is a tensor density of the weight $w = - 1$ and its contravariant components transform as follows 
\cite{MTW,Thorne,Fock,Kopeikin_Efroimsky_Kaplan,Poisson_Lecture_Notes,Weinberg,Wald} (e.g. Eq.~(4.4.4) in \cite{Weinberg}),  
\begin{eqnarray}
{\overline g}^{\prime\,\alpha\beta}\left(x^{\prime}\right) = \frac{1}{\left| J\left(x\right) \right|}\;  
\frac{\partial x^{\prime\,\alpha}}{\partial x^{\mu}}\, \frac{\partial x^{\prime\,\beta}}{\partial x^{\nu}}\, 
{\overline g}^{\mu\nu}\left(x\right)\,,   
\label{transformation_gothic_metric_5}
\end{eqnarray}

\noindent
where $J\left(x\right)$ is the determinant of the Jacobi matrix (\ref{Jacobi_Matrix}),  
\begin{eqnarray}
J = {\rm det}\left(A^{\alpha}_{\mu}\right) = {\rm e}^{{\rm Tr} \left(\ln {\rm A}^{\alpha}_{\mu}\right)}\;,  
\label{Jacobian_Determinant}
\end{eqnarray}

\noindent
where the second relation in (\ref{Jacobian_Determinant}) is a theorem which allows to compute the determinant \cite{Feynman} and which can be  
proven by Schur's matrix decomposition. One obtains 
\begin{eqnarray}
\frac{1}{\left|J\right|} &=& 1 - \varphi^{\sigma}_{\;,\,\sigma} + \frac{1}{2}\,\varphi^{\sigma}_{\;,\,\omega}\,\varphi^{\omega}_{\;,\,\sigma} 
+ \frac{1}{2}\,\varphi^{\sigma}_{\;,\,\sigma}\,\varphi^{\omega}_{\;,\,\omega} + {\cal O} \left(\varphi^3\right),  
\nonumber\\ 
\label{Jacobian_Determinant_2}
\end{eqnarray}

\noindent 
which is sufficient four our investigations in the post-linear approximation.  
The arguments on the l.h.s. and r.h.s. in  
Eq.~(\ref{transformation_gothic_metric_5}) refer to one and the same point ${\cal P}$ of the physical manifold ${\cal M}$, that means  
$x^{\prime} = x^{\prime}\left({\cal P}\right)$ and $x = x \left({\cal P}\right)$ denote the four-coordinates in both systems but  
of one and the same point ${\cal P}$ of the physical manifold, which is arbitrary: $\forall\;{\cal P}\in {\cal M}$.
By substituting (\ref{harmonic_gauge_condition_2}) into (\ref{transformation_gothic_metric_5}) and performing a series expansion 
(recall that the residual gauge transformation is infinitesimal) of the gothic metric on the l.h.s. around the old coordinates $\{x\}$, one obtains  
\begin{eqnarray}
&& {\overline g}^{\prime\,\alpha\beta} = \frac{1}{\left| J \right|}
\left({\overline g}^{\alpha\beta} \!+ \varphi^{\alpha}_{\;\; ,\,\mu}\;{\overline g}^{\mu\beta} \!+ \varphi^{\beta}_{\;\; ,\,\nu}\;{\overline g}^{\nu\alpha}
\! + \varphi^{\alpha}_{\;\; ,\,\mu}\;\varphi^{\beta}_{\;\; ,\,\nu}\;{\overline g}^{\mu\nu}\right)
\nonumber\\
&& \hspace{1.25cm} - \sum\limits_{n=1}^{\infty} \frac{1}{n!}\;{\overline g}^{\prime\,\alpha\beta}_{\;\;\;\;\;\; ,\, \mu_1 \dots \mu_n}\; \varphi^{\mu_1} \dots
\varphi^{\mu_n}\,,
\label{transformation_gothic_metric_10}
\end{eqnarray}

\noindent
where all expressions are functions of one and the same argument $x = \left(ct,\ve{x}\right)$.
It should be noticed that this relation is not general-covariant but Lorentz-covariant, in line with the fact that the exact field equations  
in harmonic coordinates (\ref{Field_Equations_10}) are only Lorentz-covariant; for some comments about the general-covariant gauge transformation   
see Section \ref{GaugeC}. The reason of why there are flat-space partial derivatives of Minkowskian coordinates in (\ref{transformation_gothic_metric_10})
is the same as described in the text below Eq.~(\ref{transformation_metric_tensor_A2}).  

Like in case of the metric tensors, the old gothic metric ${\overline g}^{\alpha\beta}\left(x\right)$ and the new gothic metric  
${\overline g}^{\prime\,\alpha\beta}\left(x\right)$ in (\ref{transformation_gothic_metric_10}) describe one and the same gravitational system; 
cf. text below Eq.~(7.14) in \cite{Carroll} and Theorem 4.5 in \cite{Blanchet_Damour1}.  
The gauge dependent degrees of freedom are redundant in the sense that the gauge terms in (\ref{transformation_gothic_metric_10}) have no impact
on physical observables. Nevertheless, the gauge-dependent terms have to be treated carefully because they allow to transform the
old gothic metric density into a considerably simpler form.

\subsection{Some comments on the general-covariant gauge transformation}\label{GaugeC}  

The gauge transformation considered above in Sections \ref{GaugeA} and \ref{GaugeB} 
is Lorentz-covariant and can therefore be expressed in terms of partial derivatives. A general-covariant gauge transformation  
must necessarily be given in terms of Lie derivatives $\pounds_{\xi}$ acting on the metric tensor along a vector field $\xi^{\mu}$ 
which is a general-covariant differential operation \cite{Footnote4}.  
Such a general-covariant gauge transformation has been developed during  
the last two decades \cite{Gauge1,Gauge2,Gauge3,Gauge4,Gauge5,Gauge6,Gauge7,Gauge8,Gauge9,Gauge10,Gauge11,Gauge12,Gauge13,Gauge14}. 
It might be constructive to make some comments about the general-covariant gauge transformation and its relation to the Lorentz-covariant residual  
gauge transformation considered in Sections \ref{GaugeA} and \ref{GaugeB}.   
\begin{figure}[!ht]
\begin{center}
\includegraphics[scale=0.12]{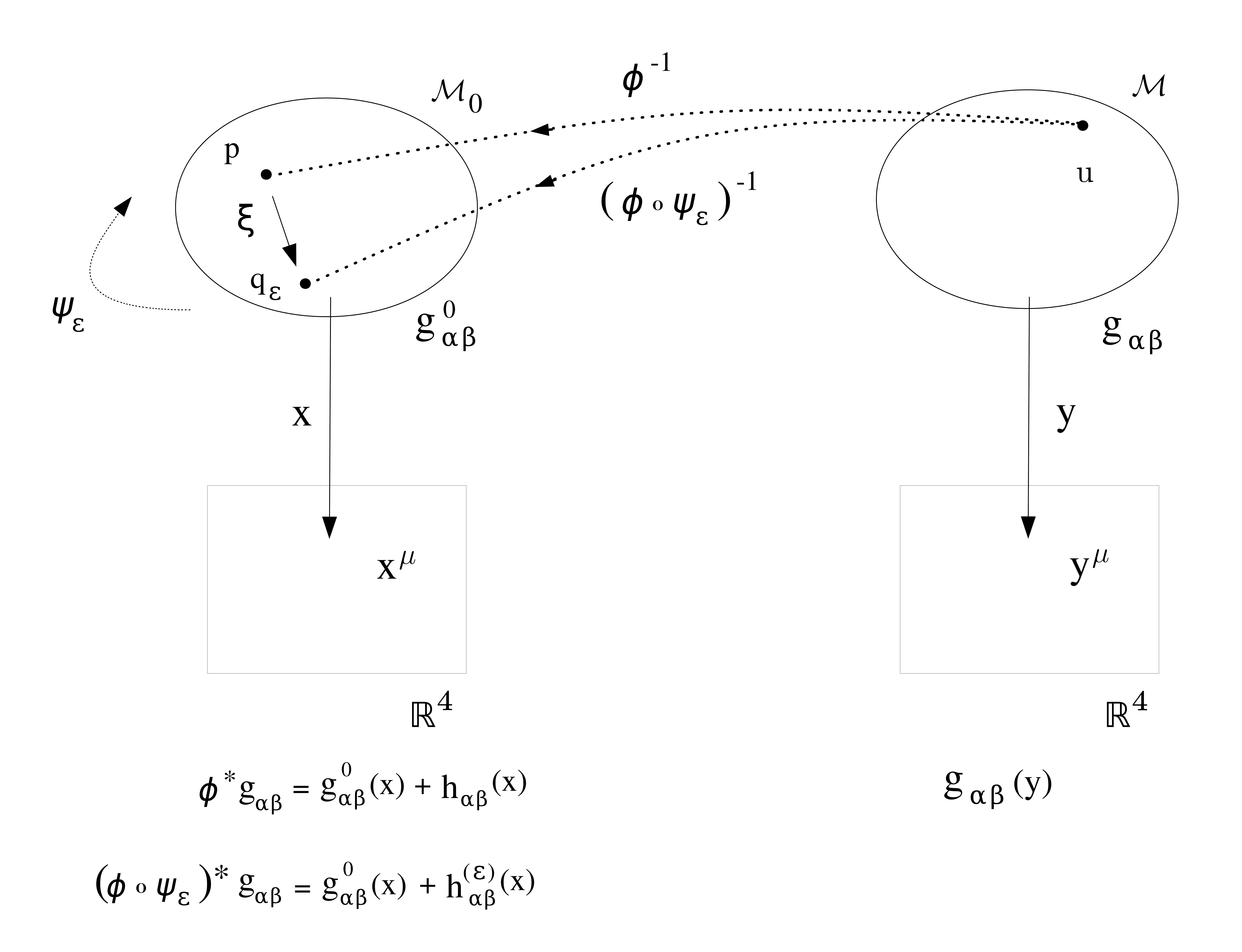}
\end{center}
	\caption{A geometrical representation of the general-covariant gauge transformation.
        The physical manifold ${\cal M}$ is covered by coordinates $\{y\}$ and endowed with the metric $g_{\alpha\beta}\left(y\right)$
        which is a solution of the exact field equations (\ref{exact_field_equations}).
        The curved background manifold ${\cal M}_0$ is covered by coordinates $\{x\}$ and endowed with the metric $g^0_{\alpha\beta}\left(x\right)$.
        The diffeomorphism $\phi: {\cal M}_0 \rightarrow {\cal M}$ (not shown in the diagram) maps the curved background mani\-fold to the physical manifold,
        e.g. a point $p \in {\cal M}_0$ to a point $u \in {\cal M}$.
        The inverse diffeomorphism $\phi^{-1}: {\cal M} \rightarrow {\cal M}_0$ maps the physical manifold to the curved background manifold, e.g.
        a point $u \in {\cal M}$ to a point $p \in {\cal M}_0$.
        The metric $g_{\alpha\beta}\left(y\right)$ of the physical manifold ${\cal M}$ is pulled back on the curved background manifold ${\cal M}_0$
        (active coordinate transformation as given by Eq.~(A.9) in \cite{Carroll}). The pulled back metric is denoted by $\phi^{\ast}\,g_{\alpha\beta}$ and
        defined by $g_{\alpha\beta}\left(x\right) = g^0_{\alpha\beta}\left(x\right) + h_{\alpha\beta}\left(x\right)$.
        The pulled back metric $g_{\alpha\beta}\left(x\right)$ on ${\cal M}_0$ is physically equivalent to the metric $g_{\alpha\beta}\left(y\right)$
        on ${\cal M}$, that means: if the metric $g_{\alpha\beta}\left(y\right)$ is a solution of the exact field equations (6) on the physical manifold
        ${\cal M}$, then $h_{\alpha\beta} = \phi^{\ast}\,g_{\alpha\beta} - g^0_{\alpha\beta}$  will be a solution of the
        exact field equations in the curved background manifold ${\cal M}_0$.
	The set of diffeomorphisms $\Phi_{\epsilon}^{-1} \equiv \left(\phi \circ \psi_{\epsilon}\right)^{-1}$ maps the same point $u \in {\cal M}$ of the
        physical space-time ${\cal M}$ to a set of points $q_{\epsilon} \in {\cal M}_0$ of the curved background space-time ${\cal M}_0$, where
        $\psi_{\epsilon}$ represents a family of diffeomorphisms which are distinguished by the parameter $\epsilon$ and which are acting on the
        curved background manifold and generated by a gauge vector field $\xi$.
        The composition of the diffeomorphisms $\psi_{\epsilon}$ with $\phi$ implies a family of pulled back metric tensors
	$\Phi_{\epsilon}^{\ast} g_{\alpha\beta} \equiv \left(\phi \circ \psi_{\epsilon}\right)^{\ast} g_{\alpha\beta}$ which reads
        $g^{(\epsilon)}_{\alpha\beta}\left(x\right) = g^0_{\alpha\beta}\left(x\right) + h^{(\epsilon)}_{\alpha\beta}\left(x\right)$ in the same chart $\{x \}$.
        The pulled back metric tensors $g_{\alpha\beta}\left(x\right)$ and $g^{(\epsilon)}_{\alpha\beta}\left(x\right)$ in ${\cal M}_0$ are related by a
        gauge transformation which can be expressed in terms of multiple Lie derivatives of $g_{\alpha\beta}\left(x\right)$ in the direction of the vector field  
	$\xi$, that means $\displaystyle g^{(\epsilon)}_{\alpha\beta}\left(x\right) = \sum\limits_{n=0}^{\infty} \frac{\epsilon^n}{n!} \, 
	{\cal L}^n_{\xi}\,g_{\alpha\beta}\left(x\right)$.}  
\label{Diagram2}
\end{figure}
In the investigations \cite{Gauge1,Gauge2,Gauge3,Gauge4,Gauge5,Gauge6,Gauge7,Gauge8,Gauge9,Gauge10,Gauge11,Gauge12,Gauge13,Gauge14}  
the metric tensor is separated in the form $g_{\alpha\beta} = g_{\alpha\beta}^0 + h_{\alpha\beta}$, which generalizes (\ref{expansion_metric_1a}) because  
the background metric $g_{\alpha\beta}^0$ of the curved background manifold ${\cal M}_0$ 
is not simply the flat Minkowskian metric, but can be the Schwarzschild metric or the Kerr metric 
or the Friedmann-{Lema{\^\i}tre-Robertson-Walker metric or some other curved space-time. The dynamical degrees of freedom, $h_{\alpha\beta}$, are governed by  
field equations which are obtained by inserting the decomposition $g_{\alpha\beta} = g_{\alpha\beta}^0 + h_{\alpha\beta}$ into Einstein's 
equations (\ref{exact_field_equations}) and describe a tensorial field $h_{\alpha\beta}$ which propagates in the 
curved background space-time ${\cal M}_0$ endowed with the background metric $g_{\alpha\beta}^0$.  

The general-covariant formalism distinguishes between the physical manifold ${\cal M}$ covered by four-coordinates $y^{\alpha}$ and endowed with 
metric $g_{\alpha\beta}$, the background manifold ${\cal M}_0$ covered by four-coordinates $x^{\alpha}$ and endowed with background metric $g_{\alpha\beta}^0$, 
and a diffeomorphism and inverse diffeomorphism between these manifolds, namely  
$\phi: {\cal M}_0 \rightarrow {\cal M}$ and $\phi^{-1}: {\cal M} \rightarrow {\cal M}_0$, respectively; see Figure~\ref{Diagram2}. 
The diffeomorphism $\phi$ maps each point $p \in {\cal M}_0$ to another point $u \in {\cal M}$ and, vice versa, the inverse  
diffeomorphism $\phi^{-1}$ maps each point $u \in {\cal M}$ to another point $p \in {\cal M}_0$ (cf. Figure 7.1 in \cite{Carroll}). 
The diffeomorphism allows to pull back the metric tensor $g_{\alpha\beta}$ from ${\cal M}$ to ${\cal M}_0$ which is given by an active coordinate transformation:  
$\displaystyle \phi^{\ast}\,g_{\alpha\beta}\!\left(x\right) =  
\frac{\partial y^{\mu}}{\partial x^{\alpha}}\,\frac{\partial y^{\nu}}{\partial x^{\beta}}\,g_{\mu\nu}\! \left(y\right)$ (cf. Eq.~(A.9) in \cite{Carroll}). 
The metric $g_{\alpha\beta}$ in ${\cal M}$ and the pulled back metric $\phi^{\ast}\,g_{\alpha\beta}$ in ${\cal M}_0$ are physically equivalent 
(cf. Section 7.1. in \cite{Carroll} and Section 7.1 in \cite{Hawking_Ellis}) and the metric perturbation is defined in the background manifold as follows:   
$h_{\alpha\beta}\!\left(x\right) = \phi^{\ast}\,g_{\alpha\beta}\!\left(x\right) - g^0_{\alpha\beta}\left(x\right)$ (cf. Eq.~(7.10) in \cite{Carroll}). 

Furthermore, the general-covariant approach of gauge transformations considers a family of  
actively constructed diffeomorphisms acting on the background manifold, $\psi_{\epsilon} : {\cal M}_0 \rightarrow {\cal M}_0$, which maps each point  
$p \in {\cal M}_0$ to another point $q \in {\cal M}_0$ (cf. Figure 7.2 in \cite{Carroll}). 
These diffeomorphisms are distinguished from each other by some parameter $\epsilon$ and they are generated by a vector field ${\xi}^{\mu}\!\left(x\right)$ 
acting on the background manifold; for explicit expressions cf. Eq.~(2.18) in \cite{Gauge8} or Eq.~(2.56) in \cite{Gauge13}. The composition of the  
family of diffeomorphisms $\psi_{\epsilon}$ with the diffeomorphism $\phi$, that is $\Phi_{\epsilon} = \phi \circ \psi_{\epsilon}$, leads to a family 
of diffeomorphisms $\Phi_{\epsilon} : {\cal M}_0 \rightarrow {\cal M}$ and its inverse $\Phi^{-1}_{\epsilon} : {\cal M} \rightarrow {\cal M}_0$.  
This family of diffeomorphisms allows to pull back the metric tensor from the physical manifold to the background manifold which implies  
a family of metric perturbations defined on the background manifold,  
$h^{\left(\epsilon\right)}_{\alpha\beta}\!\left(x\right) = \Phi_{\epsilon}^{\ast}\,g_{\alpha\beta}\!\left(x\right) - g^0_{\alpha\beta}\left(x\right)$ 
(cf. Eq.~(7.11) in \cite{Carroll}). The dependence of the metric perturbation on the parameter $\epsilon$, that reflects the dependence of the 
metric perturbation on the vector field ${\xi}^{\mu}\!\left(x\right)$, is called gauge freedom: each member of the family of metric perturbations 
$h^{\left(\epsilon\right)}_{\alpha\beta}$ is physically isometric (physically equivalent) to each other and any of them describes the same physical system
(i.e. all observables are unchanged).

This geometrical approach leads in a natural way to the gauge transformation of the metric tensor in terms of multiple Lie derivatives acting on  
the metric tensor along the gauge vector \cite{Gauge1,Gauge2,Gauge3,Gauge4,Gauge5,Gauge6,Gauge7,Gauge8,Gauge9,Gauge10,Gauge11,Gauge12,Gauge13,Gauge14}.  
The active coordinate transformation can be rewritten in terms of  
a passive coordinate transformation which relates the four-coordinates of one and the same point $q \in {\cal M}_0$ of the background manifold,
$x^{\mu}\left(q\right)$ and $x^{\prime \mu}\left(q\right)$, in two different charts of the background manifold ${\cal M}_0$;
explicit calculations and expressions up to the third-order of the perturbation theory are given, for instance, in \cite{Gauge13}.  
In this way one arrives at a general-covariant gauge transformation of the metric tensor in terms of Lie derivatives by means of a passive
coordinate transformation.

In order to make a bridge between the general-covariant approach 
in \cite{Gauge1,Gauge2,Gauge3,Gauge4,Gauge5,Gauge6,Gauge7,Gauge8,Gauge9,Gauge10,Gauge11,Gauge12,Gauge13,Gauge14} and the Lorentz-covariant approach 
described in Sections \ref{GaugeA} and \ref{GaugeB}, one would have to assume a flat background metric, $g_{\alpha\beta}^0 = \eta_{\alpha\beta}$, and one would  
have to use harmonic reference systems as well as to impose the Laplace-Beltrami condition (\ref{Laplace_Beltrami_Equation}) for the gauge vector.  
In this way one would finally arrive at a Lorentz-covariant residual gauge  
transformation in terms of Lie derivatives and based on passive coordinate transformations. But it should be emphasized that the results of such 
an approach would not differ from the Lorentz-covariant residual gauge transformation considered above, because the physical content of the gravitational  
system is comprised in the gauge-independent metric perturbation and, therefore, is independent of whether the Lorentz-covariant gauge transformation   
in terms of partial derivatives or in terms of Lie derivatives is applied. 
Here, the Lorentz-covariant residual gauge transformation in terms of Lie derivatives will 
not further be exposed, because the Multipolar Post-Minkowskian (MPM) formalism makes use of relations (\ref{transformation_metric_tensor_A2}) 
and (\ref{transformation_gothic_metric_10}) and uses   
the series expansion (\ref{harmonic_gauge_condition_7}) which, subject to the Laplace-Beltrami equation (\ref{Laplace_Beltrami_Equation}), results   
in the sequence of differential equations (\ref{sequence_gauge_1PM}) - (\ref{sequence_gauge_nPM}) for the gauge functions which will be used in what follows.

\section{The Post-Minkowskian expansion and gauge transformation}\label{Section2}  

\subsection{The Post-Minkowskian expansion of the metric tensor}\label{Section2_1}  

In the weak-field regime the old metric tensor $g_{\alpha \beta}$ in the old harmonic system $\{x^{\alpha}\}$ can be expanded 
in powers of the gravitational constant, 
\begin{eqnarray}
g_{\alpha \beta}\left(x\right) = \eta_{\alpha\beta}
+ \sum\limits_{n=1}^{\infty} G^n\,h_{\alpha \beta}^{\left({\rm n PM}\right)}\left(x\right),  
\label{expansion_metric_PM}
\end{eqnarray}

\noindent  
which is called post-Minkowskian (PM) expansion. Each individual term $h_{\alpha \beta}^{\left({\rm n PM}\right)}$ is invariant under 
Lorentz transformations; cf. text below Eq.~(3.527) in \cite{Kopeikin_Efroimsky_Kaplan}.  
The residual gauge transformation (\ref{harmonic_gauge_condition_2}) from the old harmonic coordinate system $\{x^{\alpha}\}$
to a new harmonic coordinate system $\{x^{\prime\,\alpha}\}$ is assumed to admit a series expansion in powers of the gravitational constant
(cf. Eq.~(4.23) in \cite{Blanchet_Damour1}),
\begin{eqnarray}
x^{\prime\,\alpha} = x^{\alpha} + \sum\limits_{n=1}^{\infty} G^n\, \varphi^{\alpha\,\left({\rm n PM}\right)}\left(x\right),
\label{harmonic_gauge_condition_7}
\end{eqnarray}

\noindent
where $\varphi^{\alpha\,\left({\rm n PM}\right)}_{\;\;\;,\,\beta} = {\cal O} \left(h^{\alpha\,\left({\rm n PM}\right)}_{\beta}\right)$ 
and each individual term $\varphi^{\alpha\,\left({\rm n PM}\right)}$ is a Lorentz four-vector.  
In what follows the total sum $\varphi^{\alpha}$ is  
called gauge vector, while the individual terms $\varphi^{\alpha\,\left({\rm n PM}\right)}$ are called gauge functions.
These gauge functions $\varphi^{\alpha\;\left({\rm n PM}\right)}$ to any order of the
perturbation theory are governed by a sequence of equations which are given below by Eqs.~(\ref{sequence_gauge_1PM}) - (\ref{sequence_gauge_nPM}).

The coordinate transformation (\ref{harmonic_gauge_condition_7}) transforms the old metric tensor (\ref{expansion_metric_PM}) in the old harmonic system  
$\{x^{\alpha}\}$ to the new (primed) metric tensor in the new harmonic system $\{x^{\prime\,\alpha}\}$, and its post-Minkowskian expansion reads  
\begin{eqnarray}
g^{\prime}_{\alpha \beta}\left(x^{\prime}\right) = \eta_{\alpha\beta}
+ \sum\limits_{n=1}^{\infty} G^n\,h_{\alpha \beta}^{\,\prime\,\left({\rm n PM}\right)}\left(x^{\prime}\right).
\label{expansion_metric_PM_Prime}
\end{eqnarray}

\noindent 
By inserting the post-Minkowskian expansions (\ref{expansion_metric_PM}) - (\ref{expansion_metric_PM_Prime}) into  
(\ref{transformation_metric_tensor_A1}) and performing a series expansion of (\ref{expansion_metric_PM_Prime}) around the 
four-coordinates $x^{\alpha}$, one arrives at the post-Minkowskian expansion of the gauge transformation of the metric perturbation,    
\begin{eqnarray}
&& \sum\limits_{n=1}^{\infty} G^n h^{\left({\rm n PM}\right)}_{\alpha \beta}  
\!=\! \sum\limits_{n=1}^{\infty} G^n \!\left(h^{\,\prime\,\left({\rm n PM}\right)}_{\alpha \beta}
\! + \partial \varphi^{\left({\rm n PM}\right)}_{\alpha\beta} \! + \Omega^{\left({\rm n PM}\right)}_{\alpha\beta}\right),   
\nonumber\\ 
\label{gauge_transformation_metric_PM}
\end{eqnarray}

\noindent
where all terms are given in the harmonic system $\{x\}$. 
The equation (\ref{gauge_transformation_metric_PM}) is nothing else than equation (\ref{transformation_metric_tensor_A2}) 
expressed in terms of a series expansion in powers of the gravitational constant.  
 
The gauge terms $\partial {\varphi}^{\left({\rm n PM}\right)}_{\alpha\beta}$ have the following structure,  
\begin{eqnarray}
&& \hspace{-0.5cm} \partial {\varphi}^{\left({\rm n PM}\right)}_{\alpha\beta} 
= \varphi^{\mu\;\left({\rm n PM}\right)}_{\,,\,\alpha}\,\eta_{\mu\beta}
+ \varphi^{\mu\;\left({\rm n PM}\right)}_{\,,\,\beta}\,\eta_{\mu\alpha}\,,
\label{gauge_term_metric_PM}
\end{eqnarray}

\noindent
which are called {\it linear gauge terms} since they are linear in the gauge functions.  
The gauge terms $\Omega^{\left({\rm n PM}\right)}_{\alpha\beta}$ are  
called {\it non-linear gauge terms} since they contain either products of gauge functions or products of gauge functions and metric perturbations. 
One may obtain a closed expression for $\Omega^{\left({\rm n PM}\right)}_{\alpha\beta}$ from Eq.~(\ref{transformation_metric_tensor_A2}) 
and using Eqs.~(\ref{expansion_metric_PM}) and (\ref{harmonic_gauge_condition_7}).  
Here it is sufficient to consider only the first two orders, given by  
\begin{eqnarray}
&& \hspace{-0.75cm} \Omega^{\left({\rm 1PM}\right)}_{\alpha\beta} = 0\,, 
\label{gauge_term_metric_1PM} 
\\
\nonumber\\
&& \hspace{-0.75cm} \Omega^{\left({\rm 2PM}\right)}_{\alpha\beta} = 
h^{\prime\,\left({\rm 1PM}\right)}_{\mu\beta}\,\varphi^{\mu\,{\left({\rm 1PM}\right)}}_{\,,\,\alpha}  
+ h^{\prime\,\left({\rm 1PM}\right)}_{\mu\alpha}\,\varphi^{\mu\,{\left({\rm 1PM}\right)}}_{\,,\,\beta} 
\nonumber\\ 
\nonumber\\ 
&& \hspace{0.5cm} + h^{\prime\,\left({\rm 1PM}\right)}_{\alpha\beta\,,\,\nu}\,\varphi^{\nu\,\left({\rm 1PM}\right)}  
+ \varphi^{\mu\,{\left({\rm 1PM}\right)}}_{\,,\,\alpha}\,\varphi^{\nu\,{\left({\rm 1PM}\right)}}_{\,,\,\beta}\,\eta_{\mu\nu}\,,  
\label{gauge_term_metric_2PM} 
\end{eqnarray}

\noindent
while the higher orders $n \ge 3$ are not relevant for our investigations. 
The linear 1PM term $\partial \varphi^{\left({\rm 1PM}\right)}_{\alpha\beta}$ is in agreement with Eq.~(21) in \cite{Gauge2}, while
the linear 2PM term $\partial \varphi^{\left({\rm 2PM}\right)}_{\alpha\beta}$ and the non-linear 2PM term $\Omega^{\left({\rm 2PM}\right)}_{\alpha\beta}$
are in agreement with Eq.~(22) in \cite{Gauge2} (to verify that agreement one has to adopt a flat background metric in \cite{Gauge2}).

\subsection{The Post-Minkowskian expansion of the gothic metric density}\label{Section2_2}  

The weak-field regime admits a series expansion of the gothic metric in powers of the gravitational constant 
\cite{Blanchet_Damour1,2PN_Metric1,Thorne,Kopeikin_Efroimsky_Kaplan,Poisson_Lecture_Notes} 
(cf. Eq.~(1.1) in \cite{Blanchet_Damour1}, Eq.~(9.5) in \cite{Thorne}), 
\begin{eqnarray}
\overline{g}^{\alpha \beta}\left(x\right) = 
\eta^{\alpha \beta} - \sum\limits_{n=1}^{\infty} G^n\,\overline{h}^{\alpha \beta}_{\left({\rm n PM}\right)}\left(x\right),  
\label{expansion_PM}
\end{eqnarray}

\noindent
which is called the post-Minkowskian expansion of the gothic metric.  
The post-Minkowskian expansion (\ref{expansion_PM}) implies a corresponding post-Minkowskian expansion of the expressions (\ref{metric_35})  
and (\ref{metric_40}),  
\begin{eqnarray}
\tau^{\alpha\beta} &=& T^{\alpha\beta} + \sum\limits_{n=1}^{\infty} G^n\,\tau^{\alpha \beta}_{\left({\rm n PM}\right)}\,,  
\label{metric_52}
\\
t^{\alpha\beta} &=& \sum\limits_{n=1}^{\infty} G^n\,t^{\alpha \beta}_{\left({\rm n PM}\right)}\,.   
\label{metric_53}
\end{eqnarray}

\noindent
Taking account of (\ref{expansion_metric_2a}), inserting of (\ref{expansion_PM}) - (\ref{metric_53}) into (\ref{Field_Equations_10}) yields  
a hierarchy of field equations,  
\begin{eqnarray}
&& \hspace{-0.5cm} \square\; \overline{h}_{\left({\rm 1PM}\right)}^{\alpha\beta} = - \frac{16\,\pi}{c^4}\,T^{\alpha\beta}\,, 
\label{field_equation_1PM}
\\
\nonumber\\
&& \hspace{-0.5cm} \square\; \overline{h}_{\left({\rm 2PM}\right)}^{\alpha\beta} = - \frac{16\,\pi}{c^4} 
\left(\tau_{\left({\rm 1PM}\right)}^{\alpha\beta} + t_{\left({\rm 1PM}\right)}^{\alpha\beta}\right)\,,
\label{field_equation_2PM}
\\
\nonumber\\
\vdots 
\nonumber\\ 
\nonumber\\ 
&& \hspace{-0.5cm} \square\; \overline{h}_{\left({\rm nPM}\right)}^{\alpha\beta} = - \frac{16\,\pi}{c^4}
\left(\tau_{\left({\rm (n-1) PM}\right)}^{\alpha\beta} + t_{\left({\rm (n-1) PM}\right)}^{\alpha\beta}\right)\,. 
\label{field_equation_nPM}
\end{eqnarray}

\noindent
The sequence of field equations (\ref{field_equation_1PM}) - (\ref{field_equation_nPM}) is invariant under Lorentz-transformations. 
The post-Minkowskian expansion (\ref{expansion_PM}) of the gothic metric inherits that the harmonic gauge (\ref{harmonic_gauge_condition_1})
must be satisfied order by order for the metric perturbation (cf. Eq.~(4.6b) in \cite{Blanchet_Damour1}),
\begin{eqnarray}
\left(\overline{h}^{\alpha\beta}_{\left({\rm n PM}\right)}\left(x\right)\right)_{,\,\beta} = 0 \quad {\rm for} \quad n=1,2,3, \dots \;.
\label{hierarchy_metric_gauge}
\end{eqnarray}

\noindent 
As discussed above, the harmonic gauge condition (\ref{hierarchy_metric_gauge}) still allows for a 
residual gauge transformation (\ref{harmonic_gauge_condition_7}). The post-Minkowskian expansion of the 
new gothic metric in the new harmonic coordinate system $\{ x^{\prime\,\alpha} \}$ reads  
\begin{eqnarray}
\overline{g}^{\,\prime\,\alpha \beta}\left(x^{\prime}\right) = 
\eta^{\alpha \beta} - \sum\limits_{n=1}^{\infty} G^n\,\overline{h}^{\,\prime\,\alpha \beta}_{\left({\rm n PM}\right)}\left(x^{\prime}\right).
\label{expansion_PM_prime}
\end{eqnarray}

\noindent 
By inserting the post-Minkowskian expansions (\ref{expansion_PM}) and (\ref{expansion_PM_prime}) as well as (\ref{harmonic_gauge_condition_7})  
into (\ref{transformation_gothic_metric_5}) and performing a series expansion of (\ref{expansion_PM_prime}) around the four-coordinates $x^{\alpha}$,  
one arrives at the post-Minkowskian expansion of the gauge transformation of the gothic metric perturbation,    
\begin{eqnarray}
&& \sum\limits_{n=1}^{\infty} G^n \overline{h}_{\left({\rm n PM}\right)}^{\alpha \beta} 
= \sum\limits_{n=1}^{\infty} G^n \left(\overline{h}_{\left({\rm n PM}\right)}^{\,\prime\,\alpha\beta} 
+ \partial \overline{\varphi}^{\alpha\beta}_{\left({\rm n PM}\right)} 
+ \overline{\Omega}^{\alpha\beta}_{\left({\rm n PM}\right)}\right),   
\nonumber\\ 
\label{gauge_transformation_gothic_metric_PM}
\end{eqnarray}

\noindent 
where all terms are given in the harmonic system $\{x\}$ on the flat background space-time by
Minkowskian coordinates $x= \left(ct, \ve{x}\right)$. 
The equation (\ref{gauge_transformation_gothic_metric_PM}) is nothing else than equation (\ref{transformation_gothic_metric_10}) expressed in terms 
of a series expansion in powers of the gravitational constant.  

The gauge terms $\partial \overline{\varphi}^{\alpha\beta}_{\left({\rm n PM}\right)}$ read  
\begin{eqnarray}
&& \partial \overline{\varphi}^{\alpha\beta}_{\left({\rm n PM}\right)} 
= \varphi^{\alpha\;{\left({\rm n PM}\right)}}_{\,,\,\mu}\,\eta^{\mu\beta}
+ \varphi^{\beta\;{\left({\rm n PM}\right)}}_{\,,\,\mu}\,\eta^{\mu\alpha}
- \varphi^{\mu\;{\left({\rm n PM}\right)}}_{\,,\,\mu}\,\eta^{\alpha\beta}\,,
\nonumber\\ 
\label{gauge_term_gothic_metric_PM}
\end{eqnarray}

\noindent
which are called {\it gothic linear gauge terms} since they are linear in the gauge functions. 
The gauge functions $\varphi^{\alpha\;\left({\rm n PM}\right)}$
are governed by a sequence of equations, which will be considered below; cf. Eqs.~(\ref{sequence_gauge_1PM}) - (\ref{sequence_gauge_nPM}).  
The gauge terms $\overline{\Omega}_{\left({\rm n PM}\right)}^{\alpha\beta}$ are called {\it gothic non-linear gauge terms}  
since they contain either products of gauge functions or products of gauge functions and gothic metric perturbations. 
One may obtain a closed expression for $\overline{\Omega}_{\left({\rm n PM}\right)}^{\alpha\beta}$ from Eq.~(\ref{transformation_gothic_metric_10}) 
and using Eqs.~(\ref{expansion_PM}) and (\ref{harmonic_gauge_condition_7}).
Here it is sufficient to consider the first and second order, given by  
\begin{eqnarray}
&& \overline{\Omega}_{\left({\rm 1PM}\right)}^{\alpha\beta} = 0\,,
\label{gauge_term_gothic_metric_1PM}
\\  
\nonumber\\
&& \overline{\Omega}_{\left({\rm 2PM}\right)}^{\alpha\beta} =
+ \varphi^{\alpha\,{\left({\rm 1PM}\right)}}_{\,,\,\mu}\,\varphi^{\beta\,{\left({\rm 1PM}\right)}}_{\,,\,\nu}\,\eta^{\mu\nu}
+ \left(\varphi^{\nu\,\left({\rm 1PM}\right)}
\overline{h}_{\left({\rm 1PM}\right)}^{\,\prime\,\alpha\beta}\right)_{\,,\,\nu}
\nonumber\\
&& \hspace{1.0cm} + \,\frac{1}{2}\left(\varphi^{\mu\,{\left({\rm 1PM}\right)}}_{\,,\,\nu}\,\varphi^{\nu\,{\left({\rm 1PM}\right)}}_{\,,\,\mu} 
- \varphi^{\mu\,{\left({\rm 1PM}\right)}}_{\,,\,\mu}\,\varphi^{\nu\,{\left({\rm 1PM}\right)}}_{\,,\,\nu}\right)\eta^{\alpha\beta}
\nonumber\\
&& \hspace{1.0cm}- \,\varphi^{\alpha\,{\left({\rm 1PM}\right)}}_{\,,\,\mu} 
\left(\overline{h}_{\left({\rm 1PM}\right)}^{\,\prime\,\mu\beta}+\partial\overline{\varphi}^{\mu\beta}_{\left({\rm 1PM}\right)}\right)  
\nonumber\\
&& \hspace{1.0cm} - \varphi^{\beta\,{\left({\rm 1PM}\right)}}_{\,,\,\mu} 
\left(\overline{h}_{\left({\rm 1PM}\right)}^{\,\prime\,\mu\alpha} 
+ \partial\overline{\varphi}^{\mu\alpha}_{\left({\rm 1PM}\right)}\right),  
\label{gauge_term_gothic_metric_2PM}
\end{eqnarray}
 
\noindent 
while the higher orders $n \ge 3$ are not relevant for our investigations.

\subsection{The equations for the gauge functions} 

The gauge functions are governed by Eq.~(\ref{Laplace_Beltrami_Equation}) which can also be written in the form (cf. Eq.~(4.25) in \cite{Blanchet_Damour1})  
\begin{eqnarray}
\overline{g}^{\mu\nu}\left(x\right)\;\varphi^{\alpha}_{\;\;,\,\mu \nu} \left(x\right) = 0 \,.
\label{Laplace_Beltrami_Equation_2}
\end{eqnarray}

\noindent 
By inserting the post-Minkowskian expansion of the gothic metric (\ref{expansion_PM}) and of the gauge function  
$\varphi^{\alpha} \left(x\right) = \sum\limits_{n=1}^{\infty} G^n\, \varphi^{\alpha\,\left({\rm n PM}\right)}\left(x\right)$ 
into the Laplace-Beltrami equation (\ref{Laplace_Beltrami_Equation_2}) one obtains a sequence of equations for the 
gauge functions $\varphi^{\alpha\,\left({\rm nPM}\right)}$ given by  
\begin{eqnarray}
&& \square \, \varphi^{\alpha\,\left({\rm 1PM}\right)} = 0\,, 
\label{sequence_gauge_1PM} 
\\
\nonumber\\
&& \square \, \varphi^{\alpha\,\left({\rm 2PM}\right)} = \overline{h}^{\mu\nu}_{\left({\rm 1PM}\right)}  
\,\varphi^{\alpha\,\left({\rm 1PM}\right)}_{\,,\,\mu\nu} \,,
\label{sequence_gauge_2PM}
\\
\nonumber\\
\vdots
\nonumber\\
\nonumber\\
&& \square \, \varphi^{\alpha\,\left({\rm nPM}\right)} = \sum\limits_{m=1}^{n-1} \overline{h}^{\mu\nu}_{\left({\rm (n-m)PM}\right)} 
\,\varphi^{\alpha\,\left({\rm mPM}\right)}_{\,,\,\mu\nu}\,,  
\label{sequence_gauge_nPM}
\end{eqnarray}

\noindent
where $\overline{h}^{\mu\nu}_{\left({\rm nPM}\right)}\left(x\right)$ are the terms of the post-Minkowskian  
expansion (\ref{expansion_PM}) of the old gothic metric $\overline{g}^{\alpha\beta}\left(x\right)$ in the old harmonic system $\{x^{\alpha}\}$.  
This sequence of equations allows to determine the gauge functions to any order in the post-Minkowskian expansion.

\section{The Multipolar Post-Minkowskian (MPM) formalism}\label{Section3}  

The Multipolar Post-Minkowskian (MPM) formalism represents a powerful approach in order to determine the gothic metric  
$\overline{g}^{\alpha\beta}$ external to the compact source of matter in harmonic coordinates. The MPM formalism is a considerable extension  
of previous investigations in \cite{Bonnor1,Bonnor2,Hunter_Rosenberg} and of the pioneering work \cite{Thorne}. The formalism has been developed   
within a series of articles \cite{Blanchet_Damour1,Blanchet_Damour2,Blanchet_Damour3,Blanchet_Damour4,2PN_Metric1,Multipole_Damour_2}, where the approach  
has thoroughly been described in detail; see also the descriptions of the MPM formalism in subsequent developments \cite{Blanchet4,Blanchet5,Blanchet6}. 

The fundamental concept of the MPM approach is to solve iteratively the hierarchy of field equations (\ref{field_equation_1PM}) $\dots$ (\ref{field_equation_nPM}) 
for the gothic metric density in a sequence of three steps:  
\begin{enumerate} 
\item[(i)] solving the field equations in the internal near-zone ${\cal D}_i$ of the source in the post-Newtonian  
(weak-field slow-motion) scheme; see Eq.~(1) in \cite{Blanchet6} for a concrete definition of what a post-Newtonian source is.  
The internal near-zone is defined by ${\cal D}_i = \{\left(t,\ve{x}\right) \; {\rm with}\; \left| \ve{x}\right| < r_i\}$  
where $R < r_i \ll \lambda$, where $R$ is the radius of a sphere which encloses the source and $\lambda$ is the wavelength of the gravitational  
radiation emitted by the source. So the internal near-zone is a spatial region which contains the interior of the source and a region in the exterior 
of the source but much smaller than the wavelength of the gravitational radiation emitted by the source.   
\item[(ii)] solving the field equations in the external zone ${\cal D}_e$ of the source in the post-Minkowskian (weak field) scheme. 
The external zone is defined by  
${\cal D}_e = \{\left(t,\ve{x}\right) \; {\rm with}\; \left|\ve{x}\right| > r_e\}$ where $R < r_e < r_i$. So the external zone contains the entire 
spatial region in the exterior of the source.  
\item[(iii)] performing a matching procedure of both these solutions for the metric tensor in the intermediate near-zone ${\cal D}_i \cap {\cal D}_e$ of 
the source, where both the post-Newtonian expansion and the post-Minkowskian expansion are simultaneously valid. The intermediate near-zone is defined by  
${\cal D}_i \cap {\cal D}_e = \{\left(t,\ve{x}\right) \; {\rm with}\; r_e < \left|\ve{x}\right| < r_i\}$. The definitions of the internal near-zone 
${\cal D}_i$ and external zone ${\cal D}_e$ are adjusted such that the intermediate near-zone ${\cal D}_i \cap {\cal D}_e$ is not empty. The intermediate  
near-zone is a spatial region in the exterior of the source but much smaller than the wavelength $\lambda$ of the gravitational radiation emitted by the source.  
\end{enumerate} 

\noindent 
In what follows only those fundamental results of the elaborated MPM formalism are considered which are of relevance for our analysis.  
In particular, we will not consider the specific issue related to the far-wave zone, where so-called radiative coordinates and radiative moments
$V_L$ and $U_L$ are introduced, which are uniquely related to the mass-multipoles $M_L$ and spin-multipoles $S_L$ via non-linear equations;
cf. Eqs.~(6.4a) - (6.4b) in \cite{Blanchet4}. In the far-wave zone only the transverse traceless projection of the  
metric perturbation, $h_{\alpha\beta}^{\rm TT}$, is relevant because it contains the physical degrees of freedom of the 
gravitational radiation field; cf. Eq.~(64) in \cite{Radiation_Condition}. That transverse traceless projection of the metric perturbation has been given  
in several investigations in the 1PM approximation \cite{Radiation_Condition,Blanchet6,Gauge_Transformation}  
(e.g. Eq.~(64) in \cite{Radiation_Condition}, Eq.~(66) in \cite{Blanchet6}, Eq.~(2.1) in \cite{Gauge_Transformation}); 
note that $h_{\alpha\beta\left({\rm 1PM}\right)}^{\rm TT} = \overline{h}_{\alpha\beta\left({\rm 1PM}\right)}^{\rm TT}$ (cf. Eq.(7.119) in [61]).  
Here, we will not consider the transverse-traceless gauge but emphasize that all the subsequent statements
about the gothic metric perturbation and about the metric perturbation are valid in the entire region in the exterior of the source of matter.

\subsection{The general solution of the gothic metric} 

In the MPM formalism the most general solution of the gothic metric is called {\it general gothic metric} and denoted by 
$\overline{g}^{\alpha\beta\,{\rm gen}}\left(x_{\rm gen}\right)$ given in the general harmonic reference system 
$\{x_{\rm gen}\} = \left(ct_{\rm gen},\ve{x}_{\rm gen}\right)$.  
According to Eq.~(\ref{expansion_metric_2a}) it is decomposed in the flat Minkowskian metric and a {\it general gothic metric perturbation},  
\begin{eqnarray}
\overline{g}^{\alpha\beta\,{\rm gen}}\left(x_{\rm gen}\right) = \eta^{\alpha\beta} - \overline{h}^{\alpha \beta\,{\rm gen}}\left(x_{\rm gen}\right).
\label{fundamental_theorem_0}
\end{eqnarray}

\noindent  
An important result of the MPM approach consists in a theorem (Theorem 4.2 in \cite{Blanchet_Damour1}) which states that outside the matter source the  
most general solution of the post-Minkowskian hierarchy (\ref{field_equation_1PM}) $\dots$ (\ref{field_equation_nPM}) depends on a set of altogether 
six STF multipoles \cite{Blanchet_Damour1,Blanchet5,Blanchet6,Gauge_Transformation}  
(cf. Eq.~(62) in \cite{Blanchet5}, Eq.~(50) in \cite{Blanchet6}, Eq.~(4.1) in \cite{Gauge_Transformation})  
\begin{eqnarray}
\overline{h}^{\alpha \beta\,{\rm gen}}\left(x_{\rm gen}\right) = \sum\limits_{n=1}^{\infty}  
G^n \; \overline{h}_{\left({\rm n PM}\right)}^{\alpha \beta\,{\rm gen}}\left[I_L,J_L,W_L,X_L,Y_L,Z_L\right],  
\nonumber\\ 
\label{fundamental_theorem_1}
\end{eqnarray}
 
\noindent
where the square brackets denote a functional dependence on these six STF multipoles.  
The MPM solution (\ref{fundamental_theorem_1}) is the most general solution of Einstein’s vacuum equations
outside an isolated source of matter.  
The STF multipoles in (\ref{fundamental_theorem_1}) depend on the retarded time $s_{\rm gen}$ defined by  
\begin{eqnarray}
s_{\rm gen} = t_{\rm gen} - \frac{\left|\ve{x}_{\rm gen}\right|}{c}\,,
\label{Retarded_Time_3}
\end{eqnarray}

\noindent
which is the time of retardation between some field point $\ve{x}_{\rm gen}$ and the origin of the spatial axes of the general harmonic coordinate system  
$\{x_{\rm gen}\}$ \cite{Footnote5}.  
As stated above by Eq.~(\ref{hierarchy_metric_gauge}), the harmonic gauge (\ref{harmonic_gauge_condition_1}) is satisfied order by order 
for the metric perturbation, which in terms of the MPM solution is given by (cf. Eq.~(4.6b) in \cite{Blanchet_Damour1}),  
\begin{eqnarray}
\frac{\partial}{\partial x_{\rm gen}^{\beta}}\;
\overline{h}^{\alpha\beta\,{\rm gen}}_{\left({\rm n PM}\right)}\left[I_L,J_L,W_L,X_L,Y_L,Z_L\right] = 0 \;.  
\label{hierarchy_metric_gauge_MPM}
\end{eqnarray}

\noindent 
The MPM formalism is augmented by a matching procedure described in detail in \cite{Blanchet4,Blanchet6} which 
allows to determine these six multipoles as integrals over the stress-energy tensor of the source of matter.  
For that reason these multipoles $I_L,J_L,W_L,X_L,Y_L,Z_L$ are collectively named as the {\it source multipole moments} \cite{Blanchet4}.  
In fact, such an explicit closed-form expression for the   
set of these six STF multipoles has been derived by Eqs.~(5.15) - (5.20) in \cite{Blanchet4}; see also Eqs.~(85) - (90) in \cite{Blanchet5},  
Eqs.~(123a) - (125d) in \cite{Blanchet6}.

\subsection{Residual gauge transformation of the general gothic metric} 

A further result of utmost importance of the MPM formalism (Theorem 4.5 in \cite{Blanchet_Damour1}) is that there exists a residual gauge transformation 
(cf. Eq.~(\ref{harmonic_gauge_condition_7})),  
\begin{eqnarray}
x^{\alpha}_{\rm can} = x^{\alpha}_{\rm gen} + \sum\limits_{n=1}^{\infty} G^n\,\varphi^{\alpha\,\left({\rm nPM}\right)}\left(x_{\rm gen}\right),  
\label{gen_can_transformation}
\end{eqnarray}

\noindent 
which preserves the harmonic gauge (\ref{hierarchy_metric_gauge}) and which allows to write the general metric perturbation  
in (\ref{fundamental_theorem_1}) in the following form,  
\begin{eqnarray}
&& \hspace{-0.5cm} \sum\limits_{n=1}^{\infty} \! G^n \overline{h}_{\left({\rm nPM}\right)}^{\alpha \beta\,{\rm gen}}\left[I_L,J_L,W_L,X_L,Y_L,Z_L\right] 
\nonumber\\ 
&& \hspace{-0.5cm} = \sum\limits_{n=1}^{\infty} \! G^n \! \left(\overline{h}_{\left({\rm nPM}\right)}^{\alpha \beta\,{\rm can}}\left[M_L,S_L\right] 
+ \partial \overline{\varphi}^{\alpha\beta}_{\left({\rm nPM}\right)} \!+ \overline{\Omega}^{\alpha\beta}_{\left({\rm nPM}\right)}\!\right)   
\label{general_solution_D}
\end{eqnarray}

\noindent 
where all terms depend on the four-coordinates $x^{\alpha}_{\rm gen}$ and the STF multipoles  
depend on the retarded time $s_{\rm gen}$ in (\ref{Retarded_Time_3}).  
The relation (\ref{general_solution_D}) is nothing else
than relation (\ref{gauge_transformation_gothic_metric_PM}) expressed in terms of STF multiples of the MPM formalism \cite{Footnote6}. 
The relation (\ref{general_solution_D}) states that the general gothic metric perturbation (\ref{fundamental_theorem_1}) in terms of six source multipoles
is {\it physically isometric} to the canonical gothic metric perturbation (\ref{canonical_gothic_metric_perturbation}) in terms of two canonical multipoles.
That means that the general gothic metric perturbation (\ref{fundamental_theorem_1}) contains the same
physical information as the canonical gothic metric perturbation (\ref{canonical_gothic_metric_perturbation}); see also text below Eq.~(45)
in \cite{Blanchet5}, text below Eq.~(52) in \cite{Blanchet6}, text above below Eq.~(4.26) in \cite{Gauge_Transformation}.

The term  
\begin{eqnarray}
\overline{h}^{\alpha \beta\,{\rm can}}\left(x_{\rm gen}\right) = 
\sum\limits_{n=1}^{\infty} G^n \; \overline{h}_{\left({\rm nPM}\right)}^{\alpha \beta\,{\rm can}}\left[M_L,S_L\right] 
\label{canonical_gothic_metric_perturbation}
\end{eqnarray}

\noindent 
on the r.h.s. in Eq.~(\ref{general_solution_D}) is called {\it canonical gothic metric perturbation} and  
\begin{eqnarray}
\overline{g}^{\alpha \beta\,{\rm can}}\left(x_{\rm gen}\right) = \eta^{\alpha\beta} - \overline{h}^{\alpha \beta\,{\rm can}}\left(x_{\rm gen}\right)  
\label{canonical_gothic_metric}
\end{eqnarray}

\noindent 
is the {\it canonical gothic metric}.   
The multipoles $M_L$ and $S_L$ are called {\it canonical multipoles} and they are related to the source multipoles via two non-linear equations
(cf. Eqs.~(6.1a) - (6.1b) in \cite{Blanchet4} and text below Eq.~(45) in \cite{Blanchet4}),
\begin{eqnarray}
M_L = M_L \left[I_L,J_L,W_L,X_L,Y_L,Z_L\right],
\label{general_solution_E}
\\
S_L \; = \; S_L\,\left[I_L,J_L,W_L,X_L,Y_L,Z_L\right],
\label{general_solution_F}
\end{eqnarray}

\noindent
which are of complicated structure; cf. Eqs.~(97) and (98) in \cite{Blanchet6} for the case of $L=i_1 i_2$ and $L=i_1 i_2 i_3$.
In view of the highly involved structure of the relations (\ref{general_solution_E}) - (\ref{general_solution_F}) it seems
impossible to achieve an explicit closed-form expression for the {\it canonical multipoles} $M_L,S_L$ to any order of the post-Minkowskian series expansion
\cite{Blanchet4,Blanchet5,Blanchet6}.
The gauge terms on the r.h.s. in (\ref{general_solution_D}) depend,
in the general case, on the full set of all six STF source multipoles (cf. text below Eq.~(4.23) in \cite{Blanchet_Damour1}),
\begin{eqnarray}
&& \hspace{-0.5cm} 
\partial \overline{\varphi}^{\alpha\beta}_{\left({\rm nPM}\right)}\!\left(x_{\rm gen}\right) = \partial \overline{\varphi}^{\alpha\beta}_{\left({\rm nPM}\right)}
\left[I_L,J_L,W_L,X_L,Y_L,Z_L\right],
\label{linear_gauge_term}
\\
&& \hspace{-0.5cm} \overline{\Omega}^{\alpha\beta}_{\left({\rm nPM}\right)}\!\left(x_{\rm gen}\right) \;\;= \overline{\Omega}^{\alpha\beta}_{\left({\rm nPM}\right)}
\left[I_L,J_L,W_L,X_L,Y_L,Z_L\right].
\label{non_linear_gauge_term} 
\end{eqnarray}

\noindent
The explicit structure of these gauge terms will be considered below in the linear and post-linear approximation. These gauge terms are functions of  
the gauge functions   
\begin{eqnarray}
&& \hspace{-0.75cm} \varphi^{\alpha\,\left({\rm nPM}\right)}\left(x_{\rm gen}\right) = \varphi^{\alpha\,\left({\rm nPM}\right)}\left[I_L,J_L,W_L,X_L,Y_L,Z_L\right], 
\label{gauge_vector_MPM}
\end{eqnarray}

\noindent 
governed by Eqs.~(\ref{sequence_gauge_1PM}) - (\ref{sequence_gauge_nPM}), which in terms of the STF multipoles  
of the MPM formalism read (cf. Eqs.~(4.26) - (4.27) in \cite{Blanchet_Damour1})   
\begin{eqnarray}
&& \square \, \varphi^{\alpha\,\left({\rm 1PM}\right)} = 0\,,
\label{sequence_gauge_1PM_MPM}
\\
\nonumber\\
&& \square \, \varphi^{\alpha\,\left({\rm 2PM}\right)} = \overline{h}^{\mu\nu\,{\rm gen}}_{\left({\rm 1PM}\right)}
\,\varphi^{\alpha\,\left({\rm 1PM}\right)}_{\,,\,\mu\nu}\,,
\label{sequence_gauge_2PM_MPM}
\\
\nonumber\\
\vdots
\nonumber\\
\nonumber\\
&& \square \, \varphi^{\alpha\,\left({\rm nPM}\right)} 
= \sum\limits_{m=1}^{n-1} \overline{h}^{\mu\nu\,{\rm gen}}_{\left({\rm (n-m)PM}\right)}
\,\varphi^{\alpha\,\left({\rm mPM}\right)}_{\,,\,\mu\nu}\,,  
\label{sequence_gauge_nPM_MPM}
\end{eqnarray}

\noindent 
which are given in the harmonic system $\{x_{\rm gen}\}$ and  
where the general solution of the metric perturbations is given by Eq.~(\ref{fundamental_theorem_1}). 
The sequence of differential equations for the gauge functions in (\ref{sequence_gauge_1PM_MPM}) - (\ref{sequence_gauge_nPM_MPM}) is nothing but 
the sequence of differential equations for the gauge functions in (\ref{sequence_gauge_1PM}) - (\ref{sequence_gauge_nPM}) expressed in terms 
of STF source multipoles.

\subsection{The general solution of the metric tensor} 

The most general solution of the metric tensor in the exterior of a compact source of matter is uniquely determined by the relation 
(cf. Eq.~(\ref{Appendix3_10}) in Appendix \ref{Appendix3})  
\begin{eqnarray}
g_{\alpha\beta\,{\rm gen}} = \sqrt{ - {\rm det}\left(\overline{g}^{\mu\nu\,{\rm gen}}\right)}\;\overline{g}_{\alpha \beta\,{\rm gen}}\,.
\label{Relation_gothic_density_metric_tensor}
\end{eqnarray}

\noindent
The terms on the r.h.s. of (\ref{Relation_gothic_density_metric_tensor}) are given by (\ref{fundamental_theorem_0}) - (\ref{fundamental_theorem_1})
and by the isometry relation of the gothic metric \cite{MTW,Fock,Kopeikin_Efroimsky_Kaplan} (cf. Eq.~(\ref{Appendix3_4}) in Appendix \ref{Appendix3})
\begin{eqnarray}
\overline{g}^{\alpha\sigma\,{\rm gen}}\,\overline{g}_{\sigma\beta\,{\rm gen}} = \delta^{\alpha}_{\beta}\,.
\label{isometry_relation}
\end{eqnarray}

\noindent 
According to Eq.~(\ref{expansion_metric_1a}), the general metric tensor in (\ref{Relation_gothic_density_metric_tensor}) is separated into the 
flat Minkowskian metric and the {\it general metric perturbation},  
\begin{eqnarray}
g_{\alpha \beta\,{\rm gen}}\left(x_{\rm gen}\right) = \eta_{\alpha \beta} + h_{\alpha \beta\,{\rm gen}}\left(x_{\rm gen}\right). 
\label{general_metric}
\end{eqnarray}

\noindent 
From (\ref{Relation_gothic_density_metric_tensor}) follows that the general metric perturbation formally reads  
\begin{eqnarray}
h_{\alpha \beta\,{\rm gen}}\left(x_{\rm gen}\right) = \sum\limits_{n=1}^{\infty}
G^n h^{\left({\rm n PM}\right)}_{\alpha \beta\,{\rm gen}}\left[I_L,J_L,W_L,X_L,Y_L,Z_L\right]. 
\nonumber\\ 
\label{fundamental_theorem_3}
\end{eqnarray}

\noindent 
The square brackets denote a functional dependence on the six STF source multipoles which depend on the
retarded time $s_{\rm gen}$ in Eq.~(\ref{Retarded_Time_3}).
The Eqs.~(\ref{general_metric}) and (\ref{fundamental_theorem_3}) represent the most general solution of Einstein’s vacuum equations 
outside an isolated source of matter.

\subsection{Residual gauge transformation of the general metric tensor} 

The residual gauge transformation (\ref{gen_can_transformation}) allows to transform the general metric perturbation in the following form,   
\begin{eqnarray}
&& \hspace{-0.5cm} \sum\limits_{n=1}^{\infty} \! G^n h^{\left({\rm n PM}\right)}_{\alpha \beta\,{\rm gen}}\left[I_L,J_L,W_L,X_L,Y_L,Z_L\right]
\nonumber\\ 
&& \hspace{-0.5cm} = \sum\limits_{n=1}^{\infty} \! G^n \! \left(\!h^{\left({\rm n PM}\right)}_{\alpha \beta\,{\rm can}}\left[M_L,S_L\right] 
+ \partial \varphi^{\left({\rm n PM}\right)}_{\alpha\beta} \! + \Omega^{\left({\rm n PM}\right)}_{\alpha\beta}\!\right)   
\label{general_solution_metric_PM}
\end{eqnarray}

\noindent
where all terms depend on the four-coordinates $x^{\alpha}_{\rm gen}$ and the STF multipoles
depend on the retarded time $s_{\rm gen}$ in (\ref{Retarded_Time_3}).  
The relation (\ref{general_solution_metric_PM}) is nothing else
than relation (\ref{gauge_transformation_metric_PM}) expressed in terms of STF multiples of the MPM formalism \cite{Footnote7}.  
The relation (\ref{general_solution_metric_PM}) states that
if the source multipoles and the canonical multipoles are related to each other via Eqs.~(\ref{general_solution_E}) - (\ref{general_solution_F}), then the
{\it general metric perturbation} on the l.h.s. of (\ref{general_solution_metric_PM}) and the {\it canonical metric perturbation}
on the r.h.s. of (\ref{general_solution_metric_PM}) are related by the residual coordinate transformation (\ref{gen_can_transformation}).
They are physically isometric to each other and either of them contains the entire physical information in the exterior of the gravitational source of matter.
The term  
\begin{eqnarray}
h_{\alpha\beta\,{\rm can}}\left(x_{\rm gen}\right)  
= \sum\limits_{n=1}^{\infty} G^n \; h^{\left({\rm nPM}\right)}_{\alpha \beta\,{\rm can}}\left[M_L,S_L\right]
\label{canonical_metric_perturbation}
\end{eqnarray}

\noindent
on the r.h.s. in Eq.~(\ref{general_solution_metric_PM}) is called {\it canonical metric perturbation} and 
\begin{eqnarray}
g_{\alpha \beta\,{\rm can}}\left(x_{\rm gen}\right) = \eta_{\alpha\beta} + h_{\alpha \beta\,{\rm can}}\left(x_{\rm gen}\right)
\label{canonical_metric}
\end{eqnarray}

\noindent
is the {\it canonical metric}.
The canonical multipoles $M_L$ and $S_L$ are related to the source multipoles via Eqs.~(\ref{general_solution_E}) and (\ref{general_solution_F}).  
The gauge terms on the r.h.s. in (\ref{general_solution_metric_PM}) depend, in the general case, on the full set
of all six STF source multipoles,
\begin{eqnarray}
&& \hspace{-0.5cm} \partial \varphi^{\alpha\beta}_{\left({\rm nPM}\right)}\left(x_{\rm gen}\right) = \partial \varphi^{\alpha\beta}_{\left({\rm nPM}\right)}
\left[I_L,J_L,W_L,X_L,Y_L,Z_L\right],
\label{linear_gauge_term_metric}
\\
&& \hspace{-0.5cm} \Omega^{\alpha\beta}_{\left({\rm nPM}\right)}\left(x_{\rm gen}\right) \;\;= \Omega^{\alpha\beta}_{\left({\rm nPM}\right)}
\left[I_L,J_L,W_L,X_L,Y_L,Z_L\right].
\label{non_linear_gauge_term_metric}
\end{eqnarray}

\noindent
The explicit structure of these gauge terms will be considered below in the linear and post-linear approximation. They are functionals of the  
gauge functions (\ref{gauge_vector_MPM}) which are determined by means of Eqs.~(\ref{sequence_gauge_1PM_MPM})- (\ref{sequence_gauge_nPM_MPM}).  

To simplify the notations, in all of the subsequent Sections, the four-coordinates of the general harmonic system 
$x^{\alpha}_{\rm gen} = \left(ct_{\rm gen}, \ve{x}_{\rm gen}\right)$ will be denoted by $x^{\alpha} = \left(ct,\ve{x}\right)$. 
This implies that the retarded time $s_{\rm gen}$ in (\ref{Retarded_Time_3}) is now denoted by $s = t - \left|\ve{x}\right|/c$.

\section{The gothic metric density in post-linear approximation}\label{Section4}

The post-Minkowskian expansion of the gothic metric density in the second post-Minkowskian approximation is given by (cf. Eq.~(\ref{expansion_metric_2}))  
\begin{eqnarray}
&& \overline{g}^{\alpha\beta}\left(t,\ve{x}\right) = \eta^{\alpha\beta}
- G^1\,\overline{h}_{\left({\rm 1PM}\right)}^{\alpha\beta}\left(t,\ve{x}\right)
- G^2\,\overline{h}_{\left({\rm 2PM}\right)}^{\alpha\beta}\left(t,\ve{x}\right) 
\nonumber\\
&& \hspace{+ 1.75cm} + \, {\cal O}\left(G^3\right).  
\label{expansion_gothic_metric}
\end{eqnarray}

\noindent
In this Section the linear term $\overline{h}_{\left({\rm 1PM}\right)}^{\alpha\beta}$ and 
the post-linear term $\overline{h}_{\left({\rm 2PM}\right)}^{\alpha\beta}$ are considered.

\subsection{The linear term of the gothic metric density}\label{Section4_1}  

The solution of the field equations in the first iteration (\ref{field_equation_1PM}) reads  
\begin{eqnarray}
\overline{h}^{\alpha \beta}_{\left({\rm 1PM}\right)}\left(t,\ve{x}\right) = - \frac{16\,\pi}{c^4}\, 
\left(\square_{\rm R}^{-1} \, T^{\alpha \beta}\right)\left(t,\ve{x}\right)\,,   
\label{metric_55}
\end{eqnarray}

\noindent
where $T^{\alpha \beta}$ is the stress-energy tensor of matter and $\square_{\rm R}^{-1}$ is the inverse d'Alembert operator 
defined by Eq.~(\ref{Inverse_d_Alembert_1}). The integration runs only over the finite three-dimensional volume of the compact  
source of matter \cite{Blanchet_Damour1,Blanchet_Damour2}. The integral (\ref{metric_55}) is finite and has been determined in \cite{Blanchet_Damour2}   
and has later been reconsidered in specific detail in \cite{Zschocke_Multipole_Expansion}. 

According to the fundamental theorem (\ref{fundamental_theorem_1})  
of the MPM formalism, the most general solution for the 1PM term of the gothic metric perturbation (\ref{metric_55}) in the exterior of a compact source  
of matter depends on six STF source multipoles and is denoted by $\overline{h}_{\left({\rm 1PM}\right)}^{\alpha\beta\,{\rm gen}}$. 
The residual gauge transformation (\ref{gen_can_transformation}) in 1PM approximation  
transforms the linear gothic metric perturbation $\overline{h}_{\left({\rm 1PM}\right)}^{\alpha\beta\,{\rm gen}}$ in the following form  
\cite{Blanchet_Damour1,Multipole_Damour_2,Thorne,Blanchet4,Blanchet6,Gauge_Transformation},  
\begin{eqnarray}
&& \overline{h}_{\left({\rm 1PM}\right)}^{\alpha\beta\,{\rm gen}} \left[I_L, J_L, W_L, X_L, Y_L, Z_L \right]
\nonumber\\ 
&& = \overline{h}_{\left({\rm 1PM}\right)}^{\alpha\beta\,{\rm can}}\left[M_L,S_L\right] 
+ \partial \overline{\varphi}_{\left({\rm 1PM}\right)}^{\alpha\beta}\left(t,\ve{x}\right)\,.    
\label{Gauge_1PM_C}
\end{eqnarray}

\noindent
The {\it canonical gothic metric} in 1PM approximation for one body at rest with full multipole structure is given by   
\begin{eqnarray}
&& \hspace{-1.25cm} \overline{h}_{\left({\rm 1PM}\right)}^{00\,{\rm can}}\left(t,\ve{x}\right) =
+ \frac{4}{c^2}\sum\limits_{l=0}^{\infty}\frac{\left(-1\right)^l}{l!}\,
\partial_L \frac{M_L\left(s\right)}{r}\,,
\label{1PM_Metric_A}
\\
\nonumber\\
\nonumber\\
&& \hspace{-1.25cm} \overline{h}_{\left({\rm 1PM}\right)}^{0i\,{\rm can}}\left(t,\ve{x}\right) =
- \frac{4}{c^3}\sum\limits_{l=1}^{\infty} \frac{\left(-1\right)^l}{l!} \partial_{L-1}
\frac{\dot{M}_{iL-1} \left(s\right)}{r}
\nonumber\\
&& - \frac{4}{c^3}\sum\limits_{l=1}^{\infty} \frac{\left(-1\right)^l\,l}{\left(l+1\right)!} \epsilon_{iab}\,\partial_{a L-1}
\frac{S_{bL-1} \left(s\right)}{r}\,,
\label{1PM_Metric_B}
\\
\nonumber\\
&& \hspace{-1.25cm} \overline{h}_{\left({\rm 1PM}\right)}^{ij\,{\rm can}}\left(t,\ve{x}\right) =
+ \frac{4}{c^4}\sum\limits_{l=2}^{\infty} \frac{\left(-1\right)^l}{l!} \partial_{L-2}
\frac{\ddot{M}_{ijL-2} \left(s\right)}{r} 
\nonumber\\
&& + \frac{8}{c^4}\sum\limits_{l=2}^{\infty} \frac{\left(-1\right)^l\,l}{\left(l+1\right)!}
\partial_{a L-2} \frac{\epsilon_{ab(i} \dot{S}_{j)bL-2} \left(s\right)}{r}\,.  
\label{1PM_Metric_C}
\end{eqnarray}

\noindent
The non-linear relations (\ref{general_solution_E}) and (\ref{general_solution_F}) simplify in the 1PM approximation as follows  
(cf. Eqs.~(6.2a) - (6.2b) in \cite{Blanchet4}, Eqs.~(4.25a) - (4.25b) in \cite{Gauge_Transformation}),  
\begin{eqnarray}
M_L = I_L + {\cal O}\left(G\right),
\label{M_L_PN}
\\
S_L = J_L + {\cal O}\left(G\right). 
\label{S_L_PN}
\end{eqnarray}

\noindent
The explicit expressions for the canonical multipoles $M_L$ and $S_L$ are given by Eqs.~(5.33) and (5.35) in \cite{Multipole_Damour_2} as integrals over  
the stress-energy tensor of the matter source, and they are represented by Eqs.~(\ref{M_L_Definition}) and (\ref{S_L_Definition}) in Appendix \ref{Appendix2}.  
The linear gauge term in (\ref{Gauge_1PM_C}) is given by (cf. Eq.~(\ref{gauge_term_gothic_metric_PM}))  
\begin{eqnarray}
&& \partial \overline{\varphi}_{\left({\rm 1PM}\right)}^{\alpha\beta}\left(t,\ve{x}\right) 
= \varphi^{\alpha\,{\left({\rm 1PM}\right)}}_{\,,\,\mu}\left(t,\ve{x}\right)\eta^{\mu\beta}
+ \varphi^{\beta\,{\left({\rm 1PM}\right)}}_{\,,\,\mu}\left(t,\ve{x}\right)\eta^{\mu\alpha}  
\nonumber\\ 
&& \hspace{2.5cm} -\, \varphi^{\mu\,{\left({\rm 1PM}\right)}}_{\,,\,\mu}\left(t,\ve{x}\right)\;\eta^{\alpha\beta}\,. 
\label{general_solution_C1}
\end{eqnarray}

\noindent 
The gauge function $\varphi^{\alpha\,\left({\rm 1PM}\right)}$ is determined by Eq.~(\ref{sequence_gauge_1PM_MPM}). The  
gauge function depends on four {\it source moments},  
\begin{eqnarray}
\varphi^{\alpha\,\left({\rm 1PM}\right)}\left(t,\ve{x}\right) = \varphi^{\alpha\,\left({\rm 1PM}\right)}\left[W_L,X_L,Y_L,Z_L\right], 
\label{gauge_vector_1PM}  
\end{eqnarray}

\noindent 
and is given by Eqs.~(5.31b) in \cite{Multipole_Damour_2}; see also Eqs.~(4.13a) - (4.13b) in \cite{Blanchet4} or 
Eqs.~(3.560) - (3.561) in \cite{Kopeikin_Efroimsky_Kaplan}.  

\vspace{1.0cm} 

\subsection{The post-linear term of the gothic metric density}\label{Section4_2}

The solution of the field equations in the second iteration (\ref{field_equation_2PM}) reads  
\begin{eqnarray}
\overline{h}^{\alpha \beta}_{\left({\rm 2PM}\right)}\!\left(t,\ve{x}\right) = - \frac{16\,\pi}{c^4}\,\left({\rm FP}_{B=0}\,\square_{\rm R}^{-1}
\!\left(\tau^{\alpha\beta}_1 + t^{\alpha\beta}_1\right)\right)\!\left(t,\ve{x}\right)\,,  
\nonumber\\ 
\label{metric_60}
\end{eqnarray}

\noindent
where $\tau^{\alpha\beta}_1$ and $t^{\alpha\beta}_1$ denote the first iteration of (\ref{metric_35}) and (\ref{metric_40}), respectively, 
and ${\rm FP}_{B=0}\,\square_{\rm R}^{-1}$ is the Hadamard regularized inverse d'Alembert operator defined by Eq.~(\ref{Inverse_d_Alembert_FP}); 
details of the Hadamard regularization are given in Appendix \ref{Appendix5}.   
The expression of $\tau^{\alpha\beta}_1$ follows from (\ref{metric_35}) by series expansion of the determinant. The expression of  
$t^{\alpha\beta}_1$ follows from (\ref{metric_40}) by using the 1PM approximation of the gothic metric perturbation; cf. Eq.~(3.3) in \cite{Gauge_Transformation}.  

According to the fundamental theorem (\ref{fundamental_theorem_1}) of the MPM formalism, the most general solution for the 2PM term of the gothic metric  
perturbation (\ref{metric_60}) in the exterior of a compact source of matter depends on six STF source multipoles and is denoted by  
$\overline{h}^{\alpha\beta\,{\rm gen}}_{\left({\rm 2PM}\right)}$. The residual gauge transformation (\ref{gen_can_transformation}) in 2PM approximation  
transforms the post-linear gothic metric perturbation $\overline{h}^{\alpha\beta\,{\rm gen}}_{\left({\rm 2PM}\right)}$ in the 
following form \cite{Blanchet_Damour1,Multipole_Damour_2,Blanchet4,Gauge_Transformation} (cf. Eq.~(4.26) in \cite{Gauge_Transformation})  
\vspace{-1.5cm} 
\begin{eqnarray}
&& \hspace{-0.5cm} \overline{h}_{\left({\rm 2PM}\right)}^{\alpha \beta\,{\rm gen}}\left[I_L,J_L,W_L,X_L,Y_L,Z_L\right]
\nonumber\\ 
&& \hspace{-0.5cm} = \overline{h}_{\left({\rm 2PM}\right)}^{\alpha \beta\;{\rm can}}\left[M_L,S_L\right] 
+ \partial \overline{\varphi}_{\left({\rm 2PM}\right)}^{\alpha \beta}\!\left(t,\ve{x}\right)  
+ \overline{\Omega}_{\left({\rm 2PM}\right)}^{\alpha \beta}\!\left(t,\ve{x}\right).   
\label{Gauge_2PM}
\end{eqnarray}

\noindent 
The {\it canonical gothic metric} for a source of matter with full multipole structure has not rigorously been determined in the  
second post-Minkowskian (2PM) scheme thus far, but in the following approximation  
(cf. Eqs.~(2.28a) - (2.28c) and Eq.~(2.29) together with Eqs.~(2.18a) and (2.5) in \cite{2PN_Metric1}) 
\begin{widetext}
\begin{eqnarray}
\overline{h}^{00\,{\rm can}}_{\left({\rm 2PM}\right)}\left(t,\ve{x}\right) &=& \frac{7}{c^4}
\left(\sum\limits_{l=0}^{\infty} \frac{\left(-1\right)^{l}}{l!}\,\partial_L\,\frac{M_L\left(s\right)}{r}\right)^2 + {\cal O}\left(c^{-6}\right),  
\label{gothic_metric_80}
\\
\nonumber\\
\overline{h}^{0i\,{\rm can}}_{\left({\rm 2PM}\right)}\left(t,\ve{x}\right) &=& {\cal O}\left(c^{-5}\right), 
\label{gothic_metric_85}
\\
\nonumber\\
\overline{h}^{ij\,{\rm can}}_{\left({\rm 2PM}\right)}\left(t,\ve{x}\right) &=&  
- \frac{4}{c^4}\,{\rm FP}_{B=0}\,\square_{\rm R}^{-1}
\left(\!\partial_i \sum\limits_{l=0}^{\infty}
\frac{\left(-1\right)^{l}}{l!}\,\partial_L\,\frac{M_L\left(s\right)}{r} \!\right)
\left(\!\partial_j \sum\limits_{l=0}^{\infty}
\frac{\left(-1\right)^{l}}{l!}\,\partial_L\,\frac{M_L\left(s\right)}{r} \!\right) 
\nonumber\\ 
&& + \frac{1}{c^4}\,\delta_{ij} \left(\sum\limits_{l=0}^{\infty}  
\frac{\left(-1\right)^{l}}{l!}\,\partial_L\,\frac{M_L\left(s\right)}{r}\right)^2 
+ {\cal O}\left(c^{-6}\right).   
\label{gothic_metric_90}
\end{eqnarray}
\end{widetext} 

\noindent
These expressions are also in agreement with Eqs.~(3.5a) - (3.5c) in \cite{Will_Wiseman}; the agreement of the 
MPM formalism and the Will-Wiseman approach has been explained in Section 4.3 in \cite{Blanchet6}.  
In the second line of (\ref{gothic_metric_90}) we have used the following relation \cite{Footnote8}.  
\begin{eqnarray} 
&& {\rm FP}_{B=0}\,\square_{\rm R}^{-1} \left(\partial_k \sum\limits_{l=0}^{\infty}  
\frac{\left(-1\right)^{l}}{l!} \partial_L \frac{M_L\left(s\right)}{r}\right)^2 
\nonumber\\ 
&& = \frac{1}{2} \left( \sum\limits_{l=0}^{\infty} \frac{\left(-1\right)^{l}}{l!} \partial_L \frac{M_L\left(s\right)}{r}\right)^2 
\!+ {\cal O}\left(c^{-2}\right).   
\label{Integral_Relation_1} 
\end{eqnarray}

\noindent 
On the other side, the integral in the first line of (\ref{gothic_metric_90}) is complicated because of the retarded time argument. 
Thus, while the time-time components (\ref{gothic_metric_80}) are already given in terms of multipoles, the spatial components (\ref{gothic_metric_90}) 
of the gothic metric are associated with a complicated integration procedure, ${\rm FP}_{B=0}\,\square_{\rm R}^{-1}$, consisting of the 
inverse d'Alembert operator and Hadamard's regularization, which is explained in more detail in Appendix \ref{Appendix5}.  

The spin-multipoles $S_L$ do not occur in (\ref{gothic_metric_80}) - (\ref{gothic_metric_90}) because they are terms of the order ${\cal O}\left(c^{-6}\right)$.  
A further comment should be in order. In the solution of Eqs.~(\ref{gothic_metric_80}) - (\ref{gothic_metric_90}) terms of the order  
${\cal O}\left(c^{-6}, c^{-5}, c^{-6}\right)$ are neglected \cite{Footnote9},   
while the perturbations are presented in terms of the retarded time argument, 
which is not further expanded in powers of the inverse of the speed of gravity. So the solution in (\ref{gothic_metric_80}) - (\ref{gothic_metric_90})  
is a hybrid representation in the sense that it is mixing the post-Minkowskian expansion (series in powers of $G$) and the post-Newtonian expansion 
(series in inverse powers of $c$). A good reason of such a representation is that the expressions (\ref{gothic_metric_80}) - (\ref{gothic_metric_90}) adopt  
their most simple form. But the main reason for the hybrid representation is that it permits to avoid problems regarding the convergence of the  
post-Newtonian expansion of the metric for non-compact support; cf. text below Eq.~(2.5) in \cite{2PN_Metric1}.  
In this respect we recall that the source of matter is assumed to be compact, but one has to keep in mind that the integral (\ref{metric_60})  
gets support inside and outside the matter source, that means it acquires a non-compact support; cf. text below Eq.~(\ref{Retarded_Time_2}). 
The non-linear relations (\ref{general_solution_E}) and (\ref{general_solution_F}) simplify in the corresponding approximation as follows 
(cf. Eq.~(6.3) in \cite{Blanchet4}, Eqs.~(99a) - (99b) in \cite{Blanchet6}, Eqs.~(5.11a) and (5.11b) in \cite{Gauge_Transformation}),
\begin{eqnarray}
M_L = I_L + {\cal O}\left(c^{-5}\right),
\label{M_L_2PN}
\\
S_L = J_L + {\cal O}\left(c^{-5}\right).
\label{S_L_2PN}
\end{eqnarray}

\noindent
The explicit expressions for the {\it canonical multipoles} $M_L$ and $S_L$ are given by Eqs.~(5.33) and (5.35) in \cite{Multipole_Damour_2} as integrals over  
the stress-energy tensor of the matter source, and they are represented by Eqs.~(\ref{M_L_Definition}) and (\ref{S_L_Definition}) in Appendix \ref{Appendix2}.
The linear gauge term in (\ref{Gauge_2PM}) is given by (cf. Eq.~(\ref{gauge_term_gothic_metric_PM}))  
\begin{eqnarray}
&& \partial \overline{\varphi}_{\left({\rm 2PM}\right)}^{\alpha\beta}\!\left(t,\ve{x}\right) 
= \varphi^{\alpha\,{\left({\rm 2PM}\right)}}_{\,,\,\mu}\!\left(t,\ve{x}\right)\eta^{\mu\beta}
+ \varphi^{\beta\,{\left({\rm 2PM}\right)}}_{\,,\,\mu}\!\left(t,\ve{x}\right)\eta^{\mu\alpha}
\nonumber\\ 
&& \hspace{2.5cm} - \varphi^{\mu\,{\left({\rm 2PM}\right)}}_{\,,\,\mu}\!\left(t,\ve{x}\right)\eta^{\alpha\beta}\,. 
\label{general_solution_C2}
\end{eqnarray}

\noindent
The gauge function $\varphi^{\alpha\,\left({\rm 2PM}\right)}$ is determined by Eq.~(\ref{sequence_gauge_2PM_MPM}). Its solution reads formally  
(cf. Eq.~(4.28) in \cite{Blanchet_Damour1})  
\begin{eqnarray}
\varphi^{\alpha\,\left({\rm 2PM}\right)}\left(t,\ve{x}\right) = {\rm FP}_{B=0}\,\square_{\rm R}^{-1} 
\left(\overline{h}^{\mu\nu\,{\rm gen}}_{\left({\rm 1PM}\right)}  
\;\varphi^{\alpha\,\left({\rm 1PM}\right)}_{\,,\,\mu\nu}\right)\left(t,\ve{x}\right)  
\nonumber\\ 
\label{gauge_vector_2PM_solution}
\end{eqnarray}

\noindent 
where ${\rm FP}_{B=0}\,\square_{\rm R}^{-1}$ is the Hadamard regularized inverse d'Alembertian (\ref{Inverse_d_Alembert_FP}). 
The formal solution (\ref{gauge_vector_2PM_solution}) leads to  
\begin{eqnarray}
&& \hspace{-1.0cm} \varphi^{\alpha\,\left({\rm 2PM}\right)}\left(t,\ve{x}\right) = \varphi^{\alpha\,\left({\rm 2PM}\right)}\left[I_L,J_L,W_L,X_L,Y_L,Z_L\right],   
\label{gauge_vector_2PM}
\end{eqnarray}

\noindent
that means the gauge function $\varphi^{\alpha\,\left({\rm 2PM}\right)}$ depends on the full set of the STF source moments. 
The explicit expression for the gauge function in (\ref{gauge_vector_2PM}) is complicated, but we will not pursue it here because one may show that 
\begin{eqnarray}
\partial \overline{\varphi}_{\left({\rm 2PM}\right)}^{\alpha\beta}\left(t,\ve{x}\right) = {\cal O}\left(c^{-6},c^{-5},c^{-6}\right),   
\label{order_of_2PM_gauge_term} 
\end{eqnarray}

\noindent
which is of the same order of the neglected terms in the canonical gothic metric perturbation in (\ref{gothic_metric_80}) - (\ref{gothic_metric_90}). 
The non-linear gauge term of the coordinate transformation reads (cf. Eq.~(\ref{gauge_term_gothic_metric_2PM}))  
\begin{eqnarray}
&& \hspace{-0.5cm} \overline{\Omega}_{\left({\rm 1PM}\right)}^{\alpha\beta}\left(t,\ve{x}\right) = 0\,,
\label{Gauge_2PM_A}
\\
\nonumber\\
&& \hspace{-0.5cm} \overline{\Omega}_{\left({\rm 2PM}\right)}^{\alpha\beta}\left(t,\ve{x}\right) =
\varphi^{\alpha\,{\left({\rm 1PM}\right)}}_{\,,\,\mu}\left(t,\ve{x}\right)\,
\varphi^{\beta\,{\left({\rm 1PM}\right)}}_{\,,\,\nu}\left(t,\ve{x}\right)\,\eta^{\mu\nu}
\nonumber\\
&& + \,\left(\varphi^{\nu\,\left({\rm 1PM}\right)}\left(t,\ve{x}\right)
\,\overline{h}_{\left({\rm 1PM}\right)}^{\alpha\beta\,{\rm can}}\left(t,\ve{x}\right)\right)_{\,,\,\nu}
\nonumber\\
&& - \,\varphi^{\alpha\,{\left({\rm 1PM}\right)}}_{\,,\,\mu}\left(t,\ve{x}\right)
\left(\overline{h}_{\left({\rm 1PM}\right)}^{\mu\beta\,{\rm can}}\left(t,\ve{x}\right) 
+ \partial\overline{\varphi}^{\mu\beta}_{\left({\rm 1PM}\right)}\left(t,\ve{x}\right)\right)
\nonumber\\
&& - \, \varphi^{\beta\,{\left({\rm 1PM}\right)}}_{\,,\,\mu}\left(t,\ve{x}\right)
\left(\overline{h}_{\left({\rm 1PM}\right)}^{\mu\alpha\,{\rm can}}\left(t,\ve{x}\right)
+ \partial\overline{\varphi}^{\mu\alpha}_{\left({\rm 1PM}\right)}\left(t,\ve{x}\right)\right) 
\nonumber\\
&& + \,\frac{1}{2}\,\varphi^{\mu\,{\left({\rm 1PM}\right)}}_{\,,\,\nu}\left(t,\ve{x}\right)
\,\varphi^{\nu\,{\left({\rm 1PM}\right)}}_{\,,\,\mu}\left(t,\ve{x}\right)\, \eta^{\alpha\beta}
\nonumber\\
&& - \,\frac{1}{2}\,\varphi^{\mu\,{\left({\rm 1PM}\right)}}_{\,,\,\mu}\left(t,\ve{x}\right)
\,\varphi^{\nu\,{\left({\rm 1PM}\right)}}_{\,,\,\nu}\left(t,\ve{x}\right)\,\eta^{\alpha\beta}\;. 
\label{Gauge_2PM_B}
\end{eqnarray}

\noindent
The gauge function $\varphi^{\alpha\,\left({\rm 1PM}\right)}$ on the r.h.s. in (\ref{Gauge_2PM_B}) depends on four source multipoles  
(cf. Eq.~(\ref{gauge_vector_1PM})) and is explicitly given by Eqs.~(5.31b) in \cite{Multipole_Damour_2}; see also Eqs.~(4.13a) - (4.13b) in \cite{Blanchet4} 
or Eqs.~(3.560) - (3.561) in \cite{Kopeikin_Efroimsky_Kaplan}. 
The gauge term $\partial \overline{\varphi}_{\left({\rm 1PM}\right)}^{\alpha\beta}$ is given by Eq.~(\ref{general_solution_C1}),  
while the 1PM canonical gothic metric perturbation $\overline{h}_{\left({\rm 1PM}\right)}^{\alpha\beta\,{\rm can}}$ has been given by 
Eqs.~(\ref{1PM_Metric_A}) - (\ref{1PM_Metric_C}). It has been checked that one would obtain the same non-linear
gauge term $\overline{\Omega}^{\alpha\beta}_{\left({\rm 2PM}\right)}$ as given by Eq.~(4.7a) in \cite{Gauge_Transformation} if one
would use the residual gauge transformation (25) instead of (22) and if one would series-expand the gothic metric in the system $\{ x^{\prime}\}$
which then would have to be endowed by Minkowskian coordinates; note the different sign-convention for the gothic metric perturbation.  
Here it should be emphasized again that the canonical piece of the gothic metric density
(and of the metric tensor) is gauge-independent, hence is independent of whether one uses the residual gauge transformation (\ref{harmonic_gauge_condition_2}) 
or (\ref{harmonic_gauge_condition_inverse}) (cf. text below Eq.~(\ref{Laplace_Beltrami_Equation_inverse})).

\section{The metric tensor in post-linear approximation}\label{Section5}

The post-Minkowskian expansion of the metric tensor up to terms of the order 
${\cal O}\left(G^3\right)$ is given by (cf. Eqs.~(\ref{expansion_metric_1a}) and (\ref{expansion_metric_1}))  
\begin{eqnarray}
g_{\alpha \beta}\!\left(t,\ve{x}\right) &=& \eta_{\alpha \beta} + G^1 h_{\alpha \beta}^{\left({\rm 1PM}\right)}\left(t,\ve{x}\right)  
+ G^2 h_{\alpha \beta}^{\left({\rm 2PM}\right)}\left(t,\ve{x}\right)
\nonumber\\
&& + {\cal O}\left(G^3\right).  
\label{expansion_metric}
\end{eqnarray}

\noindent
In this Section the linear term $h_{\alpha\beta}^{\left({\rm 1PM}\right)}$ and
the post-linear term $h_{\alpha\beta}^{\left({\rm 2PM}\right)}$ of the metric tensor are considered.

\subsection{The linear term of the metric tensor}\label{Section5_1}

In order to determine the linear term $h_{\alpha\beta}^{\left({\rm 1PM}\right)}$ of the metric, the following relation between   
\begin{eqnarray}
h_{\alpha\beta}^{\left({\rm 1PM}\right)}
= \overline{h}^{\mu\nu}_{\left({\rm 1PM}\right)}\, \eta_{\alpha\mu}\,\eta_{\beta\nu}
- \frac{1}{2}\,\overline{h}_{\left({\rm 1PM}\right)}\;\eta_{\alpha\beta}\,,
\label{Metric_Relation_Indices_1PM}
\end{eqnarray}

\noindent
where $\overline{h}_{\left({\rm 1PM}\right)} = \eta_{\mu\nu}\,\overline{h}^{\mu\nu}_{\left({\rm 1PM}\right)}$.  
The relation (\ref{Metric_Relation_Indices_1PM}) allows to determine the covariant components of the metric tensor from the contravariant components of the
gothic metric in 1PM approximation.  
By inserting Eq.~(\ref{Gauge_1PM_C}) into Eq.~(\ref{Metric_Relation_Indices_1PM}) with the expressions  
in (\ref{1PM_Metric_A}) - (\ref{1PM_Metric_C}) and (\ref{general_solution_C1}), one obtains the general solution for the 
metric perturbation in the 1PM approximation,  
\begin{eqnarray}
&& h^{\left({\rm 1PM}\right)}_{\alpha\beta\,{\rm gen}}\left[I_L,J_L,W_L,X_L,Y_L,Z_L\right]
\nonumber\\ 
&& = h^{\left({\rm 1PM}\right)}_{\alpha\beta\,{\rm can}}\left[M_L,S_L\right] + \partial \varphi^{\left({\rm 1PM}\right)}_{\alpha\beta}\left(t,\ve{x}\right)\,.  
\label{Metric_Gauge_1PM}
\end{eqnarray}

\noindent
The linear term of the {\it canonical metric perturbation} for one body at rest with full multipole structure is given by
\begin{eqnarray}
h^{\left({\rm 1PM}\right)}_{00\,{\rm can}}\left(t,\ve{x}\right) &=&  
+ \frac{2}{c^2}\sum\limits_{l=0}^{\infty}\frac{\left(-1\right)^l}{l!}\,
\partial_L \frac{M_L\left(s\right)}{r}\,,
\label{1PM_Metric_00}
\\
\nonumber\\
h^{\left({\rm 1PM}\right)}_{0i\,{\rm can}}\left(t,\ve{x}\right) &=&  
+ \frac{4}{c^3}\sum\limits_{l=1}^{\infty} \frac{\left(-1\right)^l}{l!} \partial_{L-1}
\frac{\dot{M}_{iL-1} \left(s\right)}{r}
\nonumber\\
&& \hspace{-1.25cm} + \frac{4}{c^3}\sum\limits_{l=1}^{\infty} \frac{\left(-1\right)^l\,l}{\left(l+1\right)!} \epsilon_{iab}\,\partial_{a L-1}
\frac{S_{bL-1} \left(s\right)}{r}\,,
\label{1PM_Metric_0i}
\\
\nonumber\\
h^{\left({\rm 1PM}\right)}_{ij\,{\rm can}}\left(t,\ve{x}\right) &=& + \frac{2}{c^2} \, \delta_{ij} \sum\limits_{l=0}^{\infty}\frac{\left(-1\right)^l}{l!}\,
\partial_L \frac{M_L\left(s\right)}{r}
\nonumber\\
&& \hspace{-1.25cm} + \frac{4}{c^4}\sum\limits_{l=2}^{\infty} \frac{\left(-1\right)^l}{l!} \partial_{L-2}
\frac{\ddot{M}_{ijL-2} \left(s\right)}{r}
\nonumber\\
&& \hspace{-1.25cm} + \frac{8}{c^4}\sum\limits_{l=2}^{\infty} \frac{\left(-1\right)^l\,l}{\left(l+1\right)!}
\partial_{a L-2} \frac{\epsilon_{ab(i} \dot{S}_{j)bL-2} \left(s\right)}{r}\,.
\label{1PM_Metric_ij}
\end{eqnarray}

\noindent
In order to get (\ref{1PM_Metric_00}) - (\ref{1PM_Metric_ij}) we made use of the property that $\ddot{M}_{ii} = 0$ since the multipoles are trace-free,  
as well as of the identity $\epsilon_{ab(i}\,\dot{S}_{i)bL-2} = 0$ due to antisymmetry of the Levi-Civita symbol and the symmetry of multipoles.
One may verify that Eqs.~(\ref{1PM_Metric_00}) - (\ref{1PM_Metric_ij}) agree with Eq.~(2) in \cite{Zschocke_Soffel};
just use the decomposition of the metric tensor as given by Eq.~(A.1) in \cite{Zschocke_Soffel} and apply the orthogonality relation of the metric tensor
in 1PM approximation. The gauge term in (\ref{Metric_Gauge_1PM}) reads (cf. Eq.~(\ref{gauge_term_metric_PM}))  
\begin{eqnarray}
\partial \varphi^{\left({\rm 1PM}\right)}_{\alpha\beta}\!\left(t,\ve{x}\right) 
= \varphi^{\mu\,{\left({\rm 1PM}\right)}}_{\,,\,\alpha}\!\left(t,\ve{x}\right) \eta_{\mu\beta} 
+ \varphi^{\mu\,{\left({\rm 1PM}\right)}}_{\,,\,\beta}\!\left(t,\ve{x}\right) \eta_{\mu\alpha}\,,
\nonumber\\ 
\label{gauge_term_1PM}
\end{eqnarray}

\noindent
where the gauge function $\varphi^{\alpha\,\left({\rm 1PM}\right)}$ on the r.h.s. in (\ref{gauge_term_1PM}) is governed by Eq.~(\ref{sequence_gauge_1PM_MPM}). 
The gauge function depends on four source multipoles (cf. Eq.~(\ref{gauge_vector_1PM})) and its explicit form is given by 
Eqs.~(5.31b) in \cite{Multipole_Damour_2};  
see also Eqs.~(4.13a) - (4.13b) in \cite{Blanchet4} or Eqs.~(3.560) - (3.561) in \cite{Kopeikin_Efroimsky_Kaplan}.

$\vspace{0.0cm}$  

\subsection{The post-linear term of the metric tensor}\label{Section5_2}

In order to determine the post-linear term $h_{\alpha\beta}^{\left({\rm 2PM}\right)}$ of the metric,  
the following relation between the metric and gothic metric is used, which is shown in Appendix \ref{Appendix4} (cf. Eq.~(1.6.3) in \cite{Poisson_Lecture_Notes}),  
\begin{eqnarray}
h_{\alpha\beta}^{\left({\rm 2PM}\right)} &=& \overline{h}^{\mu\nu}_{\left({\rm 2PM}\right)}\, \eta_{\alpha\mu}\,\eta_{\beta\nu}  
- \frac{1}{2}\,\overline{h}_{\left({\rm 2PM}\right)}\,\eta_{\alpha\beta} 
+ \frac{1}{8}\,\overline{h}^2_{\left({\rm 1PM}\right)}\,\eta_{\alpha\beta}  
\nonumber\\
&& \hspace{-1.0cm} - \frac{1}{2}\,\overline{h}_{\left({\rm 1PM}\right)}\,\overline{h}^{\mu\nu}_{\left({\rm 1PM}\right)}\,\eta_{\alpha\mu}\,\eta_{\beta\nu}
+ \overline{h}^{\rho\nu}_{\left({\rm 1PM}\right)}\,\overline{h}^{\mu\sigma}_{{\left({\rm 1PM}\right)}}
\,\eta_{\mu\nu}\,\eta_{\alpha\rho}\,\eta_{\beta\sigma} 
\nonumber\\
&& \hspace{-1.0cm} 
- \frac{1}{4}\,\overline{h}^{\mu\nu}_{\left({\rm 1PM}\right)}\,\overline{h}^{\rho\sigma}_{\left({\rm 1PM}\right)}\,
\eta_{\mu\rho}\,\eta_{\nu\sigma}\,\eta_{\alpha\beta}\;,
\label{Retartion_Gothic_Metric_2}
\end{eqnarray}

\noindent
where $\overline{h}_{\left({\rm 2PM}\right)} = \eta_{\mu\nu}\,\overline{h}^{\mu\nu}_{\left({\rm 2PM}\right)}$. The relation (\ref{Retartion_Gothic_Metric_2})  
allows to determine the covariant components of the metric tensor from the contravariant components of the gothic metric in 2PM approximation.  
By inserting Eqs.~(\ref{Gauge_1PM_C}) and (\ref{Gauge_2PM}) into (\ref{Retartion_Gothic_Metric_2}) with the expressions  
in (\ref{1PM_Metric_A}) - (\ref{1PM_Metric_C}) and (\ref{general_solution_C1}) as well as (\ref{gothic_metric_80}) - (\ref{gothic_metric_90}) 
and (\ref{general_solution_C2}) and (\ref{Gauge_2PM_B}), one obtains the general solution for the metric perturbation in the 2PM approximation,  
\begin{eqnarray}
&& \hspace{-0.75cm} h^{\left({\rm 2PM}\right)}_{\alpha\beta\,{\rm gen}}\left[I_L,J_L,W_L,X_L,Y_L,Z_L\right]
\nonumber\\ 
&& \hspace{-0.75cm} \!=\! h^{\left({\rm 2PM}\right)}_{\alpha\beta\,{\rm can}}\left[M_L,S_L\right] 
+ \partial \varphi^{\left({\rm 2PM}\right)}_{\alpha\beta}\!\left(t,\ve{x}\right) 
+ \Omega^{\left({\rm 2PM}\right)}_{\alpha\beta}\!\left(t,\ve{x}\right).
\label{Metric_Gauge_2PM}
\end{eqnarray}

\noindent
The post-linear term of the {\it canonical metric perturbation} for one body at rest with full multipole structure reads  
\begin{widetext}
\begin{eqnarray}
h_{00\,{\rm can}}^{\left({\rm 2PM}\right)}\left(t,\ve{x}\right) &=& - \frac{2}{c^4}
\left(\sum\limits_{l=0}^{\infty} \frac{\left(-1\right)^{l}}{l!}\,\partial_L\,\frac{M_L\left(s\right)}{r}\right)^2 + {\cal O}\left(c^{-6}\right),  
\label{metric_80}
\\
\nonumber\\
 h_{0i\,{\rm can}}^{\left({\rm 2PM}\right)}\left(t,\ve{x}\right) &=& {\cal O}\left(c^{-5}\right),  
\label{metric_85}
\\
\nonumber\\
h_{ij\,{\rm can}}^{\left({\rm 2PM}\right)}\left(t,\ve{x}\right) &=& - \frac{4}{c^4} \,{\rm FP}_{B=0}\,\square_{\rm R}^{-1}
\left(\partial_i \sum\limits_{l=0}^{\infty}
\frac{\left(-1\right)^{l}}{l!}\,\partial_L\,\frac{M_L\left(s\right)}{r}\right)
\left(\partial_j \sum\limits_{l=0}^{\infty}
\frac{\left(-1\right)^{l}}{l!}\,\partial_L\,\frac{M_L\left(s\right)}{r}\right)
\nonumber\\
\nonumber\\
&& + \frac{2}{c^4}\,\delta_{ij}
\left(\sum\limits_{l=0}^{\infty} \frac{\left(-1\right)^{l}}{l!}\,\partial_L\,\frac{M_L\left(s\right)}{r}\right)^2
+ {\cal O}\left(c^{-6}\right).  
\label{metric_90}
\end{eqnarray}
\end{widetext}

\noindent
The spatial components (\ref{metric_90}) of the metric tensor are associated with an  
integration procedure, ${\rm FP}_{B=0}\,\square_{\rm R}^{-1}$, consisting of the inverse d'Alembert operator and Hadamard's regularization,  
which is explained in Appendix \ref{Appendix5}. The canonical mass-multipoles $M_L$ are given by Eq.~(5.33) in \cite{Multipole_Damour_2} as  
integrals over the stress-energy tensor of the matter source; note that there are no  
spin-multipoles $S_L$ in (\ref{metric_80}) - (\ref{metric_90}) because they are terms of the order ${\cal O}\left(c^{-6}\right)$.  
The linear gauge term in (\ref{Metric_Gauge_2PM}) is given by (cf. Eq.~(\ref{gauge_term_metric_PM}))  
\begin{eqnarray}
\partial \varphi^{\left({\rm 2PM}\right)}_{\alpha\beta}\!\left(t,\ve{x}\right) 
= \varphi^{\mu\,{\left({\rm 2PM}\right)}}_{\,,\,\alpha}\!\left(t,\ve{x}\right) \eta_{\mu\beta}
+ \varphi^{\mu\,{\left({\rm 2PM}\right)}}_{\,,\,\beta}\!\left(t,\ve{x}\right) \eta_{\mu\alpha}\,,
\nonumber\\ 
\label{Gauge_2PM_B1}
\end{eqnarray}

\noindent
where the gauge function $\varphi^{\alpha\,\left({\rm 2PM}\right)}$ on the r.h.s. in (\ref{Gauge_2PM_B1}) is governed by Eq.~(\ref{sequence_gauge_2PM_MPM}).  
It's explicit form is formally given by Eq.~(\ref{gauge_vector_2PM_solution}) and depends on all six STF source multipoles; cf. Eq.~(\ref{gauge_vector_2PM}).  
That explicit expression for the gauge function in (\ref{Gauge_2PM_B1}) is complicated and we will not pursue it here because one may show that
\begin{eqnarray}
\partial \varphi^{\left({\rm 2PM}\right)}_{\alpha\beta}\left(t,\ve{x}\right) = {\cal O}\left(c^{-6},c^{-5},c^{-6}\right), 
\label{order_of_2PM_gauge_term_metric}
\end{eqnarray}

\noindent 
which is of the same order of the neglected terms in the canonical metric perturbation in (\ref{metric_80}) - (\ref{metric_90}). 
The non-linear gauge term in (\ref{Metric_Gauge_2PM}) reads (cf. Eq.~(\ref{gauge_term_metric_2PM}))  
\begin{eqnarray}
\Omega^{\left({\rm 2PM}\right)}_{\alpha\beta}\left(t,\ve{x}\right) 
&=& h^{\left({\rm 1PM}\right)}_{\mu\beta\,{\rm can}}\left(t,\ve{x}\right) \,\varphi^{\mu\,{\left({\rm 1PM}\right)}}_{\,,\,\alpha}\left(t,\ve{x}\right)  
\nonumber\\  
&& \hspace{-1.0cm} + h^{\left({\rm 1PM}\right)}_{\mu\alpha\,{\rm can}}\left(t,\ve{x}\right) 
\varphi^{\mu\,{\left({\rm 1PM}\right)}}_{\,,\,\beta}\left(t,\ve{x}\right) 
\nonumber\\  
&& \hspace{-1.0cm} + h^{\left({\rm 1PM}\right)}_{\alpha\beta\,{\rm can}\,,\,\nu}\left(t,\ve{x}\right) 
\varphi^{\nu\,\left({\rm 1PM}\right)}\left(t,\ve{x}\right)  
\nonumber\\  
&& \hspace{-1.0cm} + \varphi^{\mu\,{\left({\rm 1PM}\right)}}_{\,,\,\alpha}\left(t,\ve{x}\right)
\varphi^{\nu\,{\left({\rm 1PM}\right)}}_{\,,\,\beta}\left(t,\ve{x}\right) 
\eta_{\mu\nu}\,.  
\label{Gauge_2PM_B2}
\end{eqnarray}

\noindent
The gauge function $\varphi^{\alpha\,\left({\rm 1PM}\right)}$ on the r.h.s. in (\ref{Gauge_2PM_B2})   
is formally given by Eq.~(\ref{gauge_vector_1PM}) and explicitly given by Eqs.~(5.31b) in \cite{Multipole_Damour_2}; see also 
Eqs.~(4.13a) - (4.13b) in \cite{Blanchet4} or Eqs.~(3.560) - (3.561) in \cite{Kopeikin_Efroimsky_Kaplan}.   
The linear gauge term $\partial \varphi^{\left({\rm 1PM}\right)}_{\alpha\beta}$ is given by  
Eq.~(\ref{gauge_term_1PM}), while the canonical linear metric perturbation $h^{\left({\rm 1PM}\right)}_{\alpha\beta\,{\rm can}}$ is 
given by Eqs.~(\ref{1PM_Metric_00}) - (\ref{1PM_Metric_ij}).

\section{Stationary sources}\label{Section6}  

In many applications of general theory of relativity it is possible to neglect the 
time-dependence of the matter source and to consider a stationary source, defined by  
\begin{eqnarray} 
T^{\mu\nu}_{\;\;\;\;\,,\,0} = 0\,,  
\label{Stationary_Source_5} 
\end{eqnarray} 
 
\noindent  
that means an approximation where the stress-energy tensor is only a function of the spatial coordinates in the harmonic reference system.   
The condition (\ref{Stationary_Source_5}) does not necessarily imply that the source of matter is static. Namely, a static source implies  
that there is no motion at all inside the source of matter, while a stationary source only requires that motions of matter 
(e.g. inner circulations) have to be time-independent. Stated differently, for static sources not only Eq.~(\ref{Stationary_Source_5}) 
holds but in addition $T^{0i} = 0$, while for stationary sources $T^{0i} = {\rm const} \neq 0$ is possible.  
The metric of a stationary source is time-independent,   
\begin{eqnarray}
g_{\alpha\beta\;,\,0} = 0\,,  
\label{stationary_metric}
\end{eqnarray}

\noindent 
that means the metric tensor depends only on spatial coordinates. 
Let us notice that for a stationary metric $g_{0i} = {\rm const} \neq 0$ is possible, while for a static metric $g_{0i} = 0$ (cf. Eq.~(56.02) in \cite{Fock}).  
In the stationary case the post-Minkowskian expansion of the metric tensor up to terms of the order ${\cal O}\left(G^3\right)$  
reads (cf. Eq.~(\ref{expansion_metric_3}))
\begin{eqnarray}
g_{\alpha \beta}\left(\ve{x}\right) &=& \eta_{\alpha \beta} + G^1\,h_{\alpha \beta}^{\left({\rm 1PM}\right)}\left(\ve{x}\right)
+ G^2\,h_{\alpha \beta}^{\left({\rm 2PM}\right)}\left(\ve{x}\right) 
+ {\cal O}\left(G^3\right).
\nonumber\\ 
\label{expansion_metric_2PM_static}
\end{eqnarray}

\noindent
In this Section the linear term $h_{\alpha \beta}^{\left({\rm 1PM}\right)}$ and the post-linear term $h_{\alpha \beta}^{\left({\rm 2PM}\right)}$
are considered.

\subsection{The linear term of the metric tensor for stationary sources} 

According to Eq.~(\ref{Metric_Gauge_1PM}) the residual gauge transformation of the 1PM terms of the metric perturbation for stationary sources reads  
\begin{eqnarray}
&& \hspace{-1.0cm} h^{\left({\rm 1PM}\right)}_{\alpha\beta\,{\rm gen}} \left[I_L, J_L,W_L,X_L,Y_L,Z_L\right]
\nonumber\\ 
&& \hspace{0.25cm} 
= h^{\left({\rm 1PM}\right)}_{\alpha\beta\,{\rm can}}\left[M_L,S_L\right] + \partial \varphi^{\left({\rm 1PM}\right)}_{\alpha\beta}\left(\ve{x}\right)\,, 
\label{stationary_2PM_10}  
\end{eqnarray}

\noindent  
where the source multipoles $I_L,J_L,W_L,X_L,Y_L,Z_L$ and the canonical multipoles $M_L,S_L$ are time-independent now. 
For the time-independent canonical multipoles one obtains from Eqs.~(\ref{M_L_Definition}) and (\ref{S_L_Definition}) in Appendix \ref{Appendix2} 
(cf. Eqs.~(5.33) and (5.35) in \cite{Multipole_Damour_2}), 
\begin{eqnarray}
M_L &=& \int d^3 x \; \hat{x}_L\;\frac{T^{00} + T^{kk}}{c^2}\;, 
\label{M_L}
\\
\nonumber\\ 
S_L &=& {\rm STF}_L \int d^3 x \; \hat{x}_{L-1}\;\epsilon_{i_l j k}\;x^j\;\frac{T^{0k}}{c}\;.   
\label{S_L}
\end{eqnarray}

\noindent 
For stationary sources the canonical metric perturbation in (\ref{stationary_2PM_10}) simplifies considerably.  
From Eqs.~(\ref{1PM_Metric_00}) - (\ref{1PM_Metric_ij}) one obtains  
\begin{eqnarray}
h^{\left({\rm 1PM}\right)}_{00\,{\rm can}}\left(\ve{x}\right) &=&   
\frac{2}{c^2}\sum\limits_{l=0}^{\infty}\frac{\left(-1\right)^l}{l!}\,
\partial_L \frac{M_L}{r}\,,
\label{canonical_metric_1PN_00}
\\
\nonumber\\
 h^{\left({\rm 1PM}\right)}_{0i\,{\rm can}}\left(\ve{x}\right) &=&  
\frac{4}{c^3}\sum\limits_{l=1}^{\infty} \frac{\left(-1\right)^l\,l}{\left(l+1\right)!} \epsilon_{iab}\,\partial_{a L-1}
\frac{S_{bL-1}}{r}\,,
\label{canonical_metric_1PN_0i}
\\
\nonumber\\
 h^{\left({\rm 1PM}\right)}_{ij\,{\rm can}}\left(\ve{x}\right) &=&  
\frac{2}{c^2}\,\delta_{ij} \sum\limits_{l=0}^{\infty}\frac{\left(-1\right)^l}{l!}\,
\partial_L \frac{M_L}{r}\,.  
\label{canonical_metric_1PN_ij}
\end{eqnarray}

\noindent 
This is the linear term of the canonical metric in case of a stationary source. 
The gauge term $\partial \varphi^{\left({\rm 1PM}\right)}_{\alpha\beta}$ in (\ref{stationary_2PM_10}) is given by  
\begin{eqnarray}
\partial \varphi^{\left({\rm 1PM}\right)}_{\alpha\beta}\left(\ve{x}\right) 
= \varphi^{\mu\,{\left({\rm 1PM}\right)}}_{\,,\,\alpha}\left(\ve{x}\right)\eta_{\mu\beta}
+ \varphi^{\mu\,{\left({\rm 1PM}\right)}}_{\,,\,\beta}\left(\ve{x}\right)\eta_{\mu\alpha}\,,
\nonumber\\ 
\label{gauge_term_1PM_stationary}
\end{eqnarray}

\noindent  
which depends on the gauge function $\varphi^{\alpha\,\left({\rm 1PM}\right)}$ determined by  
\begin{eqnarray}
\Delta\,\varphi^{\alpha\,\left({\rm 1PM}\right)}\left(\ve{x}\right) = 0\,,
\label{sequence_gauge_1PM_stationary}
\end{eqnarray}

\noindent 
which follow from (\ref{sequence_gauge_1PM_MPM}) in the case of time-independence of the gauge functions; $\Delta = \partial_k\,\partial_k$ 
is the flat Laplace operator and the gauge function is time-independent: $\varphi^{\alpha\,\left({\rm 1PM}\right)}_{\,,\,0} = 0$.  
This gauge function is formally given by 
\begin{eqnarray}
\varphi^{\alpha\,\left({\rm 1PM}\right)}\left(\ve{x}\right) = \left[W_L,X_L,Y_L,Z_L\right]\,,
\label{gauge_vector_1PM_stationary}
\end{eqnarray}

\noindent 
that means it depends on four STF source multipoles which are time-independent now. An explicit expression of (\ref{gauge_vector_1PM_stationary})  
can be deduced from Eqs.~(5.31b) in \cite{Multipole_Damour_2} by taking the limit of vanishing time argument.

\subsection{The post-linear term of the metric tensor for stationary sources} 

According to Eq.~(\ref{Metric_Gauge_2PM}) the residual gauge transformation of the 2PM terms of the metric perturbation for stationary sources reads  
\begin{eqnarray}
&& \hspace{-0.75cm} h^{\left({\rm 2PM}\right)}_{\alpha\beta\,{\rm gen}}\left[I_L,J_L,W_L,X_L,Y_L,Z_L\right]
\nonumber\\ 
&& \hspace{-0.5cm} = h^{\left({\rm 2PM}\right)}_{\alpha\beta\,{\rm can}}\left[M_L,S_L\right] 
+ \partial \varphi^{\left({\rm 2PM}\right)}_{\alpha\beta}\left(\ve{x}\right) + \Omega^{\left({\rm 2PM}\right)}_{\alpha\beta}\left(\ve{x}\right),  
\label{stationary_2PM_20}
\end{eqnarray}

\noindent 
where the source multipoles $I_L,J_L,W_L,X_L,Y_L,Z_L$ and the canonical multipoles $M_L,S_L$ are time-independent now.  
For stationary sources the canonical metric perturbation in (\ref{stationary_2PM_20}) simplifies considerably.  
From Eqs.~(\ref{metric_80}) - (\ref{metric_90}) one obtains  
\begin{widetext}
\begin{eqnarray}
h_{00\,{\rm can}}^{\left({\rm 2PM}\right)}\left(\ve{x}\right) &=& - \frac{2}{c^4}
\left(\sum\limits_{l=0}^{\infty} \frac{\left(-1\right)^{l}}{l!}\,\partial_L \frac{M_L}{r}\right)^2
+ {\cal O}\left(c^{-6}\right), 
\label{canonical_metric_2PN_00}
\\
\nonumber\\
h_{0i\,{\rm can}}^{\left({\rm 2PM}\right)}\left(\ve{x}\right) &=& {\cal O}\left(c^{-5}\right), 
\label{canonical_metric_2PN_0i}
\\
\nonumber\\
h_{ij\,{\rm can}}^{\left({\rm 2PM}\right)}\left(\ve{x}\right) &=&  
- \frac{4}{c^4} \,{\rm FP}_{B=0}\,\Delta^{-1}
\left(\frac{\partial}{\partial x_i}\sum\limits_{l=0}^{\infty}
\frac{\left(-1\right)^{l}}{l!}\,\partial_L\,\frac{M_L}{r}\right)
\left(\frac{\partial}{\partial x_j}\sum\limits_{l=0}^{\infty}
\frac{\left(-1\right)^{l}}{l!}\,\partial_L\,\frac{M_L}{r}\right)
\nonumber\\
&& + \frac{2}{c^2}\,\delta_{ij}
\left(\sum\limits_{l=0}^{\infty} \frac{\left(-1\right)^{l}}{l!}\,\partial_L \frac{M_L}{r}\!\right)^2 + \, {\cal O}\left(c^{-6}\right). 
\label{canonical_metric_2PN_ij}
\end{eqnarray}
\end{widetext}

\noindent 
This is the post-linear term of the canonical metric in case of a stationary source.  
The spatial components of the 
canonical post-linear metric in (\ref{canonical_metric_2PN_ij}) are associated with an integration procedure via the Hadamard regularized 
inverse Laplace operator, ${\rm FP}_{B=0}\,\Delta^{-1}$, defined by Eq.~(\ref{Inverse_Laplacian_FP}).  
The gauge term $\partial \varphi^{\left({\rm 2PM}\right)}_{\alpha\beta}$ reads 
\begin{eqnarray}
\partial \varphi^{\left({\rm 2PM}\right)}_{\alpha\beta}\left(\ve{x}\right) 
= \varphi^{\mu\,{\left({\rm 2PM}\right)}}_{\,,\,\alpha}\left(\ve{x}\right)\eta_{\mu\beta}
+ \varphi^{\mu\,{\left({\rm 2PM}\right)}}_{\,,\,\beta}\left(\ve{x}\right)\eta_{\mu\alpha}\,,
\nonumber\\ 
\label{gauge_term_2PM_stationary}
\end{eqnarray}

\noindent 
which is time-independent, that means $\varphi^{\alpha\,\left({\rm 2PM}\right)}_{\,,\,0} = 0$. 
The gauge term $\partial \varphi^{\left({\rm 2PM}\right)}_{\alpha\beta}$ depends on the gauge function  
$\varphi^{\alpha\,\left({\rm 2PM}\right)}$ which is determined by the equation 
\begin{eqnarray}
\Delta\, \varphi^{\alpha\,\left({\rm 2PM}\right)}\left(\ve{x}\right) = \overline{h}^{ij\,{\rm gen}}_{\left({\rm 1PM}\right)}\left(\ve{x}\right)
\,\varphi^{\alpha\,\left({\rm 1PM}\right)}_{\,,\,ij}\left(\ve{x}\right) \,,
\label{sequence_gauge_2PM_stationary}
\end{eqnarray}

\noindent  
as it follows from (\ref{sequence_gauge_2PM_MPM}) in the limit of time-independence of the gauge functions; 
$\Delta = \partial_k\,\partial_k$ is the flat Laplace operator. A formal solution is provided by 
\begin{eqnarray}
\varphi^{\alpha\,\left({\rm 2PM}\right)}\left(\ve{x}\right) = {\rm FP}_{B=0}\, \Delta^{-1}   
\left(\overline{h}^{ij\,{\rm gen}}_{\left({\rm 1PM}\right)}
\,\varphi^{\alpha\,\left({\rm 1PM}\right)}_{\,,\,ij}\right)\left(\ve{x}\right)\,,
\nonumber\\ 
\label{gauge_vector_2PM_solution_stationary}
\end{eqnarray}
 
\noindent
where ${\rm FP}_{B=0}\,\Delta^{-1}$ is the Hadamard regularized inverse Laplacian defined by Eq.~(\ref{Inverse_Laplacian_FP}).  
This gauge function is formally given by 
\begin{eqnarray}
\varphi^{\alpha\,\left({\rm 2PM}\right)}\left(\ve{x}\right) = \left[I_L,J_L,W_L,X_L,Y_L,Z_L\right]\,,
\label{gauge_vector_2PM_stationary}
\end{eqnarray}

\noindent 
that means it depends on all the six STF source multipoles which are time-independent now. The explicit expression for the gauge function  
in (\ref{gauge_vector_2PM_stationary}) is complicated but not relevant because of (\ref{order_of_2PM_gauge_term_metric}).  
The non-linear gauge term $\Omega^{\left({\rm 2PM}\right)}_{\alpha\beta}$ in (\ref{stationary_2PM_20}) reads 
\begin{eqnarray}
\Omega^{\left({\rm 2PM}\right)}_{\alpha\beta}\left(\ve{x}\right)
&=& h^{\left({\rm 1PM}\right)}_{\mu\beta\,{\rm can}}\left(\ve{x}\right) \,\varphi^{\mu\,{\left({\rm 1PM}\right)}}_{\,,\,\alpha}\left(\ve{x}\right)
\nonumber\\
&& \hspace{-2.0cm} + h^{\left({\rm 1PM}\right)}_{\mu\alpha\,{\rm can}}\left(\ve{x}\right)\,\varphi^{\mu\,{\left({\rm 1PM}\right)}}_{\,,\,\beta}\left(\ve{x}\right) 
+ h^{\left({\rm 1PM}\right)}_{\alpha\beta\,{\rm can}\,,\,\nu}\left(\ve{x}\right) \, \varphi^{\nu\,\left({\rm 1PM}\right)}\left(\ve{x}\right)
\nonumber\\
&& \hspace{-2.0cm} + \varphi^{\mu\,{\left({\rm 1PM}\right)}}_{\,,\,\alpha}\left(\ve{x}\right)\,\varphi^{\nu\,{\left({\rm 1PM}\right)}}_{\,,\,\beta}\left(\ve{x}\right)
\, \eta_{\mu\nu}\,, 
\label{Gauge_2PM_B2_stationary}
\end{eqnarray}

\noindent  
and depends on the gauge function $\varphi^{\alpha\,\left({\rm 1PM}\right)}\left(\ve{x}\right)$ which is formally given by Eq.~(\ref{gauge_vector_1PM_stationary});  
for an explicit expression see text below that equation. Let us recall that the metric perturbations and the gauge functions in (\ref{Gauge_2PM_B2_stationary}) are 
time-independent: $h^{\left({\rm 1PM}\right)}_{\alpha\beta\,{\rm can}\,,\,0} = 0$ and $\varphi^{\alpha\,\left({\rm 1PM}\right)}_{\,,\,0} = 0$.  

In summary of this Section, Eq.~(\ref{stationary_2PM_10}) with (\ref{canonical_metric_1PN_00}) - (\ref{canonical_metric_1PN_ij}) and Eq.~(\ref{stationary_2PM_20})  
with (\ref{canonical_metric_2PN_00}) - (\ref{canonical_metric_2PN_ij}) represent the metric perturbation in the second post-Minkowskian approximation for  
one body at rest as function of the time-independent multipoles $M_L$ and $S_L$. The mass-multipoles $M_L$ in (\ref{M_L}) allow to describe an arbitrary shape  
and inner structure of the body, while the spin-multipoles $S_L$ in (\ref{M_L}) allow to account for stationary currents of matter, like circulations of  
matter inside the body or stationary rotational motions of the body as a whole.

\subsection{Monopole and spin and quadrupole terms of 2PM metric for stationary sources} 

The metric in 2PM approximation of an arbitrarily shaped body is considered, where the monopole and spin and quadrupole terms of the metric tensor  
are taken into account. The expressions for the monopole and spin and quadrupole follow from (\ref{M_L}) and (\ref{S_L}), viz.    
\begin{eqnarray}
M &=& \int d^3 x \,\frac{T^{00} + T^{kk}}{c^2}\,, 
\label{Mass}
\\
S_i &=& \int d^3 x \;\epsilon_{i j k}\;x^j\;\frac{T^{0k}}{c}\,, 
\label{Spin}
\\
M_{ab} &=& \int d^3 x \,\hat{x}_{ab}\,\frac{T^{00} + T^{kk}}{c^2}\,, 
\label{Quadrupole}
\end{eqnarray}

\noindent
where the integrals run over the three-dimensional volume of the body, 
and $\hat{x}_{ab} = x_a x_b - \frac{1}{3}\,r^2\,\delta_{ab}$.  
The mass-dipole terms are not considered here, because they can be eliminated, $M_i = 0$, by an appropriate choice of the coordinate system
(origin of the spatial axes are tied to the center of mass of the source; cf. comment below Eq.~(\ref{sigma}).

\subsubsection{The 1PM terms of canonical metric perturbation} 

From Eqs.~(\ref{canonical_metric_1PN_00}) - (\ref{canonical_metric_1PN_ij}) one immediately obtains the 
1PM terms of the canonical metric perturbation: 
\begin{eqnarray}
h^{\left({\rm 1PM}\right)}_{00\,{\rm can}}\left(\ve{x}\right) &=& \frac{2}{c^2}\,\frac{M}{r} 
+ \frac{3}{c^2}\,\frac{\hat{n}_{ab}\, M_{ab}}{r^3}\,,  
\label{metric_1PM_00}
\\
\nonumber\\
h^{\left({\rm 1PM}\right)}_{0i\,{\rm can}}\left(\ve{x}\right) &=& \frac{2}{c^3}\, \epsilon_{iab}\,n_a\,\frac{S_b}{r^2}\,,  
\label{metric_1PM_0i}
\\
\nonumber\\
h^{\left({\rm 1PM}\right)}_{ij\,{\rm can}}\left(\ve{x}\right) &=& \frac{2}{c^2}\,\frac{M}{r}\,\delta_{ij} + \frac{3}{c^2}\,\frac{\hat{n}_{ab}\,M_{ab}}{r^3}  
\,\delta_{ij}\,.
\label{metric_1PM_ij}
\end{eqnarray}

\noindent 
These expressions are in agreement with Eqs.~(1) - (2) in \cite{Klioner_1991}. 
It should be noticed that $\hat{n}_{ab}\,M_{ab} = n_{ab}\,M_{ab}$ because of the STF structure of the multipoles.  

\subsubsection{The 2PM terms of canonical metric perturbation} 

From Eqs.~(\ref{canonical_metric_2PN_00}) - (\ref{canonical_metric_2PN_ij}) one obtains the 
2PM terms of the canonical metric perturbation: 
\begin{widetext}
\begin{eqnarray}
h^{\left({\rm 2PM}\right)}_{00\,{\rm can}}\left(\ve{x}\right) &=& - \frac{2}{c^4}\,\frac{M^2}{r^2}
- 6\,\frac{M\,M_{ab}}{c^4\;r^4}\,\hat{n}_{ab}
- \frac{M_{ab}\,M_{cd}}{c^4\;r^6} \left(\frac{3}{5}\,\delta_{ac}\,\delta_{bd}  
+ \frac{18}{7}\,\delta_{ac}\,\hat{n}_{bd} + \frac{9}{2}\,\hat{n}_{abcd}\right) + {\cal O}\left(c^{-6}\right),  
\label{metric_2PM_00}
\\
\nonumber\\
h^{\left({\rm 2PM}\right)}_{0i\,{\rm can}}\left(\ve{x}\right) &=& {\cal O}\left(c^{-5}\right),  
\label{metric_2PM_0i}
\\
\nonumber\\
h^{\left({\rm 2PM}\right)}_{ij\,{\rm can}}\left(\ve{x}\right) &=& \frac{1}{c^4}\,\frac{M^2}{r^2}\left(\frac{4}{3}\,\delta_{ij} + \hat{n}_{ij}\right)
+ \frac{M\,M_{ab}}{c^4\;r^4}\left(\frac{15}{2}\,\hat{n}_{ijab} + \frac{32}{7}\,\delta_{ij}\,\hat{n}_{ab}  
- \frac{12}{7}\,\delta_{a\,(i} \hat{n}_{j\,) b} \right)  
\nonumber\\
\nonumber\\
&& + \frac{M_{ab}\,M_{cd}}{c^4\,r^6}
\bigg(\frac{75}{4}\,\hat{n}_{ijabcd} - \frac{90}{11}\,\hat{n}_{ijac}\,\delta_{bd} + \frac{27}{11}\,\hat{n}_{abcd}\,\delta_{ij}
- \frac{25}{84}\,\hat{n}_{ij}\,\delta_{ac}\,\delta_{bd}
\nonumber\\
\nonumber\\
&& + \frac{83}{42}\,\hat{n}_{ad}\,\delta_{bc}\,\delta_{ij}
 + \frac{16}{35}\,\delta_{ac}\,\delta_{bd}\,\delta_{ij} + \frac{18}{11}\,\hat{n}_{acd(i}\,\delta_{j)b}
- \frac{5}{21}\,\hat{n}_{a(i}\,\delta_{j)c}\,\delta_{bd}
\nonumber\\
\nonumber\\
&& + \frac{10}{21}\,\delta_{ci}\,\delta_{dj}\,\hat{n}_{ab}
- \frac{23}{42}\,\delta_{b(i}\,\delta_{j)c}\,\hat{n}_{ad} - \frac{6}{35}\,\delta_{ad}\,\delta_{b(i}\,\delta_{j)c} \bigg)  
+ {\cal O}\left(c^{-6}\right), 
\label{metric_2PM_ij}
\end{eqnarray}
\end{widetext}

\noindent
where the details of the calculations are relegated to Appendices \ref{Appendix7} and \ref{Appendix8};  
for various careful checks see text below Eq.~(\ref{Appendix_metric_2PM_ij_10}) in Appendix \ref{Appendix8}.

\section{Summary}\label{Section8} 

In this article four issues have been considered: 

1. A coherent exposition of the Multipolar Post-Minkowskian (MPM) formalism has been presented, 
where we have focused on those results of the MPM formalism which are relevant for our investigations. Special care 
has been taken about the gauge transformation of the metric tensor and metric density. 
It has been emphasized that the canonical piece of the metric tensor and
metric density is gauge-independent, hence is independent of whether one uses the residual gauge transformation (\ref{harmonic_gauge_condition_2}) 
or (\ref{harmonic_gauge_condition_inverse}). The Lorentz covariant gauge transformation and the general-covariant 
gauge transformation and how they are related to each other has been expounded in some detail.  

2. The MPM formalism has been used in order to obtain the metric coefficients in harmonic coordinates in the post-linear approximation 
in the exterior of a compact source of matter,  
\begin{eqnarray}
&& \hspace{-0.15cm} g_{\alpha\beta}\left(x\right) = \eta_{\alpha\beta} 
+ G^1 h^{\left({\rm 1PM}\right)}_{\alpha\beta\,{\rm can}}\left(x\right)
+ G^1 \partial \varphi^{\left({\rm 1PM}\right)}_{\alpha\beta}\!\left(x\right) 
\nonumber\\ 
&& \hspace{1.0cm} + \, G^2 h^{\left({\rm 2PM}\right)}_{\alpha\beta\,{\rm can}}\left(x\right)
+ G^2 \partial \varphi^{\left({\rm 2PM}\right)}_{\alpha\beta}\!\left(x\right) 
+ \, G^2 \Omega^{\left({\rm 2PM}\right)}_{\alpha\beta}\!\left(x\right) 
\nonumber\\ 
&& \hspace{1.0cm} + {\cal O}\left(G^3\right),  
\label{Summary_1}
\end{eqnarray}

\noindent 
where $x= \left(ct,\ve{x}\right)$. 
The linear canonical metric perturbation $h^{\left({\rm 1PM}\right)}_{\alpha\beta\,{\rm can}}$ is given by Eqs.~(\ref{1PM_Metric_00}) - (\ref{1PM_Metric_ij})
and the post-linear canonical metric perturbation $h^{\left({\rm 2PM}\right)}_{\alpha\beta\,{\rm can}}$ is given by Eqs.~(\ref{metric_80}) - (\ref{metric_90})
up to terms of the order ${\cal O}\left(c^{-6},c^{-5},c^{-6}\right)$.
The gauge term $\partial \varphi^{\left({\rm 1PM}\right)}_{\alpha\beta}$ is given by Eq.~(\ref{gauge_term_1PM}), while
the gauge terms $\partial \varphi^{\left({\rm 2PM}\right)}_{\alpha\beta}$ and $\Omega^{\left({\rm 2PM}\right)}_{\alpha\beta}$ are given by
Eqs.~(\ref{Gauge_2PM_B1}) and (\ref{Gauge_2PM_B2}).  
The canonical metric perturbations depend on the canonical mass and spin multipoles, $M_L$ and $S_L$, which are functions of the retarded time 
$s = t - \left|\ve{x}\right|/c$. These multipoles are given by  
Eqs.~(\ref{M_L_Definition}) and (\ref{S_L_Definition}), allowing to account for arbitrary shape, inner structure, oscillations, and rotational motions of the  
source of matter. The metric tensor $g_{\alpha\beta}\left(t,\ve{x}\right)$ in (\ref{Summary_1}) represents the most general solution  
for the spatial region in the exterior of a compact source of matter. 

3. Furthermore, the metric of a stationary source of matter has been considered, 
\begin{eqnarray}
&& \hspace{-0.15cm} g_{\alpha\beta}\left(\ve{x}\right) = \eta_{\alpha\beta}
+ G^1 h^{\left({\rm 1PM}\right)}_{\alpha\beta\,{\rm can}}\left(\ve{x}\right)
+ G^1 \partial \varphi^{\left({\rm 1PM}\right)}_{\alpha\beta}\!\left(\ve{x}\right)
\nonumber\\
&& \hspace{1.0cm} + \, G^2 h^{\left({\rm 2PM}\right)}_{\alpha\beta\,{\rm can}}\left(\ve{x}\right)
+ G^2 \partial \varphi^{\left({\rm 2PM}\right)}_{\alpha\beta}\!\left(\ve{x}\right)
+ \, G^2 \Omega^{\left({\rm 2PM}\right)}_{\alpha\beta}\!\left(\ve{x}\right)
\nonumber\\
&& \hspace{1.0cm} + {\cal O}\left(G^3\right).
\label{Summary_2}
\end{eqnarray}

\noindent
The canonical linear metric perturbation
$h^{\left({\rm 1PM}\right)}_{\alpha\beta\,{\rm can}}$ is given by Eqs.~(\ref{canonical_metric_1PN_00}) - (\ref{canonical_metric_1PN_ij})
and the canonical post-linear metric perturbation $h^{\left({\rm 2PM}\right)}_{\alpha\beta\,{\rm can}}$ is given by
Eqs.~(\ref{canonical_metric_2PN_00}) - (\ref{canonical_metric_2PN_ij}) up to terms of the order ${\cal O}\left(c^{-6},c^{-5},c^{-6}\right)$.
The gauge term $\partial \varphi^{\left({\rm 1PM}\right)}_{\alpha\beta}$ is given by Eq.~(\ref{gauge_term_1PM_stationary}), while  
the gauge terms $\partial \varphi^{\left({\rm 2PM}\right)}_{\alpha\beta}$ and $\Omega^{\left({\rm 2PM}\right)}_{\alpha\beta}$ are given by
Eqs.~(\ref{gauge_term_2PM_stationary}) and (\ref{Gauge_2PM_B2_stationary}). 
The canonical metric perturbation depends on the canonical mass and spin multipoles, $M_L$ and $S_L$, which are time-independent. 
These multipoles are given by Eqs.~(\ref{M_L}) and (\ref{S_L}),  
allowing to account for arbitrary shape and inner structure as well as inner stationary currents of the source of matter.  
The metric tensor $g_{\alpha\beta}\left(\ve{x}\right)$ in (\ref{Summary_2}) represents the most general solution  
for the spatial region in the exterior of a stationary compact source of matter.

4. The spatial components of the canonical post-linear metric perturbation are associated with an integration procedure:  
in (\ref{Summary_1}) by the inverse d'Alembertian (\ref{Inverse_d_Alembert}) and in (\ref{Summary_2}) by the inverse Laplacian (\ref{Inverse_Laplacian}).  
That integration procedure has been performed explicitly in (\ref{Summary_2}), where the first multipoles  
(monopole and quadrupole) are taken into account. The linear and post-linear metric coefficients are given by 
Eqs.~(\ref{metric_1PM_00}) - (\ref{metric_1PM_ij}) and (\ref{metric_2PM_00}) - (\ref{metric_2PM_ij}), respectively. 

The investigations are motivated by the rapid progress in astrometric science,  
which has recently succeeded in making the giant step from the milli-arcsecond level \cite{Hipparcos,Hipparcos1,Hipparcos2} 
to the micro-arcsecond level \cite{GAIA,GAIA1,GAIA2,GAIA4,GAIA_DR2_1} in angular measurements of celestial objects, like stars and quasars.  
A fundamental issue in relativistic astrometry concerns the precise modeling of the trajectories of light signals emitted by some celestial light source and 
propagating through the curved space-time of the solar system. The light trajectories are governed by the geodesic equation, which implies the  
knowledge of the metric coefficients for solar system bodies. Accordingly, interpreting the compact source of matter just as 
some massive body of arbitrary shape and inner structure, the post-linear metric coefficients allow to determine the light trajectory 
in the gravitational field of such a massive solar system body in the post-linear approximation.  
Thus far, the impact of higher multipoles on the light trajectories in the post-linear approximation is unknown. So the results  
of this investigation are a fundamental requirement in order to determine the impact of higher multipoles on the light trajectory in the post-linear approximation.

\section{Acknowledgment}

This work was funded by the German Research Foundation (Deutsche Forschungsgemeinschaft DFG) under grant number 263799048.
Sincere gratitude is expressed to Prof. Sergei A. Klioner and Prof. Michael H. Soffel for kind encouragement and enduring support.
The author also wish to thank Dr. Alexey G. Butkevich, Prof. Laszlo P. Csernai, Prof. Burkhard K\"ampfer, PD Dr. G\"unter Plunien, 
and Prof. Ralf Sch\"utzhold  
for inspiring discussions and conversations about the general theory of relativity and astro\-metric science  
during recent years. Dr. Sebastian Bablok and Dipl.-Inf. Robin Geyer are kindly acknowledged for computer assistance.

\appendix

\section{Notation}\label{Appendix0}

The notation of the standard literature \cite{Multipole_Damour_2,MTW,Thorne} is used:  

\begin{enumerate}
\item[$\bullet$] $G$ is the Newtonian constant of gravitation.  
\item[$\bullet$] $c$ is the speed of light in flat space-time which equals the speed of gravitational waves in a flat background manifold.  
\item[$\bullet$] lower case Latin indices take values $1,2,3$.
\item[$\bullet$] lower case Greek indices take values 0,1,2,3. 
\item[$\bullet$] $\delta_{ij} = \delta^{ij} = {\rm diag} \left(+1,+1,+1\right)$ is the Kronecker delta.
\item[$\bullet$] $\varepsilon_{ijk} = \varepsilon^{ijk}$ with $\varepsilon_{123} = + 1$ is the three-dimensional Levi-Civita symbol.  
\item[$\bullet$] $\varepsilon_{\alpha\beta\mu\nu} = \varepsilon^{\alpha\beta\mu\nu}$ with $\varepsilon_{0123} = + 1$ is the four-dimensional Levi-Civita symbol.  
\item[$\bullet$] metric of Minkowskian space-time is $\eta_{\alpha\beta} = \eta^{\alpha\beta} = {\rm diag}\left(-1,+1,+1,+1\right)$.  
\item[$\bullet$] covariant components of metric of Riemann space-time are $g_{\alpha\beta}$.  
\item[$\bullet$] contravariant components of metric of Riemann space-time are $g^{\alpha\beta}$.  
\item[$\bullet$] the metric signature is $\left(-,+,+,+\right)$. 
\item[$\bullet$] $g = {\rm det}\left(g_{\alpha\beta}\right)$ is the determinant of the covariant components of the metric tensor  
\item[] that means: 
$\displaystyle g = \frac{1}{4!}\,
\varepsilon^{\alpha\beta\gamma\delta}\,\varepsilon^{\mu\nu\rho\sigma}\,g_{\alpha\mu}\,g_{\beta\nu}\,g_{\gamma\rho}\,g_{\delta\sigma}$.  
\item[$\bullet$] covariant and contravariant components of three-vectors: $a_i = a^i = \left(a^{\,1},a^2,a^3\right)$.
\item[$\bullet$] $n! = n \left(n-1\right)\left(n-2\right)\cdot\cdot\cdot 2 \cdot 1$ is the factorial for positive integer ($0! = 1$).   
\item[$\bullet$] $n!! = n \left(n-2\right) \left(n-4\right)\cdot\cdot\cdot \left(2\;{\rm or}\;1\right)$ is the double factorial  
for positive integer ($0!! = 1$).  
\item[$\bullet$] $L=i_1 i_2 ...i_l$ and $M=i_1 i_2 ...i_m$ are Cartesian multi-indices of a given tensor $T$, that means
$T_L \equiv T_{i_1 i_2 \,.\,.\,.\,i_l}$ and  $T_M \equiv T_{i_1 i_2 \,.\,.\,.\,i_m}$, respectively.
\item[$\bullet$] two identical multi-indices imply summation:
$A_L\,B_L \equiv \sum\limits_{i_1\,.\,.\,.\,i_l}\,A_{i_1\,.\,.\,.\,i_l}\,B_{i_1\,.\,.\,.\,i_l}$. 
\item[$\bullet$] triplet of spatial coordinates (three-vectors) are in boldface: e.g. $\ve{a}$, $\ve{b}$. 
\item[$\bullet$] the absolute value of three-vector is determined by $\left|\ve{a}\right| = \sqrt{\delta_{ij} a^i a^j}$. 
\item[$\bullet$] covariant components of four-vectors: $a_{\mu} = \left(a_0,a_1,a_2,a_3\right)$.
\item[$\bullet$] contravariant components of four-vectors: $a^{\mu} = \left(a^0,a^1,a^2,a^3\right)$.  
\item[$\bullet$] $\displaystyle \partial_i = \frac{\partial}{\partial x^i}$ is partial derivative w.r.t. $x^i$. 
\item[$\bullet$] $f_{\,,\,i}$ is partial derivative of $f$ w.r.t. $x^i$. 
\item[$\bullet$] $\partial_L = \partial_{a_1\,...\,a_l}$ denotes $l$ partial derivatives w.r.t. $x^{a_1} \dots x^{a_l}$.  
\item[$\bullet$] $f_{\,,\,a_1\,...\,a_l}$ denotes $l$ partial derivatives of $f$ w.r.t. $x^{a_1} \dots x^{a_l}$.
\item[$\bullet$] $\displaystyle \partial_{\alpha} = \frac{\partial}{\partial x^{\alpha}}$ is partial derivative w.r.t. $x^{\alpha}$.  
\item[$\bullet$] $f_{\,,\,\alpha}$ is partial derivative of $f$ w.r.t. $x^{\alpha}$.  
\item[$\bullet$] $f_{\,,\,\mu_1\,...\,\mu_n}$ denotes $n$ partial derivatives of $f$ w.r.t. $x^{\mu_1} \dots x^{\mu_n}$.  
\item[$\bullet$] $\displaystyle \dot{f} = \frac{d f}{d t}$ is total time-derivative of $f$.  
\item[$\bullet$] $\displaystyle \ddot{f} = \frac{d^2 f}{d t^2}$ is double total time-derivative of $f$. 
\item[$\bullet$] $A^{\alpha}_{\,;\,\mu} = A^{\alpha}_{\,,\,\mu} + \Gamma^{\alpha}_{\mu\nu}\,A^{\nu}$ is covariant derivative of first rank tensor.
\item[$\bullet$] $B^{\alpha\beta}_{\;\;\;\;;\,\mu} = B^{\alpha\beta}_{\;\;\;\;,\,\mu} + \Gamma^{\alpha}_{\mu\nu}\,B^{\nu\beta} + \Gamma^{\beta}_{\mu\nu}\,B^{\alpha\nu}$
is covariant derivative of second rank tensor.
\item[$\bullet$] repeated indices are implicitly summed over ({\it Einstein's} sum convention).
\end{enumerate}

\section{Some useful relations of Cartesian tensors}\label{Appendix1}

The irreducible Cartesian tensor technique has been developed in \cite{Coope1,Coope2,Coope3} and is a very useful tool of the MPM formalism.  
Here we summarize some relevant relations of the Cartesian tensor technique. 

The symmetric part of a Cartesian tensor $T_L$ is, cf. Eq.~(2.1) in \cite{Thorne}:
\begin{eqnarray}
 \hspace{1.0cm} T_{\left(L\right)} = T_{\left(i_1 ... i_l \right)} = \frac{1}{l!} \sum\limits_{\sigma}
A_{i_{\sigma\left(1\right)} ... i_{\sigma\left(l\right)}}\,,
\label{symmetric_1}
\end{eqnarray}

\noindent
where $\sigma$ is running over all permutations of $\left(1,2,...,l\right)$.

The symmetric trace-free part of a Cartesian tensor $T_L$ is denoted as $\hat{T}_L$ and given by (cf. Eq.~(2.2) in \cite{Thorne})  
\begin{eqnarray}
&& T_{<L>} = T_{<i_1 \dots i_l>} = \hat{T}_L  
\nonumber\\ 
&& = \sum_{k=0}^{\left[l/2\right]} a_{l k}\,\delta_{(i_1 i_2 \dots} \delta_{i_{2k-1} i_{2k}}\,
S_{i_{2k+1 \dots i_l) \,a_1 a_1 \dots a_k a_k}}\,,
\label{anti_symmetric_1}
\end{eqnarray}

\noindent
where $\left[l/2\right]$ means the largest integer less than or equal to $l/2$, and $S_L \equiv T_{\left(L\right)}$
abbreviates the symmetric part of tensor $T_L$. 
The coefficient in (\ref{anti_symmetric_1}) is given by
\begin{eqnarray}
a_{l k} = \left(-1\right)^k \frac{l!}{\left(l - 2 k\right)!}\,
\frac{\left(2 l - 2 k - 1\right)!!}{\left(2 l - 1\right)!! \left(2k\right)!!}\,.
\label{coefficient_anti_symmetric}
\end{eqnarray}

\noindent
STF tensors vanish whenever two of their indices are equal,  
\begin{eqnarray}
T_{<i_1 \dots a \dots a \dots i_l>} = \sum\limits_{a=1}^{3} T_{<i_1 \dots a \dots a \dots i_l>} = 0\,, 
\label{Appendix1_STF5}
\end{eqnarray}

\noindent
because a summation of these indices is implied according to Einstein's sum convention; of course, the individual components of  
STF tensors do not vanish, e.g. $T_{<i_1 \dots 2 \dots 2 \dots i_l>} \neq 0$.  
Further STF relations can be found in \cite{Blanchet_Damour1,Thorne,Coope1,Coope2,Coope3,Soffel_Hartmann}.
As instructive examples of (\ref{anti_symmetric_1}) let us consider the cases $l=2$, $l=3$, and $l=4$: 
\begin{eqnarray}
T_{<ij>} &=& T_{\left(ij\right)} - \frac{1}{3}\,\delta_{ij}\,T_{\left(aa\right)} \;,  
\label{STF_Tensor_1}
\\
\nonumber\\
T_{<ijk>} &=& T_{\left(ijk\right)} - \frac{1}{5}\left(\delta_{ij}\,T_{\left(kaa\right)} + \delta_{ik}\,T_{\left(jaa\right)} 
+ \delta_{jk}\,T_{\left(iaa\right)}\right),  
\nonumber\\
\label{STF_Tensor_2}
\\
T_{<ijkl>} &=& T_{\left(ijkl\right)} - \frac{1}{7} \left(\delta_{ij}\,T_{\left(klaa\right)} + \delta_{ik}\,T_{\left(jlaa\right)}\right)  
\nonumber\\ 
&&\hspace{-1.0cm} - \,\frac{1}{7} \left(\delta_{il}\,T_{\left(jkaa\right)} + \delta_{jk}\,T_{\left(ilaa\right)} + \delta_{jl}\,T_{\left(ikaa\right)}  
+ \delta_{kl}\,T_{\left(ijaa\right)} \right)  
\nonumber\\ 
&& \hspace{-1.0cm} + \,\frac{1}{35} \left(\delta_{ij} \delta_{kl} \,T_{\left(aabb\right)} + \delta_{ik} \delta_{jl} \,T_{\left(aabb\right)} 
+ \delta_{il} \delta_{jk} \,T_{\left(aabb\right)}\right),  
\nonumber\\
\label{STF_Tensor_3}
\end{eqnarray}

\noindent 
where the expressions in (\ref{STF_Tensor_1}) and (\ref{STF_Tensor_2}) were also given by Eqs.~(2.3) and (2.4) in \cite{Multipole_Damour_2}.  
Especially, the following Cartesian tensor is of primary importance, which just consists of products of unit three-vectors,
\begin{eqnarray}
n_L = \frac{x_{i_1}}{r}\,\frac{x_{i_2}}{r}\,\dots\,\frac{x_{i_l}}{r} \quad {\rm where} \quad r = \left|\ve{x}\right|\;. 
\label{Appendix_Cartesian_Tensor}
\end{eqnarray}

\noindent
The symmetric trace-free part of the  Cartesian tensor (\ref{Appendix_Cartesian_Tensor}) reads  
\begin{eqnarray}
\hat{n}_L = \frac{x_{< i_1}}{r}\,\frac{x_{i_2}}{r}\,\dots\,\frac{x_{i_l >}}{r} \,. 
\label{Appendix_Cartesian_Tensor_n_L}
\end{eqnarray}

\noindent  
Using Eq.~(A 20a) in \cite{Blanchet_Damour1}, we present the explicit structure of the following STF Cartesian tensors,  
\begin{eqnarray}
\hat{n}_{a b} &=& n_{a b} - \frac{1}{3}\,\delta_{a b}\;,
\label{Appendix1_STF2}
\\
\nonumber\\
\hat{n}_{a b c} &=& n_{a b c} - \frac{1}{5}\left(\delta_{a b}\,n_c + \delta_{a c}\,n_b + \delta_{b c}\,n_a\right)\,,
\label{Appendix1_STF3}
\\
\nonumber\\
\hat{n}_{a b c d} &=& n_{a b c d}
\nonumber\\
&& \hspace{-1.25cm} - \frac{1}{7} \left(\delta_{a b} n_{c d} + \delta_{a c} n_{b d} + \delta_{a d} n_{b c} + \delta_{b c} n_{a d}
+ \delta_{b d} n_{a c} + \delta_{c d} n_{a b} \right)
\nonumber\\
&& \hspace{-1.25cm} + \frac{1}{35}\left(\delta_{a b}\,\delta_{c d} + \delta_{a c}\,\delta_{b d} + \delta_{a d}\,\delta_{b c}\right)\,,
\label{Appendix1_STF4}
\end{eqnarray}

\noindent
which were also given by Eqs.~(1.8.2) and (1.8.4) in \cite{Poisson_Lecture_Notes}.  
Frequently, the following relations are needed, which convert a non-STF tensor into a STF tensor,  
\begin{eqnarray}
n_a\,\hat{n}_{L} &=& \hat{n}_{a L} + \frac{l}{2 l + 1} \, \delta_{a < a_l}\;\hat{n}_{L-1 >}\;,
\label{Appendix1_Transformation_STF_1}
\\
\nonumber\\
n_a\,\hat{n}_{a L} &=& \frac{l+1}{2\,l + 1}\,\hat{n}_{L}\;,
\label{Appendix1_Transformation_STF_2}
\end{eqnarray}

\noindent
where (\ref{Appendix1_Transformation_STF_1}) has been given by Eq.~(A 22a) in \cite{Blanchet_Damour1}
(see also Eq.~(2.7) in \cite{Multipole_Damour_2}, Eq.~(A7) in \cite{Soffel_Hartmann}), while (\ref{Appendix1_Transformation_STF_2})
is given by Eq.~(A 23) in \cite{Blanchet_Damour1} (see also Eq.~(A.8) in \cite{Soffel_Hartmann}).
Some explicit examples of (\ref{Appendix1_Transformation_STF_1}) are noticed which are of relevance for our investigations,
\begin{widetext}
\begin{eqnarray}
n_a\,\hat{n}_{b c} &=& \hat{n}_{a b c}
+ \frac{1}{5} \left(\delta_{a c}\,\hat{n}_b + \delta_{a b}\,\hat{n}_c\right) - \frac{2}{15}\,\delta_{b c}\,\hat{n}_a\;,  
\label{Appendix1_Transformation_STF_3}
\\
\nonumber\\
n_a\,\hat{n}_{b c d} &=& \hat{n}_{a b c d}
+ \frac{1}{7} \left(\delta_{a b}\,\hat{n}_{c d} + \delta_{a c}\,\hat{n}_{b d} + \delta_{a d}\,\hat{n}_{b c}\right)
- \frac{2}{35} \left(\delta_{b c}\,\hat{n}_{a d} + \delta_{c d}\,\hat{n}_{a b} + \delta_{b d}\,\hat{n}_{a c}\right), 
\label{Appendix1_Transformation_STF_4}
\\
\nonumber\\
 n_a\,\hat{n}_{b c d e} &=& \hat{n}_{a b c d e} + \frac{1}{9} \left(\delta_{ae}\,\hat{n}_{bcd} + \delta_{ab}\,\hat{n}_{cde}
+ \delta_{ac}\,\hat{n}_{bde} + \delta_{ad}\,\hat{n}_{bce}\right)
\nonumber\\
&& \hspace{1.0cm} - \,\frac{2}{63} \left(\delta_{be}\,\hat{n}_{acd} + \delta_{cd}\,\hat{n}_{abe} + \delta_{ce}\,\hat{n}_{abd}
+ \delta_{de}\,\hat{n}_{abc} + \delta_{bc}\,\hat{n}_{ade} + \delta_{bd}\,\hat{n}_{ace} \right), 
\label{Appendix1_Transformation_STF_5}
\\
\nonumber\\
n_a\,\hat{n}_{b c d e f} &=& \hat{n}_{a b c d e f} + \frac{1}{11} \left(\delta_{af}\,\hat{n}_{bcde} + \delta_{ab}\,\hat{n}_{cdef}
+ \delta_{ac}\,\hat{n}_{bdef} + \delta_{ad}\,\hat{n}_{bcef} + \delta_{ae}\,\hat{n}_{bcdf} \right)
\nonumber\\
&& \hspace{1.15cm} - \, \frac{2}{99} \big(\delta_{bf}\,\hat{n}_{acde} + \delta_{cf}\,\hat{n}_{abde} + \delta_{df}\,\hat{n}_{abce} +
\delta_{ef}\,\hat{n}_{abcd} + \delta_{bc}\,\hat{n}_{adef}
\nonumber\\
&& \hspace{2.00cm} + \, \delta_{bd}\,\hat{n}_{acef} + \delta_{be}\,\hat{n}_{acdf} + \delta_{cd}\,\hat{n}_{abef} + \delta_{ce}\,\hat{n}_{abdf}
+ \delta_{de}\,\hat{n}_{abcf} \big).
\label{Appendix1_Transformation_STF_6}
\end{eqnarray}
\end{widetext}

\noindent
We recall that $\hat{n}_a = n_a$ but, nevertheless, we keep the notation in (\ref{Appendix1_Transformation_STF_3}) as is,
in order to emphasize that, according to the meaning of relation (\ref{Appendix1_Transformation_STF_1}), there are
STF tensors on the r.h.s. of each of these relations (\ref{Appendix1_Transformation_STF_3}) - (\ref{Appendix1_Transformation_STF_6}).
We also need the following relations, which are specific cases of the general relation given by Eq.~(2.13) in \cite{Multipole_Damour_2}:
\begin{eqnarray}
n_{ab}\,\hat{n}_{ijcd}
&=& \hat{n}_{ijabcd} + \frac{4}{11} \left(\hat{n}_{a < ijc}\,\delta_{d > b} + \hat{n}_{b < ijc}\,\delta_{d > a}\right)
\nonumber\\
&& + \frac{12}{63}\,\hat{n}_{< ij} \,\delta^a_{c}\,\delta^b_{d >} + \frac{1}{11}\,\delta_{ab}\,\hat{n}_{ijcd}\;,
\label{appendix_T2_30}
\\
\nonumber\\
n_{ab}\,\hat{n}_{cd}
&=& \hat{n}_{abcd} + \frac{2}{7} \left(\hat{n}_{a < c}\,\delta_{d > b} + \hat{n}_{b < c}\,\delta_{d > a}\right)
\nonumber\\
&& + \frac{2}{15}\,\delta_{< c}^{\;\;\,a}\,\delta_{d >}^b + \frac{1}{7}\,\delta_{ab}\,\hat{n}_{cd}\;. 
\label{appendix_T2_35}
\end{eqnarray}

\noindent 
Finally, we notice  
\begin{eqnarray}
\partial_L \frac{1}{r} = \left(-1\right)^l\;\frac{\left(2\,l - 1\right)!!}{r^{l + 1}}\,\hat{n}_L\;,
\label{Appendix1_Differentiation_A}
\end{eqnarray}

\noindent
which agrees with Eq.~(A 34) in \cite{Blanchet_Damour1}.

\section{The STF mass-multipoles and spin-multipoles}\label{Appendix2}

In the MPM formalism the solution of the gothic metric is given in terms of irreducible symmetric and trace-free (STF) Cartesian tensors.
It is a fundamental result of the MPM formalism \cite{Blanchet_Damour1,Blanchet_Damour3,Multipole_Damour_2} that in the  
exterior of a source of matter the gothic metric (\ref{expansion_PM}) to any order in $G$ depends 
on a set of only two kind of
irreducible STF tensors (Theorem 4.5 in \cite{Blanchet_Damour1}, see also Eq.~(3.1) and (3.2) in \cite{Blanchet_Damour3}), namely
mass-type multipoles $M_L$ and current-type multipoles $S_L$ \cite{Multipole_Damour_2,Thorne}.
The explicit expressions for these multipoles up to terms of the order ${\cal O}\left(G^2\right)$ are given by Eqs.~(5.33) and (5.35)
in \cite{Multipole_Damour_2} and read
\begin{eqnarray}
M_L &=& \int d^3 x \int \limits_{-1}^{+1} d z
\nonumber\\
&& \hspace{-1.0cm}\times \left[\delta_l\left(z\right)\,\hat{x}_L\,\sigma + a_l\,\delta_{l+1}\left(z\right)\,\hat{x}_{iL}\,
\dot{\sigma}^i + b_l\,\delta_{l+2}\left(z\right)\,\hat{x}_{ijL}\,\ddot{\sigma}^{ij} \right],
\nonumber\\
\label{M_L_Definition}
\\
\nonumber\\
S_L &=& \int d^3 x \int \limits_{-1}^{+1} d z\;{\rm STF}_{\rm L}\;
\nonumber\\
&& \hspace{-1.0cm} \times \left[\delta_l\left(z\right)\,\hat{x}_{L-1}\,\epsilon_{i_l j k}\,x^j\,\sigma^k  
+ c_l\,\delta_{l+1}\left(z\right)\,\epsilon_{i_l j k}\,\hat{x}_{j s L}\,\dot{\sigma}^{k s} \right],
\nonumber\\
\label{S_L_Definition}
\end{eqnarray}

\noindent
where the integrals (\ref{M_L_Definition}) and (\ref{S_L_Definition}) run only over the finite three-dimensional space of the compact source of matter, and where  
\begin{eqnarray}
\delta_l(z) &=& \frac{\left(2\,l + 1\right)!!}{2^{l+1} l!} \!\left(1 - z^2\right)^l \,{\rm with}\, \int \limits_{-1}^{+1} d z \delta_l\left(z\right) = 1\,,
\label{delta_l}
\\
\nonumber\\
a_l &=& - \frac{4 \left(2 l + 1\right)}{c^2\;\left(l + 1\right)\left(2l + 3\right)}\,,
\label{coefficients_a}
\\
b_l &=& \frac{2 \left(2 l + 1\right)}{c^4\;\left(l + 1\right)\left(l+2\right) \left(2l+5\right)}\,,
\label{coefficients_b}
\\
c_l &=& - \frac{\left(2l + 1\right)}{c^2\;\left(l + 2\right) \left(2 l + 3 \right)}\,,
\label{coefficients_c}
\end{eqnarray}

\noindent
and where
\begin{eqnarray}
\sigma = \frac{T^{00} + T^{kk}}{c^2}\;, \quad \sigma^i = \frac{T^{0i}}{c}\;,\quad \sigma^{ij} = T^{ij}\;,
\label{sigma}
\end{eqnarray}

\noindent
with $T^{\alpha \beta}$ being the energy-momentum tensor of the isolated system taken at the time-argument
$t - \left|\ve{x}\right|/c + z\,\left|\ve{x}\right|/c$, and a dot in (\ref{M_L_Definition}) and (\ref{S_L_Definition}) means
partial derivative with respect to coordinate time. The mass-type multipoles $M_L$ and the spin-type multipoles $S_L$ are
STF tensors, but we adopt the notation as frequently used in the literature and do not write the multipoles with a hat, say
$M_L \equiv \hat{M}_L$ and $S_L \equiv \hat{S}_L$.

It should be noticed that the multipoles are functions of time, except the mass-monopole $M$, mass-dipole $M_i$,  
and spin-dipole $S_i$, which are strictly conserved quantities, that means $\dot{M} = \dot{M}_i = \dot{S}_i = 0$. The system  
may emit gravitational radiation which would change the mass $M$ and the mass-dipole $M_i$ and the spin-dipole $S_i$ of the compact source of matter, 
but these effects occur at higher order beyond the 1PM and 2PM approximation. Furthermore, if the  
origin of the spatial coordinate axes is located at the center-of-mass of the source, then the mass-dipole vanishes, i.e. $M_i = 0$.
 
A further note is in order about the mass-type multipoles $M_L$ and current-type $S_L$ as given by Eq.~(\ref{M_L_Definition}) and Eq.~(\ref{S_L_Definition}), 
respectively. Usually, for practical applications their explicit form as given by Eqs.~(\ref{M_L_Definition}) and (\ref{S_L_Definition}) is not needed.  
Instead, these multipoles can be related to observables of the massive bodies of the solar system, and can be determined by fitting 
astrometric observations.

\section{Relations between metric tensor and gothic metric density}\label{Appendix3}  

The contravariant and covariant components of the gothic metric density are defined by \cite{MTW,Fock,Kopeikin_Efroimsky_Kaplan} 
(e.g. text below Eq.~(3.506) in \cite{Kopeikin_Efroimsky_Kaplan})
\begin{eqnarray}
\overline{g}^{\alpha\beta} = \sqrt{ - {\rm det}\left(g_{\mu\nu}\right)}\;g^{\alpha \beta}\,,
\label{Appendix3_1}
\end{eqnarray}

\noindent 
and 
\begin{eqnarray}
\overline{g}_{\alpha\beta} = \frac{1}{\sqrt{ - {\rm det}\left(g_{\mu\nu}\right)}}\;g_{\alpha \beta}\,. 
\label{Appendix3_2}
\end{eqnarray}

\noindent
The orthogonality relation of the metric tensor
\begin{eqnarray}
g^{\alpha\sigma}\,g_{\sigma\beta} = \delta^{\alpha}_{\beta}\,,
\label{Appendix3_3}
\end{eqnarray}

\noindent
implies, subject to (\ref{Appendix3_1}) and (\ref{Appendix3_2}), the orthogonality relation of the gothic metric density 
(e.g. text below Eq.~(3.506) in \cite{Kopeikin_Efroimsky_Kaplan}),
\begin{eqnarray}
\overline{g}^{\alpha\sigma}\,\overline{g}_{\sigma\beta} = \delta^{\alpha}_{\beta}\,, 
\label{Appendix3_4}
\end{eqnarray}

\noindent 
which is sometimes called isomorphism identity. 
From (\ref{Appendix3_3}) one gets  
\begin{eqnarray}
{\rm det}\left(g_{\mu\nu}\right) = \frac{1}{{\rm det}\left(g^{\mu\nu}\right)}\,,  
\label{Appendix3_5}
\end{eqnarray}
 
\noindent
while from (\ref{Appendix3_4}) one gets  
\begin{eqnarray}
{\rm det}\left(\overline{g}_{\mu\nu}\right) = \frac{1}{{\rm det}\left(\overline{g}^{\mu\nu}\right)}\,.   
\label{Appendix3_6}
\end{eqnarray}

\noindent  
By calculating the determinant of (\ref{Appendix3_1}) one finds that the determinant of the contravariant components of the gothic metric 
equals the determinant of the covariant components of the metric tensor \cite{MTW,Fock,Kopeikin_Efroimsky_Kaplan,Poisson_Lecture_Notes} 
(e.g. Eqs.~(D.67) and (D.68) in \cite{Fock})  
\begin{eqnarray}
{\rm det}\left(\overline{g}^{\mu\nu}\right) = {\rm det}\left(g_{\mu\nu}\right).  
\label{Appendix3_7}
\end{eqnarray}

\noindent 
The relations (\ref{Appendix3_5}) - (\ref{Appendix3_7}) imply that the determinant of the covariant components of the gothic metric 
equals the determinant of the contravariant components of the metric tensor,
\begin{eqnarray}
{\rm det}\left(\overline{g}_{\mu\nu}\right) = {\rm det}\left(g^{\mu\nu}\right), 
\label{Appendix3_8}
\end{eqnarray}
 
\noindent 
which can also be obtained by calculating the determinant of (\ref{Appendix3_2}) and by means of (\ref{Appendix3_5}).  
These relations (\ref{Appendix3_7}) and (\ref{Appendix3_8}) 
allow to derive from (\ref{Appendix3_1}) and (\ref{Appendix3_2}) the following relations between the metric tensor and gothic metric density, 
\begin{eqnarray}
g^{\alpha\beta} = \frac{1}{\sqrt{ - {\rm det}\left(\overline{g}^{\mu\nu}\right)}}\;\overline{g}^{\alpha \beta}\,,
\label{Appendix3_9}
\end{eqnarray}

\noindent
and 
\begin{eqnarray}
g_{\alpha\beta} = \sqrt{- {\rm det}\left(\overline{g}^{\mu\nu}\right)}\;\overline{g}_{\alpha \beta}\,. 
\label{Appendix3_10}
\end{eqnarray}

\section{Derivation of Eqs.~(\ref{Metric_Relation_Indices_1PM}) and (\ref{Retartion_Gothic_Metric_2})}\label{Appendix4}  

In what follows the relation (\ref{Appendix3_10}) between the covariant components of the metric tensor and the 
covariant components of the gothic metric density is important,  
\begin{eqnarray}
g_{\alpha\beta} = \sqrt{ - {\rm det}\left(\overline{g}^{\mu\nu}\right)}\;\overline{g}_{\alpha \beta}\,.   
\label{Appendix4_5}
\end{eqnarray}

\noindent
Let us consider the evaluation of the determinant in the r.h.s. of (\ref{Appendix4_5}). 
By taking the Minkowskian metric tensor as factor in front, we rewrite (\ref{expansion_metric_2a}) as follows, 
\begin{eqnarray}
\overline{g}^{\mu\nu} = \eta^{\mu\sigma}\, C^{\nu}_{\sigma}  
\label{Appendix4_10}
\end{eqnarray}

\noindent
where 
\begin{eqnarray}
C^{\nu}_{\sigma} = \delta^{\nu}_{\sigma} - \overline{h}^{\rho\nu}\,\eta_{\rho \sigma}\,.  
\label{Appendix4_15}
\end{eqnarray}

\noindent 
The determinant in (\ref{Appendix4_10}) is calculated by means of the product law of determinants and the theorem in (\ref{Jacobian_Determinant}), 
\begin{eqnarray}
{\rm det}\left(\overline{g}^{\mu\nu}\right) = {\rm det}\left(\eta^{\mu\sigma}\right)\;{\rm det}\left(C^{\nu}_{\sigma}\right)  
= -\, {\rm e}^{{\rm Tr}\left({\rm ln}\,C^{\nu}_{\sigma}\right)}\,,  
\label{Appendix4_25}
\end{eqnarray}

\noindent 
where ${\rm det}\left(\eta^{\mu\sigma}\right) = - 1$ has been taken into account. 
Using (\ref{Appendix4_25}) one obtains  
\begin{eqnarray}
{\rm det}\left(\overline{g}^{\mu\nu}\right) &=& - 1 + \overline{h} 
+ \frac{1}{2}\,\overline{h}^{\mu\nu}\,\eta_{\mu\rho}\;\overline{h}^{\rho\sigma}\,\eta_{\sigma\nu} 
- \frac{1}{2}\,\overline{h}^2 + {\cal O}\left(G^3\right),   
\nonumber\\ 
\label{Appendix4_40}
\end{eqnarray}

\noindent
where $\overline{h} = \overline{h}^{\mu\nu}\,\eta_{\mu\nu}$.  
By inserting (\ref{Appendix4_40}) into (\ref{Appendix4_5}) one obtains by series expansion of the square root  
\begin{eqnarray}
g_{\alpha\beta} = \left[1 - \frac{1}{2}\overline{h}
- \frac{1}{4} \overline{h}^{\mu\nu}\eta_{\mu\rho}\overline{h}^{\rho\sigma}\eta_{\sigma\nu}
+ \frac{1}{8} \overline{h}^2\right] \overline{g}_{\alpha \beta} + {\cal O}\left(G^3\right).  
\nonumber\\ 
\label{Appendix4_45}
\end{eqnarray}

\noindent 
For the covariant components of the metric tensor $g_{\alpha\beta}$ we have (cf. Eq.~(\ref{expansion_metric_1}))  
\begin{eqnarray}
g_{\alpha\beta} = \eta_{\alpha\beta} + G^1 h_{\alpha \beta}^{\left({\rm 1PM}\right)} + G^2 h_{\alpha \beta}^{\left({\rm 2PM}\right)} 
+ {\cal O}\left(G^3\right).   
\label{Appendix4_50}
\end{eqnarray}

\noindent
The post-Minkowskian series expansion of the contravariant components of the gothic metric density, $\overline{g}^{\alpha\beta}$, is given by 
Eqs.~(\ref{expansion_metric_2a}) and (\ref{expansion_PM}). What we also need is the post-Minkowskian series expansion of the 
covariant components of the gothic metric density, $\overline{g}_{\alpha\beta}$, which is defined by  
\begin{eqnarray}
\overline{g}_{\alpha \beta} = \eta_{\alpha \beta} + G^1\,\overline{h}_{\alpha \beta}^{\left({\rm 1PM}\right)} 
+ G^2\,\overline{h}_{\alpha \beta}^{\left({\rm 2PM}\right)} + {\cal O}\left(G^3\right).  
\label{Appendix4_55}
\end{eqnarray}

\noindent
Here we emphasize that (\ref{Appendix4_55}) is a definition of the covariant components of the perturbations of the gothic metric density. That 
means, the relations between the contravariant and covariant components of the perturbations of the gothic metric density follow  
from the isomorphism identity (\ref{Appendix3_4}) of the gothic metric density. 
These relations are given by Eqs.~(\ref{Appendix4_70}) - (\ref{Appendix4_75}) in 2PM approximation.  
Inserting (\ref{Appendix4_50}) and (\ref{Appendix4_55}) into (\ref{Appendix4_45}) and equating the powers of the gravitational constant yields 
for the terms proportional to $G^1$:  
\begin{eqnarray}
h^{\left({\rm 1PM}\right)}_{\alpha\beta} = \overline{h}^{\left({\rm 1PM}\right)}_{\alpha\beta} 
- \frac{1}{2}\,\overline{h}_{\left({\rm 1PM}\right)}\,\eta_{\alpha\beta}\,,  
\label{Appendix4_60}
\end{eqnarray}

\noindent
where $\overline{h}_{\left({\rm 1PM}\right)} = \overline{h}_{\left({\rm 1PM}\right)}^{\mu\nu}\,\eta_{\mu\nu}$. Similarly, 
for the terms proportional to $G^2$ one obtains:  
\begin{eqnarray}
h^{\left({\rm 2PM}\right)}_{\alpha\beta} &=& \overline{h}^{\left({\rm 2PM}\right)}_{\alpha\beta}
- \frac{1}{2}\,\overline{h}_{\left({\rm 2PM}\right)}\,\eta_{\alpha\beta} 
- \frac{1}{2}\,\overline{h}_{\left({\rm 1PM}\right)}\,\overline{h}^{\left({\rm 1PM}\right)}_{\alpha\beta}  
\nonumber\\
&& + \frac{1}{8}\,\overline{h}^{\,2}_{\left({\rm 1PM}\right)}\,\eta_{\alpha\beta}  
- \frac{1}{4}\,\overline{h}_{\left({\rm 1PM}\right)}^{\mu\nu} \eta_{\mu\rho}\;
\overline{h}_{\left({\rm 1PM}\right)}^{\rho\sigma} \eta_{\sigma\nu}\;\eta_{\alpha\beta}\,, 
\nonumber\\
\label{Appendix4_65}
\end{eqnarray}

\noindent
where $\overline{h}_{\left({\rm 2PM}\right)} = \overline{h}_{\left({\rm 2PM}\right)}^{\mu\nu}\,\eta_{\mu\nu}$. 
However, because the MPM formalism determines the contravariant components of the gothic metric, we have to express 
the relations (\ref{Appendix4_60}) and (\ref{Appendix4_65}) solely in terms of the contravariant components of the gothic metric. 
From the above mentioned isomorphism identity of the gothic metric density (\ref{Appendix3_4}) 
and from the series expansion (\ref{Appendix4_55}) as well as the series expansion (\ref{expansion_PM}) (with Eq.~(\ref{expansion_metric_2a})), 
we obtain the following relations,
\begin{eqnarray}
\overline{h}^{\left({\rm 1PM}\right)}_{\alpha\beta} &=& \overline{h}_{\left({\rm 1PM}\right)}^{\mu\nu}\,\eta_{\alpha\mu}\,\eta_{\beta\nu}\,,  
\label{Appendix4_70}
\\
\overline{h}^{\left({\rm 2PM}\right)}_{\alpha\beta} &=& \overline{h}_{\left({\rm 2PM}\right)}^{\mu\nu}\,\eta_{\alpha\mu}\,\eta_{\beta\nu} 
+ \overline{h}_{\left({\rm 1PM}\right)}^{\mu\nu}\,\overline{h}_{\left({\rm 1PM}\right)}^{\rho\sigma}\,\eta_{\mu\alpha}\,\eta_{\rho\nu}\,\eta_{\sigma\beta}\,, 
\nonumber\\
\label{Appendix4_75}
\end{eqnarray}
 
\noindent
which allow to determine the covariant components of the gothic metric from the contravariant components of the gothic metric.  
Finally, by inserting (\ref{Appendix4_70}) into (\ref{Appendix4_60}) as well as inserting (\ref{Appendix4_70}) and (\ref{Appendix4_75}) 
into (\ref{Appendix4_65}), one confirms (\ref{Metric_Relation_Indices_1PM}) and (\ref{Retartion_Gothic_Metric_2}), respectively.

\section{Hadamard regularization of the inverse d'Alembertian}\label{Appendix5}  

The spatial components (\ref{gothic_metric_90}) of the post-linear terms of the gothic metric density as well as the  
spatial components (\ref{metric_90}) of the post-linear terms of the metric tensor are associated with an integration procedure, which is 
abbreviated by the symbol ${\rm FP}_{B=0}\,\square_{\rm R}^{-1}$. In this Appendix some details of that integration procedure will be given.  

The symbol $\square_{\rm R}^{-1}$ denotes the inverse d'Alembertian   
acting on some function $f$, which is defined by (cf. Eq.~(\ref{Inverse_d_Alembert_1}))  
\begin{eqnarray}
\left(\square^{-1}_{\rm R} f \right)\left(t,\ve{x}\right) = - \frac{1}{4\,\pi} \int 
\frac{d^3 x^{\prime}}{\left|\ve{x}-\ve{x}^{\prime}\right|}\,f\left(t - \frac{\left|\ve{x}-\ve{x}^{\prime}\right|}{c}\,,\,\ve{x}^{\prime}\right),  
\nonumber\\ 
\label{Inverse_d_Alembert}
\end{eqnarray}

\noindent
for notational conventions see also text below Eq.~(3.4) in \cite{Blanchet_Damour1}. 
The inverse d'Alembert operator (\ref{Inverse_d_Alembert}) is standard in the  
literature \cite{Blanchet_Damour1,Blanchet_Damour2,Blanchet_Damour3,2PN_Metric1,2PN_Metric2,DSX1,DSX2,Book_Gravitational_Waves}.  
As explained in detail in the original work of the MPM formalism \cite{Blanchet_Damour1}, the integral in (\ref{Inverse_d_Alembert}) is 
not well-defined in general because, depending of the behavior of function $f$, the integral might become singular at $r^{\prime} \rightarrow 0$, 
where $r^{\prime} = \left|\ve{x}^{\prime}\right|$.  
As it has already been described in the original work of the MPM formalism \cite{Blanchet_Damour1}, the reason for this difficulty is caused
by the fact that the gothic metric density (\ref{gothic_metric_80}) - (\ref{gothic_metric_90}) as well as the metric tensor (\ref{metric_80}) - (\ref{metric_90}) 
are only valid in the exterior of the source of matter, while the integration of the inverse d'Alembertian
extends over the entire three-dimensional space, hence includes the inner region of the matter source, where the multipole decomposition
becomes infinite at the origin. This issue is of course a pure mathematical problem and not a physical one.

A way out of this problem is found by the fact that the limit $r^{\prime} \rightarrow 0$ is impossible because each real body is of finite size 
while the gothic metric and the metric tensor are strictly valid only in the exterior of the body. Therefore, the inverse d'Alembert operator 
is replaced by the Hadamard regularized inverse d'Alembert operator,   
\begin{eqnarray}
&& {\rm FP}_{B=0} \left(\square^{-1}_{\rm R} f \right)\left(t,\ve{x}\right) 
\nonumber\\
&=& - \lim_{B \rightarrow 0}\,\frac{1}{4\,\pi}  
\int \left(\frac{r^{\prime}}{r_0}\right)^B  
\frac{d^3 x^{\prime}}{\left|\ve{x}-\ve{x}^{\prime}\right|} 
f\left(t - \frac{\left|\ve{x}-\ve{x}^{\prime}\right|}{c}\,,\,\ve{x}^{\prime}\right),
\nonumber\\
\label{Inverse_d_Alembert_FP}
\end{eqnarray}

\noindent 
where a factor $\left(r^{\prime}/r_0\right)^B$ is imposed and where $B \in \mathbb{C}$ is some complex number and $r_0$ is an auxiliary real constant 
with the dimension of a length. The abbreviation FP denotes Hadamard's {\it partie finie} of the integral.  
If the real part of $B$, denoted by $\Re \left(B\right)$, is large enough,  
then all singularities at $r^{\prime} = 0$ are cancelled. So the procedure to determine the finite part (FP) consists of three consecutive steps:
\begin{enumerate} 
\item[] (i) computation of the integral (\ref{Inverse_d_Alembert_FP}) with sufficiently large real part of $B$,  
\item[] (ii) inserting the limits of integration,  
\item[] (iii) performing the limit $B \rightarrow 0$. 
\end{enumerate}

\noindent 
The final results of that procedure are equivalent to Hadamard's technique of {\it partie finie} \cite{Hadamard}; for   
mathematical rigor of Hadamard's procedure we refer to Section 3 in \cite{Blanchet_Damour1} and the article \cite{Hadamard_Blanchet_Faye}, 
where the approach has been described in specific detail.  

In many subsequent investigations of the MPM formalism that approach for determining the finite part has been 
applied \cite{Blanchet_Damour3,2PN_Metric1,2PN_Metric2,Blanchet4,Blanchet5,Blanchet6}.  
In particular, we refer to the important relations (4.24) in \cite{Blanchet_Damour3} or Eq.~(A.11) in \cite{2PN_Metric2},  
which are very useful in order to perform that integration procedure, where these effects occur in association with gravitational
radiation (tail effects, retardation effects, divergencies at spatial infinity, etc.) which have been
elaborated in \cite{Blanchet_Damour1,Blanchet_Damour2,2PN_Metric1,2PN_Metric2,Thorne,Book_Gravitational_Waves}.

Hadamard regularization leads to consistent results in different approaches up to 2.5PN order. Later it has been discovered that  
from the 3PN order on the Hadamard concept is not sufficient and the {\it gauge invariant dimensional regularization approach} is introduced.  
This approach and its implementation in the MPM formalism was a serious work over a longer period of time \cite{DR1,DR2,DR3,DR4}.  
In this investigation we are interested in the metric up to terms of the order ${\cal O}\left(c^{-6},c^{-5},c^{-6}\right)$ and will not consider that 
specific issue of the MPM formalism. But it should be kept in mind that for higher orders of the post-Minkowskian or post-Newtonian expansion  
the Hadamard concept has to be replaced by the {\it gauge invariant dimensional regularization}.

\section{Hadamard regularization of the inverse Laplacian}\label{Appendix6}

In this Section Hadamard's concept for the case of time-independent integrals will be considered in some more detail. 
In case of stationary sources, the 2PM metric perturbations in (\ref{canonical_metric_2PN_00}) and (\ref{canonical_metric_2PN_ij}) are 
associated with an inverse Laplace operator,   
\begin{eqnarray}
\left(\Delta^{-1} f \right)\left(\ve{x}\right) = - \frac{1}{4\,\pi} \int 
\frac{d^3 x^{\prime}}{\left|\ve{x}-\ve{x}^{\prime}\right|}\,f\left(\ve{x}^{\prime}\right),
\label{Inverse_Laplacian}
\end{eqnarray}

\noindent
where Hadamard's regularization of the inverse Laplacian is given by  
\begin{eqnarray}
&& {\rm FP}_{B=0} \left(\Delta^{-1} f \right)\left(\ve{x}\right) 
\nonumber\\ 
&=& - \lim_{B \rightarrow 0}\,\frac{1}{4\,\pi}
\int \left(\frac{r^{\prime}}{r_0}\right)^B \frac{d^3 x^{\prime}}{\left|\ve{x}-\ve{x}^{\prime}\right|}\,f\left(\ve{x}^{\prime}\right),  
\label{Inverse_Laplacian_FP}
\end{eqnarray}

\noindent 
where $r^{\prime} = \left|\ve{x}^{\prime}\right|$ and $r_0$ is an auxiliary real constant with the dimension of a length, 
and $B \in \mathbb{C}$ is some complex number.  
Because there is no time-dependence, the integration procedure in (\ref{Inverse_Laplacian_FP}) is considerably simpler than (\ref{Inverse_d_Alembert_FP}).  

According to the 2PM metric perturbations in (\ref{canonical_metric_2PN_00}) and (\ref{canonical_metric_2PN_ij}), we need to determine the following integral,  
\begin{eqnarray}
{\rm FP}_{B=0}\,\Delta^{-1}\;\frac{\hat{n}_L}{r^k}
= - \frac{1}{4\,\pi} \int \frac{d^3 x^{\prime}}{\left|\ve{x} - \ve{x}^{\prime}\right|}\,\left(\frac{r^{\prime}}{r_0}\right)^B \,
\frac{\hat{n}^{\prime}_L}{\left(r^{\prime}\right)^k}\;,
\nonumber\\ 
\label{Appendix_Integral_15}
\end{eqnarray}

\noindent
where the abbreviated notation $\hat{n}^{\prime}_L$ means   
\begin{equation}
\hat{n}^{\prime}_L \equiv \hat{n}^{\prime}_L\left(\varphi^{\prime},\vartheta^{\prime}\right) 
= \frac{x^{\prime}_{< i_1}\,x^{\prime}_{i_2}\, \dots x^{\prime}_{i_l >}}{\left(r^{\prime}\right)^l}\;.
\label{Appendix_Integral_10}
\end{equation}

\noindent
The integral (\ref{Appendix_Integral_15}) is for sufficiently large values of the real part $\Re\left(B\right)$ of the complex number $B$ well-defined.  
In order to determine that integral the following expansion of the denominator is used (cf. Eq.~(8.188) in \cite{Mathematical_Methods})  
\begin{equation}
\frac{1}{\left|\ve{x}-\ve{x}^{\prime}\right|}
= \begin{array}[c]{l}
\displaystyle
\left \{ \begin{array}[c]{l}
\displaystyle
\frac{1}{r}\,\sum\limits_{m=0}^{\infty} \;P_m\!\left(\cos \theta\right)\; \left(\frac{r^{\prime}}{r}\right)^m
\quad \mbox{if} \quad r > r^{\prime} \\
\\
\displaystyle
\frac{1}{r^{\prime}}\,\sum\limits_{m=0}^{\infty} \;P_m\!\left(\cos \theta\right)\; \left(\frac{r}{r^{\prime}}\right)^m
\quad \mbox{if} \quad r^{\prime} > r
\end{array} \right \} \;,
\end{array}
\label{Appendix_Integral_20}
\end{equation}

\noindent
where $P_m$ are the Legendre polynomials and $\theta$ is the angle between $\ve{x}$ and $\ve{x}^{\prime}$. 
By inserting (\ref{Appendix_Integral_20}) into (\ref{Appendix_Integral_15}) one encounters the following angular integration 
\begin{equation}
I = \int\limits_{0}^{2 \pi} d \varphi^{\prime} \int \limits_{0}^{\pi} d \vartheta^{\prime}\;\sin \vartheta^{\prime}\;
\hat{n}^{\prime}_L\left(\varphi^{\prime},\vartheta^{\prime}\right)\, 
P_m\!\left(\cos \theta\right)
\label{Appendix_Integral_21}
\end{equation}

\noindent 
which deserves special attention. The addition theorem for Legendre polynomial states (cf. Eq.~(8.189) in \cite{Mathematical_Methods})   
\begin{equation}
P_m\!\left(\cos \theta\right) = \frac{4 \pi}{2 m + 1} \sum\limits_{n=-m}^{m} Y_{mn}\left(\varphi,\vartheta\right) \; 
Y^{\ast}_{mn}\left(\varphi^{\prime},\vartheta^{\prime}\right),   
\label{Appendix_Integral_22a}
\end{equation}

\noindent
where $Y_{mn}$ and $Y^{\ast}_{mn}$ are the spherical harmonics and complex conjugated spherical harmonics, respectively. 
The spherical harmonics can be expanded in terms of the STF tensor in (\ref{Appendix_Cartesian_Tensor_n_L}), which reads  
(cf. Eq.~(2.11) in \cite{Thorne}) 
\begin{equation}
Y_{mn}\left(\varphi,\vartheta\right) = \hat{Y}^{mn}_M \; \hat{n}_M\left(\varphi,\vartheta\right),  
\label{Appendix_Integral_22b}
\end{equation}

\noindent 
where the coefficients $\hat{Y}^{mn}_M$ are independent of the angles $\varphi$ and $\vartheta$. Their explicit expressions are given by  
Eq.~(2.12) in \cite{Thorne} or by Eq.~(2.20) in \cite{Soffel_Hartmann}, but they are not needed here, because we use the following relation  
(cf. Eq.~(2.23) in \cite{Soffel_Hartmann}),  
\begin{equation}
\sum\limits_{n=-m}^{m} \hat{Y}^{mn}_M \;Y^{\ast}_{mn}\left(\varphi^{\prime},\vartheta^{\prime}\right) = \frac{\left(2 m + 1\right)!!}{4\,\pi\,m!}\;
\hat{n}^{\prime}_M\left(\varphi^{\prime},\vartheta^{\prime}\right)\;. 
\label{Appendix_Integral_22c}
\end{equation}

\noindent
By inserting (\ref{Appendix_Integral_22a}) - (\ref{Appendix_Integral_22c}) into (\ref{Appendix_Integral_21}) one obtains for the angular integration  
\begin{eqnarray} 
I &=& \frac{\left(2 m - 1\right)!!}{m!}\;\hat{n}_M\left(\varphi,\vartheta\right)  
\nonumber\\ 
&& \hspace{-0.5cm} \times \int\limits_{0}^{2 \pi} d \varphi^{\prime} \int \limits_{0}^{\pi} d \vartheta^{\prime}\;\sin \vartheta^{\prime}\;
\hat{n}^{\prime}_L\left(\varphi^{\prime},\vartheta^{\prime}\right)\;\hat{n}^{\prime}_M\left(\varphi^{\prime},\vartheta^{\prime}\right)\,. 
\label{Appendix_Integral_23a}
\end{eqnarray} 

\noindent
The angular integration (\ref{Appendix_Integral_23a}) yields (cf. Eq.~(2.5) in \cite{Thorne}) 
\begin{equation}
\int\limits_{0}^{2 \pi} d \varphi^{\prime} \! \int \limits_{0}^{\pi} d \vartheta^{\prime}\sin \vartheta^{\prime}
\hat{n}^{\prime}_L\left(\varphi^{\prime},\vartheta^{\prime}\right)\hat{n}^{\prime}_M\left(\varphi^{\prime},\vartheta^{\prime}\right) 
= \frac{4\,\pi\,m!}{\left(2 m + 1\right)!!}\,\delta_{lm} \,.
\label{Appendix_Integral_23b}
\end{equation}

\noindent 
Inserting (\ref{Appendix_Integral_23b}) into (\ref{Appendix_Integral_23a}) yields finally for the 
angular integration in (\ref{Appendix_Integral_21}) the following result (cf. Eq.~(B.3) in \cite{Dixon}),  
\begin{eqnarray}
I &=& \int\limits_{0}^{2 \pi} d \varphi^{\prime} \int \limits_{0}^{\pi} d \vartheta^{\prime}\;\sin \vartheta^{\prime} 
\hat{n}_L^{\prime}\left(\varphi^{\prime},\vartheta^{\prime}\right)
P_m\!\left(\cos \theta^{\prime}\right)
\nonumber\\ 
&=& \frac{4\,\pi}{2 m + 1} \,\hat{n}_L\, \delta_{lm}\,,
\label{Appendix_Integral_25}
\end{eqnarray}

\noindent 
where
\begin{equation}
\hat{n}_L \equiv \hat{n}_L\left(\varphi,\vartheta\right) = \frac{x_{< i_1}\,x_{i_2}\, \dots x_{i_l >}}{\left(r\right)^l}\;. 
\label{Appendix_Integral_11}
\end{equation}

\noindent 
It is important to realize that relation (\ref{Appendix_Integral_25}) necessitates the irreducible STF tensor $\hat{n}_L$ as integrand. If the
integrand would not be of irreducible STF structure, then relation (\ref{Appendix_Integral_25}) would not be valid. 
Accordingly, by means of (\ref{Appendix_Integral_20}) and owing to relation (\ref{Appendix_Integral_25}) one obtains for the integral (\ref{Appendix_Integral_15}),  
\begin{eqnarray}
&& \hspace{-1.0cm} {\rm FP}_{B=0}\,\Delta^{-1}\;\frac{\hat{n}_L}{r^k}
\nonumber\\ 
&& \hspace{-1.00cm} = - \frac{\hat{n}_L}{2 l + 1} \frac{1}{r}\; \int\limits_0^{r} d r^{\prime} \left(r^{\prime}\right)^2 \left(\frac{r^{\prime}}{r_0}\right)^B 
\left(\frac{r^{\prime}}{r}\right)^l \left(\frac{1}{r^{\prime}}\right)^k
\nonumber\\
&& \hspace{-0.6cm} - \, \frac{\hat{n}_L}{2 l + 1} \int\limits_{r}^{\infty} d r^{\prime} r^{\prime} 
\left(\frac{r^{\prime}}{r_0}\right)^B \left(\frac{r}{r^{\prime}}\right)^l \left(\frac{1}{r^{\prime}}\right)^k\,.
\label{Appendix_Integral_30}
\end{eqnarray}

\noindent
The radial integration yields
\begin{eqnarray}
&& \hspace{-1.0cm}{\rm FP}_{B=0}\,\Delta^{-1}\;\frac{\hat{n}_L}{r^k}
\nonumber\\
&& \hspace{-0.5cm} = - \frac{\hat{n}_L}{2 l + 1}  \left(\frac{1}{r_0}\right)^B \frac{1}{r^{l+1}}\!
\left[\frac{\left(r^{\prime}\right)^{B+l-k+3}}{B+l-k+3}\right]_{r^{\prime} = 0}^{r^{\prime} = r}
\nonumber\\
&& \hspace{-0.15cm} - \, \frac{\hat{n}_L}{2 l + 1}  \left(\frac{1}{r_0}\right)^B \;r^{l}\;\,
\left[\frac{\left(r^{\prime}\right)^{B-l-k+2}}{B-l-k+2}\right]_{r^{\prime} = r}^{r^{\prime} = \infty}\;.
\label{Appendix_Integral_35}
\end{eqnarray}

\noindent
For $\Re \left(B\right) + l - k + 3 > 0 > \Re \left(B\right) - l - k + 2$, the lower integration constant $r^{\prime} = 0$ in the first line and
the upper integration constant $r^{\prime} = \infty$ in the second line do not contribute and one arrives at
\begin{eqnarray}
&& {\rm FP}_{B=0}\,\Delta^{-1}\;\frac{\hat{n}_L}{r^k} 
\nonumber\\ 
&& = \frac{\hat{n}_L}{2 l + 1} \left(\!\frac{r}{r_0}\right)^B
r^{2 - k}\left(\frac{1}{B - l - k + 2} - \frac{1}{B + l - k + 3}\!\right).
\nonumber\\
\label{Appendix_Integral_40}
\end{eqnarray}

\noindent
The limit $B \rightarrow 0$ yields finally
\begin{eqnarray}
&& \hspace{-0.75cm} {\rm FP}_{B=0}\,\Delta^{-1}\,\frac{\hat{n}_L}{r^k}
= \frac{1}{\left(k + l - 2\right) \left(k - l - 3\right)} \frac{\hat{n}_L}{r^{k - 2}}\,,  
\label{Appendix_Integral_45}
\end{eqnarray}

\noindent
which is meaningful for $k\ge 3$ as well as $k \neq l + 3$; note that always $l \ge 0$. The solution of the integral in (\ref{Appendix_Integral_45}) is  
a specific case of the integrals given by Eqs.~(A.11) and (A.16) in \cite{2PN_Metric2}, respectively, and has been presented within several investigations,  
for instance by Eq.~(3.9) in \cite{Blanchet_Damour1} and by Eq.~(3.9) in \cite{Dixon} and by Eq.~(3.42) in \cite{Dissertation_Kokcu}.   

One might wonder about the global sign of the solution (\ref{Appendix_Integral_45}).  
For instance, if one considers the case $l = 0$ and $k \ge 4$, then (\ref{Appendix_Integral_45}) is a
positive-valued expression, irrespective of the negative-valued integral in (\ref{Appendix_Integral_15}). In order to understand the global sign
in (\ref{Appendix_Integral_45}), one has to realize that the {\it partie finie} procedure in (\ref{Appendix_Integral_15}) implies that the
lower integration constant $r^{\prime} = 0$ in the first line in (\ref{Appendix_Integral_35}) does not contribute. Stated differently,
in case of $l = 0$ and $k \ge 4$ the {\it partie finie} procedure eliminates a (infinitely) large negative term from the entire expression,
so that the final result becomes positive-valued in the case under consideration.

\section{The proof of Eqs.~(\ref{metric_2PM_00})}\label{Appendix7}  

In this Appendix some details of the computation of the matrix coefficients in Eqs.~(\ref{metric_2PM_00}) are given.    
Accounting for monopole and quadrupole, one gets from Eqs.~(\ref{canonical_metric_2PN_00})   
\begin{eqnarray}
h^{\left({\rm 2PM}\right)}_{00\,{\rm can}}\!\left(\ve{x}\right) = - \frac{2}{c^4}  
\left(\!\frac{M}{r} + \frac{\partial_{ab}}{2!} \frac{M_{ab}}{r}\!\right)^2 + {\cal O}\left(c^{-6}\right).  
\label{Appendix_metric_2PM_00_1}
\end{eqnarray}

\noindent 
By means of (\ref{Appendix1_Differentiation_A}) one obtains  
\begin{eqnarray}
&& \hspace{-1.25cm} h^{\left({\rm 2PM}\right)}_{00\,{\rm can}}\!\left(\ve{x}\right) = 
- \frac{2}{c^4}\left(\frac{M^2}{r^2} + 3\,M\,M_{ab}\,\frac{\hat{n}_{ab}}{r^4}\right)
\nonumber\\
&& \hspace{0.75cm} - \frac{2}{c^4}\left(\frac{9}{4} M_{ab}\,M_{cd} \frac{\hat{n}_{ab} \hat{n}_{cd}}{r^6}\right)  
+ {\cal O}\left(c^{-6}\right).  
\label{Appendix_metric_2PM_00_2}
\end{eqnarray}

\noindent 
The last term is rewritten in the form $M_{ab}\,M_{cd}\,\hat{n}_{ab}\,\hat{n}_{cd} = M_{ab}\,M_{cd}\,n_{abcd}$, then relations (\ref{Appendix1_STF4}) 
and (\ref{Appendix1_STF2}) are applied; note $M_{ab}\,\delta_{ab} = M_{cd}\,\delta_{cd} = 0$. One arrives at  
\begin{eqnarray}
h^{\left({\rm 2PM}\right)}_{00\,{\rm can}}\!\left(\ve{x}\right) &=& - \frac{2}{c^4}\,\frac{M^2}{r^2} 
- \frac{6}{c^4}\,M\,M_{ab}\, \frac{\hat{n}_{ab}}{r^4} 
\nonumber\\
\nonumber\\
&& \hspace{-1.0cm} - \,\frac{3}{5}\,\frac{1}{c^4}\,M_{ab}\,M_{cd}\,\frac{\delta_{ac} \delta_{bd}}{r^6}  
- \frac{18}{7}\,\frac{1}{c^4}\,M_{ab}\,M_{cd}\,\frac{\delta_{ac} \hat{n}_{bd}}{r^6} 
\nonumber\\
&& \hspace{-1.0cm} - \,\frac{9}{2}\,\frac{1}{c^4}\,M_{ab}\,M_{cd}\,\frac{\hat{n}_{abcd}}{r^6} + {\cal O}\left(c^{-6}\right).  
\label{Appendix_metric_2PM_00_5}
\end{eqnarray}

\section{The proof of Eqs.~(\ref{metric_2PM_ij})}\label{Appendix8} 

In this Appendix some details of the computation of the matrix coefficients in Eqs.~(\ref{metric_2PM_ij}) are given.                          
Accounting for the monopole and quadrupole term, one obtains from Eqs.~(\ref{canonical_metric_2PN_ij})                                  
\begin{eqnarray}
h^{\left({\rm 2PM}\right)}_{ij\,{\rm can}}\!\left(\ve{x}\right) &=& 
+ \frac{2}{c^4}\,\delta_{ij}
\left(\frac{M}{r} + \frac{\partial_{ab}}{2!} \frac{M_{ab}}{r}\right)^2
\nonumber\\ 
&& \hspace{-1.75cm} - \,\frac{4}{c^4} \,{\rm FP}_{B=0}\,\Delta^{-1}\!
\left(\!\partial_i \frac{M}{r} + \frac{\partial_{iab}}{2!} \frac{M_{ab}}{r}\!\right)\! 
\left(\!\partial_j \frac{M}{r} + \frac{\partial_{jab}}{2!} \frac{M_{ab}}{r}\!\right) 
\nonumber\\ 
&& \hspace{-1.75cm} + \, {\cal O}\left(c^{-6}\right)\!.
\label{Appendix_metric_2PM_ij_1}
\end{eqnarray}

\noindent  
By means of (\ref{Appendix1_Differentiation_A}) one obtains  
\begin{widetext}
\begin{eqnarray}
 h^{\left({\rm 2PM}\right)}_{ij\,{\rm can}}\!\left(\ve{x}\right) &=& 
+ \frac{2}{c^4}\,\delta_{ij} \left(\frac{M^2}{r^2} + 3\,M\,M_{ab}\,\frac{\hat{n}_{ab}}{r^4}
+ \frac{9}{4}\,M_{ab}\,M_{cd}\,\frac{\hat{n}_{ab}\,\hat{n}_{cd}}{r^6} \right)
\nonumber\\
\nonumber\\
&& - \,\frac{4}{c^4}\,{\rm FP}_{B=0}\,\Delta^{-1}
\left(M^2\,\frac{n_{ij}}{r^4} + 15\,M\,M_{ab}\,\frac{n_{(i}\,\hat{n}_{j) ab}}{r^6}
+ \frac{225}{4}\,M_{ab}\,M_{cd}\, \frac{\hat{n}_{iab}\,\hat{n}_{jcd}}{r^8} \right) + {\cal O}\left(c^{-6}\right).
\label{Appendix_metric_2PM_ij_2}
\end{eqnarray}
\end{widetext}

\noindent 
Before relation (\ref{Appendix_Integral_45}) can be applied, one has to express the enumerator in the second line of (\ref{Appendix_metric_2PM_ij_2})  
in terms of irreducible STF tensors. The first term of the second line of (\ref{Appendix_metric_2PM_ij_2}) is rewritten in terms of  
irreducible STF tensors by means of relation (\ref{Appendix1_STF2}), 
\begin{eqnarray}
n_{ij} = \hat{n}_{ij} + \frac{1}{3}\,\delta_{ij}\;, 
\label{Appendix_metric_2PM_ij_3a}
\end{eqnarray}

\noindent
while for the second term of the second line of (\ref{Appendix_metric_2PM_ij_2}) one obtains
\begin{eqnarray}
M_{ab}\,n_{(i}\,\hat{n}_{j) ab} = M_{ab} \left(\hat{n}_{ijab} + \frac{1}{7}\,\delta_{ij}\,\hat{n}_{ab} 
+ \frac{6}{35}\,\delta_{a\,(i} \hat{n}_{j\,) b} \right),  
\nonumber\\ 
\label{Appendix_metric_2PM_ij_3}
\end{eqnarray}

\noindent
where relation (\ref{Appendix1_Transformation_STF_4}) and the STF structure of the multipole $M_{ab}$ has been used. The last term in the second line  
of (\ref{Appendix_metric_2PM_ij_2}) is expressed in terms of irreducible STF tensors by means of relations (\ref{Appendix1_STF3}) and 
(\ref{Appendix1_Transformation_STF_4}) as well as (\ref{appendix_T2_30}) and (\ref{appendix_T2_35}). After some steps one obtains  
the following expression in terms of irreducible STF tensors,  
\begin{eqnarray}
&& M_{ab}\,M_{cd}\,\hat{n}_{iab}\,\hat{n}_{jcd}   
\nonumber\\ 
\nonumber\\ 
&& = M_{ab}\,M_{cd} \left[\hat{n}_{ijabcd} + \frac{8}{11}\,\hat{n}_{a < ijc}\,\delta_{d > b} 
+ \frac{12}{63}\,\hat{n}_{< ij} \,\delta^a_{c}\,\delta^b_{d >}\right]
\nonumber\\
\nonumber\\
 && + \frac{1}{7}\,M_{ab}\,M_{cd}\,\delta_{ij}
\left[\hat{n}_{abcd} + \frac{4}{7} \,\hat{n}_{a < c}\,\delta_{d > b} 
+ \frac{2}{15}\,\delta_{< c}^{\;\;\,a}\,\delta_{d >}^b\right]
\nonumber\\
\nonumber\\
&& + \frac{2}{7}\,M_{ab}\,M_{cd}\,\delta_{ic}
\left[\hat{n}_{abjd} + \frac{4}{7} \,\hat{n}_{a < j}\,\delta_{d > b} 
+ \frac{2}{15}\,\delta_{< j}^{\;\;\,a}\,\delta_{d >}^b\right]
\nonumber\\
\nonumber\\
&& - \frac{4}{35}\,M_{ab}\,M_{cd}\,\delta_{jc}
\left[\hat{n}_{abid} + \frac{4}{7} \, \hat{n}_{a < i}\,\delta_{d > b} 
+ \frac{2}{15}\,\delta_{< i}^{\;\;\,a}\,\delta_{d >}^b\right]
\nonumber\\
\nonumber\\
&& - \frac{2}{5} M_{ab} M_{cd} \delta_{ib}\!\left[\hat{n}_{jacd} + \frac{1}{7}\hat{n}_{cd} \delta_{aj}
+ \frac{2}{7} \hat{n}_{jc} \delta_{ad} - \frac{4}{35} \hat{n}_{ac} \delta_{jd}\right]\! .
\nonumber\\ 
\label{Appendix_metric_2PM_ij_4}
\end{eqnarray}

\noindent
Taking account for the STF structure of the quadrupoles, one may combine the second term and the fourth term of the last line,  
but here we keep these terms as given. 
The r.h.s. of Eq.~(\ref{Appendix_metric_2PM_ij_4}) has now been expressed in terms of irreducible STF tensors. But the structure of these terms is  
presented in a rather compact notification. A more explicit form is arrived with the aid of relations (\ref{STF_Tensor_1}) and  
(\ref{STF_Tensor_3}), by means of which one obtains  
\begin{eqnarray}
&& \hspace{-0.5cm} M_{ab}\;M_{cd}\;\hat{n}_{a < ijc} \delta_{d > b} = + \frac{1}{2}\,M_{ab}\;M_{cd}\,\hat{n}_{acij}\,\delta_{bd}
\nonumber\\
&& \hspace{-0.5cm} + \frac{3}{14}\,M_{ab}\;M_{cd}\,\hat{n}_{acd\,(i}\,\delta_{j\,)\,b}
- \frac{1}{14}\,M_{ab}\;M_{cd}\,\hat{n}_{abcd} \,\delta_{ij}\;,
\label{appendix_metric_160}
\end{eqnarray}

\begin{eqnarray}
&& \hspace{-0.5cm} M_{ab}\,M_{cd}\;\hat{n}_{< ij} \delta^a_{c} \delta^b_{d >} = 
+ \frac{1}{6}\,M_{ab}\,M_{cd}\,\hat{n}_{ij}\,\delta_{ac}\,\delta_{bd}
\nonumber\\
&& \hspace{-0.5cm} + \frac{10}{21}\,M_{ab}\,M_{cd}\,\hat{n}_{c\,(i}\,\delta_{j\,)\,a}\, \delta_{bd}
+ \frac{43}{210}\,M_{ab}\,M_{cd}\,\hat{n}_{ab}\,\delta_{ic}\,\delta_{jd}
\nonumber\\
&& \hspace{-0.5cm} - \frac{2}{21}\,M_{ab}\,M_{cd}\,\delta_{ij}\,\delta_{bd}\,\hat{n}_{ac}
- \frac{4}{21}\,M_{ab}\,M_{cd}\,\delta_{ic}\,\delta_{bj}\,\hat{n}_{ad}\;,
\label{appendix_metric_165}
\end{eqnarray}

\begin{eqnarray}
&& \hspace{-0.5cm} M_{ab}\;M_{cd}\;\hat{n}_{a < i}\,\delta_{d > b} = + \frac{1}{2}\,M_{ab}\;M_{cd}\,\hat{n}_{ai}\,\delta_{db}  
\nonumber\\
&& \hspace{-0.5cm} + \frac{1}{2}\,M_{ab}\;M_{cd}\,\hat{n}_{ad}\,\delta_{bi} 
- \frac{1}{3}\,M_{ab}\;M_{cd}\,\hat{n}_{ab}\,\delta_{di} \,,
\label{appendix_metric_190}
\end{eqnarray}

\begin{eqnarray}
M_{ab}\;M_{cd}\;\delta_{< i}^{\;\;\,a}\,\delta_{d >}^b &=& + M_{ab}\;M_{cd}\;\delta_{ad}\;\delta_{bi}\;.
\label{appendix_metric_195}
\end{eqnarray}

\noindent 
By inserting (\ref{Appendix_metric_2PM_ij_3a}) - (\ref{Appendix_metric_2PM_ij_4}) into (\ref{Appendix_metric_2PM_ij_2}) 
by taking into account the relations (\ref{appendix_metric_160}) - (\ref{appendix_metric_195}) as well as  
the solution for the integrals (\ref{Appendix_Integral_45}), one finally arrives at 
\begin{widetext}
\begin{eqnarray}
h^{\left({\rm 2PM}\right)}_{ij\,{\rm can}}\!\left(\ve{x}\right) &=& \frac{1}{c^4}\,\frac{M^2}{r^2}\left(\frac{4}{3}\,\delta_{ij} + \hat{n}_{ij}\right)  
\nonumber\\ 
\nonumber\\ 
&& + \frac{15}{2}\,\frac{1}{c^4}\,M\,M_{ab}\,\frac{\hat{n}_{ijab}}{r^4} 
+ \frac{32}{7}\,\frac{\delta_{ij}}{c^4}\,M\,M_{ab}\,\frac{\hat{n}_{ab}}{r^4} 
- \frac{12}{7}\,M\,M_{ab}\,\frac{\delta_{a\,(i} \hat{n}_{j\,) b}}{r^4}  
\nonumber\\
\nonumber\\
&& + \frac{M_{ab}\,M_{cd}}{c^4\,r^6}
\bigg(\frac{75}{4}\,\hat{n}_{ijabcd} - \frac{90}{11}\,\hat{n}_{ijac}\,\delta_{bd} + \frac{27}{11}\,\hat{n}_{abcd}\,\delta_{ij}
- \frac{25}{84}\,\hat{n}_{ij}\,\delta_{ac}\,\delta_{bd}
\nonumber\\
\nonumber\\
&& + \frac{83}{42}\,\hat{n}_{ad}\,\delta_{bc}\,\delta_{ij}
 + \frac{16}{35}\,\delta_{ac}\,\delta_{bd}\,\delta_{ij} + \frac{18}{11}\,\hat{n}_{acd(i}\,\delta_{j)b}
- \frac{5}{21}\,\hat{n}_{a(i}\,\delta_{j)c}\,\delta_{bd}
\nonumber\\
\nonumber\\
&& + \frac{10}{21}\,\delta_{ci}\,\delta_{dj}\,\hat{n}_{ab}
- \frac{23}{42}\,\delta_{b(i}\,\delta_{j)c}\,\hat{n}_{ad} - \frac{6}{35}\,\delta_{ad}\,\delta_{b(i}\,\delta_{j)c} \bigg) + {\cal O}\left(c^{-6}\right). 
\label{Appendix_metric_2PM_ij_10}
\end{eqnarray}
\end{widetext}

\noindent
Several careful checks have been performed in order to be certain about the correctness of these metric coefficients.  
For instance, one may see that the terms proportional to $M^2$ are in agreement with the same terms of Eq.~(25) in \cite{Article_Zschocke1}. 
Furthermore, it has been checked that inserting the gothic metric coefficients (\ref{gothic_metric_stationary_1PM_00}) - (\ref{gothic_metric_stationary_2PM_ij})
in Appendix \ref{Appendix9} into (\ref{Retartion_Gothic_Metric_2}) yields the same metric coefficients as presented by Eqs.~(\ref{metric_2PM_00}) - (\ref{metric_2PM_ij}). 
In addition, each metric coefficient has been determined in different ways and assisted by the computer algebra system {\it Maple} \cite{Maple}.

\section{Monopole and spin and quadrupole terms of 2PM gothic metric for stationary sources}\label{Appendix9}

In case of stationary source the post-linear gothic metric (\ref{expansion_metric_2}) simplifies as follows, 
\begin{eqnarray}
\overline{g}^{\alpha\beta}\left(\ve{x}\right) \!=\! \eta^{\alpha\beta}
- G^1 \overline{h}_{\left({\rm 1PM}\right)}^{\alpha \beta}\left(\!\ve{x}\!\right)
- G^2 \overline{h}_{\left({\rm 2PM}\right)}^{\alpha \beta}\left(\!\ve{x}\!\right) + {\cal O}\left(G^3\right). 
\nonumber\\ 
\label{gothic_metric_stationary_5a}
\end{eqnarray}

\noindent
The gauge transformation (\ref{harmonic_gauge_condition_7}) leads, up to terms of the order ${\cal O}\left(G^3\right)$, to  
\begin{eqnarray}
&& \hspace{-0.5cm} \overline{g}^{\alpha\beta}\left(\ve{x}\right) = \eta^{\alpha\beta}
- G^1 \overline{h}_{\left({\rm 1PM}\right)}^{\alpha \beta {\rm can}}\left(\ve{x}\right) 
- \partial \overline{\varphi}_{\left({\rm 1PM}\right)}^{\alpha\beta}\left(\ve{x}\right)  
\nonumber\\ 
&& - G^2 \overline{h}_{\left({\rm 2PM}\right)}^{\alpha \beta {\rm can}}\left(\ve{x}\right) 
- \partial \overline{\varphi}_{\left({\rm 2PM}\right)}^{\alpha\beta}\left(\ve{x}\right)  
- \overline{\Omega}_{\left({\rm 2PM}\right)}^{\alpha\beta}\left(\ve{x}\right),  
\label{gothic_metric_stationary_5b}
\end{eqnarray}

\noindent
where the gauge terms are time-independent and formally given by Eqs.~(\ref{general_solution_C1}) and Eqs.~(\ref{general_solution_C2}) and (\ref{Gauge_2PM_B}), 
respectively. Accounting for the monopole and spin and quadrupole terms, one arrives at  
\begin{eqnarray}
\overline{h}_{\left({\rm 1PM}\right)}^{00\,{\rm can}} &=& 4\,\frac{M}{c^2 r} + 6\,\frac{M_{ab}}{c^2\,r^3}\,\hat{n}_{ab}\,,  
\label{gothic_metric_stationary_1PM_00}
\\
\nonumber\\
\overline{h}_{\left({\rm 1PM}\right)}^{0i\,{\rm can}} &=& - \frac{2}{c^3}\,\epsilon_{iab}\,n_a\,\frac{S_b}{r^2}\,,  
\label{gothic_metric_stationary_1PM_0i}
\\
\nonumber\\
\overline{h}_{\left({\rm 1PM}\right)}^{ij\,{\rm can}} &=& 0\,,  
\label{gothic_metric_stationary_1PM_ij}
\end{eqnarray}

\noindent 
for the linear coefficients, and 
\begin{widetext}
\begin{eqnarray}
\overline{h}_{\left({\rm 2PM}\right)}^{00\,{\rm can}} &=& 7\,\frac{M^2}{c^4 r^2} + 21\,\frac{M\,M_{ab}}{c^4\,r^4}\,\hat{n}_{ab} 
+ \frac{63}{4}\,\frac{M_{ab}\,M_{cd}}{c^4\,r^6}\,n_{abcd} + {\cal O}\left(c^{-6}\right),  
\label{gothic_metric_stationary_2PM_00}
\\
\nonumber\\
 \overline{h}_{\left({\rm 2PM}\right)}^{0i\,{\rm can}} &=& {\cal O}\left(c^{-5}\right),  
\label{gothic_metric_stationary_2PM_0i}
\\
\nonumber\\
 \overline{h}_{\left({\rm 2PM}\right)}^{ij\,{\rm can}} &=& \frac{M^2}{c^4}\,\frac{1}{r^2}\,n_{ij} 
+ \frac{M\,M_{ab}}{c^4\,r^4}
\left(\frac{15}{2}\,n_{ijab} + \frac{1}{2}\,\delta_{ij}\,n_{ab} - 6\,n_{a (i}\delta_{j)b} + \delta_{ai}\,\delta_{bj}\right) 
\nonumber\\ 
\nonumber\\ 
&& + \frac{M_{ab}\,M_{cd}}{c^4\,r^6}
\bigg(\frac{75}{4}\,\hat{n}_{ijabcd} - \frac{90}{11}\,\hat{n}_{ijac}\,\delta_{bd} + \frac{9}{44}\,\hat{n}_{abcd}\,\delta_{ij}
- \frac{25}{84}\,\hat{n}_{ij}\,\delta_{ac}\,\delta_{bd}
\nonumber\\
\nonumber\\
&& + \frac{29}{42}\,\hat{n}_{ad}\,\delta_{bc}\,\delta_{ij}
 + \frac{11}{70}\,\delta_{ac}\,\delta_{bd}\,\delta_{ij} + \frac{18}{11}\,\hat{n}_{acd(i}\,\delta_{j)b}
- \frac{5}{21}\,\hat{n}_{a(i}\,\delta_{j)c}\,\delta_{bd}
\nonumber\\
\nonumber\\
&& + \frac{10}{21}\,\delta_{ci}\,\delta_{dj}\,\hat{n}_{ab}
- \frac{23}{42}\,\delta_{b(i}\,\delta_{j)c}\,\hat{n}_{ad} - \frac{6}{35}\,\delta_{ad}\,\delta_{b(i}\,\delta_{j)c} \bigg) + {\cal O}\left(c^{-6}\right),   
\label{gothic_metric_stationary_2PM_ij}
\end{eqnarray}
\end{widetext}

\noindent
for the post-linear coefficients. These gothic metric coefficients in (\ref{gothic_metric_stationary_1PM_00}) - (\ref{gothic_metric_stationary_2PM_ij}) 
have been calculated by the same approach as presented in the previous Appendix \ref{Appendix8}; the last term in (\ref{gothic_metric_stationary_2PM_00}) and 
the first line in (\ref{gothic_metric_stationary_2PM_ij}) are not expressed in terms of irreducible STF multipoles, but it could be done by means of 
relations (\ref{Appendix1_STF2}) and (\ref{Appendix1_STF4}).  

The quadrupole-quadrupole gothic metric density for a time-dependent compact source of matter has been determined in \cite{2PN_Metric2} which allows  
to deduce the gothic metric coefficients (\ref{gothic_metric_stationary_1PM_00}) - (\ref{gothic_metric_stationary_2PM_ij}).  
Furthermore, it should be mentioned that these gothic metric coefficients (\ref{gothic_metric_stationary_1PM_00}) - (\ref{gothic_metric_stationary_2PM_ij})
have also been presented by Eq.~(16) in \cite{Frutos_Soffel1}; the incorrect coefficient $z_0^6$ of Eq.~(16) in \cite{Frutos_Soffel1}  
has later been corrected by Eq.~(21) in \cite{Frutos_Soffel2}.


\begin{thebibliography}{99}
\bibitem{Footnote1}
The contracted Bianchi identities for the Einstein tensor on the l.h.s. in (\ref{exact_field_equations}) implies the local energy-momentum
conservation $T^{\alpha\beta}_{\;\;\;;\,\beta} = 0$ on the r.h.s. in (\ref{exact_field_equations}).
This situation is similar to the classical electrodynamics in flat space-time, where the four field equations
$\displaystyle \square\,A_{\mu} - A^{\nu}_{\;\;,\,\mu\nu} = - \frac{4\,\pi}{c}\,J_{\mu}\;\left({\rm ED1}\right)$ for the four-potential $A_{\mu}$ are reduced
to only three independent field equations because of the conservation law $J^{\mu}_{\;\;,\,\mu} = 0$ of the electromagnetic four-current $J_{\mu}$
which reflects the conservation of the electric charge.
\bibitem{Footnote3}
The harmonic gauge (\ref{alternative_gauge_condition}) which permits a residual gauge transformation (\ref{harmonic_gauge_condition_2}) if the gauge
functions obey the homogeneous Laplace-Beltrami equation (\ref{Laplace_Beltrami_Equation}) resembles the Lorenz gauge of the four-potentials in
classical electrodynamics, $\;A^{\mu}_{\;\;,\,\mu}\left(x\right) = 0\;$, which permits a residual gauge transformation of the four-potentials,
$\;A^{\prime\,\alpha}\left(x\right) = A^{\alpha}\left(x\right) + \xi^{\alpha}\left(x\right)\;$, if the gauge functions $\xi^{\alpha}$ obey
the homogeneous d'Alembert equation $\eta^{\mu\nu}\,\xi^{\alpha}_{\;\;,\,\mu\nu}\left(x\right) = 0$ \cite{Jackson}.
\bibitem{Footnote5}
The retarded time $u$ in Eq.~(\ref{Retarded_Time_2}) is denoted by $s$ in Eq.~(3.526) in \cite{Kopeikin_Efroimsky_Kaplan}. And vice
versa, the retarded time $s_{\rm gen}$ in Eq.~(\ref{Retarded_Time_3}) is denoted by $u$ in the text below Eq.~(3.534) in \cite{Kopeikin_Efroimsky_Kaplan}. This
opposite kind of notation for the retarded time in Eqs.~(\ref{Retarded_Time_2}) and (\ref{Retarded_Time_3}) is frequently used in the theory of light propagation
and will certainly not cause any kind of confusion here.
\bibitem{Footnote6}
Just identify the general gothic metric perturbations $\overline{h}_{\left({\rm nPM}\right)}^{\alpha\beta\,{\rm gen}}$
and the canonical gothic metric perturbations $\overline{h}_{\left({\rm nPM}\right)}^{\alpha\beta\,{\rm can}}$ in (\ref{general_solution_D})
with the old gothic metric perturbations $\overline{h}_{\left({\rm nPM}\right)}^{\alpha\beta}$ and the new gothic metric perturbations
$\overline{h}_{\left({\rm nPM}\right)}^{\,\prime\,\alpha\beta}$ in (\ref{gauge_transformation_gothic_metric_PM}), respectively.
\bibitem{Footnote7}
Just identify the general metric perturbations $h^{\left({\rm nPM}\right)}_{\alpha\beta\,{\rm gen}}$
and the canonical metric perturbations $h^{\left({\rm nPM}\right)}_{\alpha\beta\,{\rm can}}$ in (\ref{general_solution_metric_PM})
with the old metric perturbations $h^{\left({\rm nPM}\right)}_{\alpha\beta}$ and the new metric perturbations
$h^{\prime\,\left({\rm nPM}\right)}_{\alpha\beta}$ in (\ref{gauge_transformation_metric_PM}), respectively.
\bibitem{Footnote8}
In the last line of (\ref{gothic_metric_90}) and in (\ref{Integral_Relation_1}) the shorthand notation
$\left(\partial_k f\right)^2 = \sum\limits_{k=1}^3 \left(\partial_k f\right) \left(\partial_k f\right)$ is used.
The relation (\ref{Integral_Relation_1}) can be shown by means of
$\left(\partial_k f\right)^2 = \frac{1}{2}\,\Delta f^2 - f \Delta f$, where $\Delta = \partial_k\,\partial_k$ is
the flat Laplace operator, and then using ${\rm FP}_{B=0}\,\square_{\rm R}^{-1} = {\rm FP}_{B=0}\,\Delta^{-1} + {\cal O}\left(c^{-2}\right)$
(cf. Eq.~(3.11) in \cite{Blanchet_Damour1}), where ${\rm FP}_{B=0}\,\Delta^{-1}$ is the Hadamard regularized inverse Laplace operator defined by
Eq.~(\ref{Inverse_Laplacian_FP}); note that $\Delta f$ yields terms of the order ${\cal O}\left(c^{-6}\right)$ in the metric in the exterior of
the source of matter.
\bibitem{Footnote9}
The notation ${\cal O}\left(c^{-6}, c^{-5}, c^{-6}\right)$ means that the time-time-components, space-time components and space-space-components
of the gothic metric perturbation (or metric perturbation) are of the order ${\cal O}\left(c^{-6}\right)$, ${\cal O}\left(c^{-5}\right)$,
and ${\cal O}\left(c^{-6}\right)$, respectively.
\end{thebibliography}
\end{document}